\documentclass[a4paper,12pt]{article}
\usepackage[colorlinks,pdfpagelabels,pdfstartview=FitH,bookmarksopen=true,bookmarksnumbered=true,linkcolor=blue,plainpages=false,hypertexnames=false,citecolor=blue, urlcolor=blue]{hyperref}
\usepackage{amsfonts}
\usepackage{amsmath}
\usepackage{graphicx}
\usepackage{amssymb}
\usepackage{longtable}
\usepackage{authblk}
\usepackage{pdflscape}
\usepackage{rotating}
\usepackage[round]{natbib}
\usepackage{booktabs}
\usepackage[margin=1in]{geometry}
\usepackage{setspace}
\usepackage{chngcntr}
\usepackage{verbatim}
\usepackage{multirow}
\usepackage{subcaption}
\usepackage{chngcntr}
\usepackage[title]{appendix}
\usepackage[gen]{eurosym}
\usepackage{comment}
\usepackage[bottom]{footmisc}
\usepackage{gensymb}
\usepackage{textcomp,mathcomp}
\usepackage{wrapfig}
\usepackage{pdflscape}
\usepackage{rotating}
\usepackage{epstopdf}
\usepackage{graphicx}
\usepackage{caption}
\usepackage{subcaption}
\usepackage{longtable,tabularx,ltxtable,ragged2e}
\usepackage{array}
\usepackage{pdflscape}
\usepackage{siunitx}
\begin{document}
\pagenumbering{arabic}
\author[1] {Dario Caldara}
\author[2] {Haroon Mumtaz}
\author[1] {Molin Zhong}
\affil[1]{Board of Governors of the Federal Reserve System}
\affil[2]{School of Economics and Finance, Queen Mary University of London}
\title{ Risk in a Data-Rich Model\footnote{Dario Caldara (\href{mailto:dario.caldara@frb.gov}{dario.caldara@frb.gov}); Haroon Mumtaz (\href{mailto:h.mumtaz@qmul.ac.uk}{h.mumtaz@qmul.ac.uk}); Molin Zhong (\href{molin.zhong@frb.gov}{molin.zhong@frb.gov}). We thank for helpful comments Todd Clark, Domenico Giannone, Francesca Loria, Matteo Luciani, Michele Modugno, seminar participants at the Adam Smith Business School, the Federal Reserve Board, 2024  BSE Summer Forum on ``Advances in Structural Shocks Identification,'' 2024 workshop on ``The Economics of Risk'' Turin, 2024 Kansas Econometrics Workshop, 2024 So.Fi.E. workshop on ``Monitoring and Forecasting Macroeconomic and Financial Risk'', 5th DC-VA-MD Econometrics Workshop, 2024 System Econometrics Conference, 2025 Nowcasting Workshop at PSE, 2025 IAAE Annual Conference,  2025 NBER-NSF Time Series, and 3rd NLMacro Workshop, Lancaster University. We also would like to thank Lilliana Wells, Sofie Grouws, and Charlotte Singer for outstanding research assistance. The views expressed in this paper are solely the responsibility of the authors and should not be interpreted as reflecting the views of the Board of Governors of the Federal Reserve System or of anyone else associated with the Federal Reserve System. }}

\maketitle

\begin{center}

\end{center}
\vspace{-2.5em}
\begin{abstract}

We characterize asymmetric tail risk across over one hundred U.S. macroeconomic and financial variables using a dynamic factor model with stochastic volatility. A single mechanism unifies growth-at-risk, inflation-at-risk, and sectoral risk heterogeneity: common factors and their volatilities move together, while heterogeneous loadings transmit the resulting asymmetry unevenly across variables. We find that asymmetric tail risk is pervasive but heterogeneous. The heterogeneity is systematic: factor exposures, especially to financial conditions and inflation, explain over half of the cross-sectional variation in tail asymmetry across variables. These exposures determine where in the economy vulnerabilities concentrate and how the balance of tail risks shifts over time.

\vspace{0.25cm}

\end{abstract}

\clearpage
\pagenumbering{arabic} 
\setcounter{page}{2}

\section{\label{intro}Introduction}

Some sectors of the economy contract sharply during recessions while others remain stable. During the 2008 financial crisis, for instance, motor vehicle production and durable goods consumption plummeted, while electricity generation declined only marginally and food consumption remained steady. All sectors experienced the same macroeconomic environment: the same aggregate shocks, the same monetary policy response, the same aggregate financial market stress. Yet their tail risk profiles differed dramatically.\footnote{Throughout, we characterize risk by the tails of a variable's conditional predictive distribution: downside risk is its 5th percentile and upside risk its 95th. Tail risk is asymmetric when one tail varies more over time than the other---for instance, when the lower tail of industrial production swings widely across the business cycle while its upper tail stays comparatively stable.} This heterogeneity is not isolated to the global financial crisis (GFC), nor to real activity. Across episodes, some sectoral activity, price, and financial variables consistently exhibit extreme risks while others remain relatively stable, with some sectors shifting from symmetric to highly asymmetric risk. Standard measures of aggregate uncertainty and risk rise uniformly during crises, offering no explanation for this variation. Where does this heterogeneity come from? Is it idiosyncratic developments, or does it reflect systematic propagation of common shocks?

Using a dynamic factor model with endogenous stochastic volatility applied to over one hundred U.S. variables, we give a two-part answer. First, tail risk is organized by the same common macroeconomic dynamics that drive business cycles. Decades of research have established that a small number of common factors summarizing these dynamics explains the bulk of fluctuations in the \textit{level} of activity and prices across sectors \citep{stock-watson-02, sw2016}. We show that the same structure organizes risk. A sector's factor exposures explain 55 percent of the cross-sectional variation in tail asymmetry across 116 variables (Table~\ref{tab:asym_loadings}): which sectors face severe downside risk, which face upside risk, and which remain stable. This finding was not guaranteed: had tail risk been idiosyncratic, factor exposures would have carried no such explanatory power. Randomly reshuffling factor exposures across variables indeed destroys the pattern.

Second, connecting macroeconomic dynamics to tail risk requires a mechanism absent from standard factor models. In a linear model with symmetric shocks, no factor structure, however rich, can generate asymmetric risk. Asymmetry arises when a factor and its volatility move together: for instance, adverse financial shocks that also raise uncertainty deepen the lower tail of real activity, while inflationary shocks that raise uncertainty stretch the upper tail of prices. This is the leverage effect documented in asset markets \citep{Black1976, Christie1982}, generalized here to the macroeconomy, where the feedback runs in both directions: adverse shocks raise uncertainty, and elevated uncertainty in turn depresses activity \citep{LMN2020}. The correlation between factors' level and volatility is strong in our estimates, and heterogeneous factor loadings (each variable's exposure to each factor) transmit it unevenly across the economy. What matters is where a sector is exposed: loadings on the aggregate consumption factor are among the largest in the panel yet contribute little to tail asymmetry, while exposures to financial conditions and inflation account for most of it.

These dynamics unify three empirical phenomena. Two have been studied in isolation. First, the growth-at-risk behavior documented by \citet{ABG19}, where GDP growth's lower tail is more volatile than its upper tail, reflects these dynamics operating in real activity variables. Second, inflation-at-risk documented by \citet{lopezsalido-loria-2020} reflects the same mechanism with the opposite sign: rising inflation accompanied by rising uncertainty generates upside risk to prices. The third has not been systematically documented: the substantial heterogeneity in tail risk across activity, price, and financial variables. Some cyclical industries display extreme asymmetry while stable sectors are close to symmetric. \citet{odf2025} document this for manufacturing growth, but not across the economy as a whole. We show that risk heterogeneity reflects how different variables respond to common underlying shocks. Rather than requiring separate models for growth-at-risk, inflation-at-risk, and sectoral risk heterogeneity, a unified framework with common macroeconomic fluctuations simultaneously generates tail risk behavior across over one hundred variables.

Our methodological contribution is a dynamic factor model that formalizes these dynamics through three interconnected mechanisms that allow factor levels and volatilities to interact. First, past volatility directly affects current factor dynamics, creating channels from uncertainty to real outcomes. Second, factor movements influence volatility evolution, making uncertainty endogenous to economic conditions rather than only driven by exogenous shocks. Third, factor levels and their volatilities are contemporaneously correlated, so disturbances simultaneously affect both. Combined with flexibly estimated factor loadings, these features enable the model to generate the rich heterogeneity in tail risk documented above while maintaining a parsimonious low-dimensional factor structure. The model produces full predictive distributions for all variables, not just point forecasts or volatility estimates, allowing us to characterize time-varying asymmetries in upside and downside risks across the entire macroeconomy.

We estimate the model using 116 monthly U.S. macroeconomic and financial time series from 1973 to 2023, covering major crises including the Great Inflation, the GFC, and the pandemic. The dataset spans real activity, prices, credit, and financial conditions with substantial sectoral disaggregation across industrial production, consumption categories, and equity portfolios. This granularity is essential for documenting the heterogeneity in tail risk and linking it to observable sectoral characteristics.

We establish four empirical findings. First, we construct three indices---for growth, financial conditions, and inflation---that aggregate tail risk information across many variables. This approach provides stable measures of economy-wide tail risk and is more robust than examining individual variables in isolation. The indices reveal meaningful time variation consistent with regime shifts. Growth risk tilts strongly toward adverse events during financial crises, when falling activity and rising uncertainty reinforce each other. Inflation risk shifts direction across episodes: upward during supply-shock periods such the 1970s, downward during demand-driven recessions.\footnote{Throughout, we use supply- and demand-driven as descriptive shorthand. A demand or financial-stress episode is one in which prices, real activity, and financial conditions deteriorate jointly; a supply episode is one in which prices and real activity variables move in opposite directions. This characterization is not based on structural identification of shocks.}

Second, we examine heterogeneity at the sectoral level. Cyclical, capital-intensive industries like construction materials exhibit extreme downside asymmetry, while regulated utilities show modest, nearly symmetric risk. Durable goods consumption displays strong growth-at-risk, while nondurable goods like food and necessities remain relatively stable. We validate these patterns against evidence external to the model: linking industrial production risk to Census Bureau microdata on job flows and establishment dynamics, sectors with more left-skewed job creation and net establishment entry rates also exhibit more pronounced tail risk asymmetries. The risk profiles we estimate capture economic fundamentals, not statistical artifacts.

Third, we analyze the estimated factor structure. The latent factors capture financial conditions, consumption, credit, and inflation dynamics. Mean-volatility correlations vary both within and across factors. Within factors, credit shows negative correlations (declining credit raises volatility), while financial and inflation factors show positive correlations. Across factors, stress in one domain amplifies uncertainty in others. Combined with heterogeneous factor loadings---for instance construction materials and durables load heavily on financial factors, while utilities and food show weak loadings---these correlation patterns generate the heterogeneous tail risk documented above.

Fourth, we probe the model's mechanisms through counterfactual analysis. A financial shock during the GFC generates severe downside risk for credit-sensitive sectors like construction materials and durables, while stable sectors like utilities remain unaffected, closely matching observed 2008-09 patterns. The model also captures risk dynamics across historical episodes: comparing the Great Inflation (1978-82) to the GFC reveals that the same factor structure produces both inflation-at-risk and growth-at-risk when supply dynamics dominate, while generating primarily growth-at-risk when demand dynamics dominate.

We assess model fit and validate the sources of heterogeneity through three exercises. First, we show that common factors explain substantial variation in the data, around 80~percent for some key series. Second, we confirm that heterogeneity in tail risk arises from factor loadings rather than idiosyncratic noise: shutting off idiosyncratic shocks preserves the structure of risk across growth, inflation, and financial variables, but randomly reshuffling factor loadings destroys it. Third, comparing conditional distributions to those from quantile regressions \citep{ABG19} reveals that our factor model outperforms in the tails of inflation and financial variables, while the two models perform comparably for real activity variables. 

Our work contributes to three literatures. First, we extend research on macroeconomic uncertainty \citep{Bloom2009, JLN2013, carriero2016} by shifting focus from symmetric volatility to asymmetric, directional tail risks. Existing uncertainty measures tell us when dispersion is elevated but not whether risks tilt toward adverse or favorable outcomes, a distinction that is critical for assessing the balance of risks. In our framework, asymmetry and directionality are inseparable: the correlation between a variable's conditional mean and its conditional volatility determines both how asymmetric its risks are and in which direction they tilt.

Second, we contribute to the growth-at-risk literature initiated by \citet{ABG19} and extended by \citet{lopezsalido-loria-2020}, \citet{kiley-2022}, and \citet{am2025}, which documents asymmetric tail behavior for individual variables using quantile regressions, and \citet{odf2025}, which does so across manufacturing industries. We show this phenomenon is pervasive across the economy and provide a unifying mechanism: endogenous mean-volatility interactions in common factors alongside heterogeneous loadings. Rather than separate models for each phenomenon, a single framework moves the literature from documenting patterns in risk to understanding their common origins. More broadly, the granular origins literature shows how idiosyncratic shocks to individual units aggregate into economy-wide fluctuations \citep{gabaix2011}. We document a channel running in the opposite direction: common macroeconomic dynamics disaggregate into sharply heterogeneous tail risk across sectors.

Third, we extend dynamic factor models \citep{stock-watson-02, bernanke-boivin-eliasz-05, sw2016} by incorporating endogenous stochastic volatility. \citet{canova-ravn-2000} and \citet{canova-ciccarelli-2013} demonstrate the importance of modeling sectoral and cross-country heterogeneity in panel frameworks, yet typically assume constant volatility. While \citet{delnegro-otrok-08} and \citet{ms2012} introduce time-varying volatility in factor models, it evolves exogenously in their frameworks---volatility shocks do not affect factor levels, nor do factor movements influence volatility. \citet{carriero2016} and \cite{dhk2025} model common volatility with mean effects but specify it as a separate latent factor rather than allowing factors' own volatility to respond to their movements. \citet{GORODNICHENKO201752} demonstrate the importance of separately identifying level and volatility factors in large panels through indirect extraction methods. We extend this insight by modeling factor volatility explicitly through stochastic processes that respond endogenously to economic conditions. To our knowledge, this is the first macroeconomic factor model estimated on a large panel to incorporate correlated level and volatility shocks.

Beyond these methodological and empirical contributions, the framework provides empirical targets for validating dynamic models with nonlinear mechanisms. Recent work on occasionally binding constraints \citep{guerrieri-lorenzoni-2017} and financial frictions predicts that binding episodes should generate asymmetric tail dynamics: severe downside risk when constraints tighten, modest upside risk when they relax. Models incorporating stochastic volatility \citep{fernandez-villaverde-guerron-rubio-ramirez-2011} predict asymmetric real effects of uncertainty shocks. Multi-sector models \citep{canova-ravn-2000} further suggest these effects should vary systematically across industries. Our framework offers not only distributional targets but also interpretable factor dynamics and counterfactual capabilities, illustrated through an historical decomposition of risks during the Great Inflation versus the GFC, that can discipline specific transmission mechanisms, extending validation beyond traditional likelihood-based and moment-matching methods \citep{fernandez-villaverde-rubio-ramirez-2007, canova-sala-2009}.

\section{\label{sec:model}The Model}

In this section, we present the dynamic factor model (DFM) with stochastic volatility that forms the foundation of our analysis. We begin by presenting the model specification and then explain why standard DFMs cannot capture the risk dynamics we document. We show how our modeling choices, particularly the correlation between factor levels and their volatilities, generate asymmetric tail risk behavior across many variables. Finally, we present the estimation algorithm and details about the data and empirical specification.

\subsection{\label{subsec:empirical-model}A Dynamic Factor Model with Endogenous Stochastic Volatility}

The DFM is represented in a state-space form, where we specify a small number of latent factors $F_t$ that drive the common dynamics of $n$ observed variables $X_t$, where $n$ is a large number (116 in our application). The observation equation links the observed variables to the latent factors, while the transition equation governs the evolution of these factors over time, incorporating both level and volatility shocks.

\textbf{The observation equation} is written as:

\begin{equation}
    X_t = B F_t + v_t,  \label{eq:measurement}  
\end{equation}

where $X_t$ is an $n \times 1$ vector of observed variables, $F_t$ is an $N \times 1$ vector of latent factors, and $v_t$ represents idiosyncratic components. The factor loadings are captured in the $B$ matrix, and the idiosyncratic shocks follow an autoregressive (AR) process:

\begin{equation}
    v_{it} = \sum_{l=1}^{L} \rho_{i,l} v_{it-l} + u_{it}, \quad u_{it} \sim N(0, R_t),    \label{eq:measurement-AR-idiosyncratic}
\end{equation}

\noindent where the diagonal covariance matrix $R_t = \text{diag}(e^{r_t})$ models time-varying idiosyncratic volatility, with $r_t$ evolving as independent random walks. We assume idiosyncratic volatilities evolve independently (i.e., $R_t$ is diagonal) to isolate the common risk dynamics driven by the factors. This structure ensures that individual shocks exhibit heteroscedasticity without direct interactions across variables or with the common factors.

\textbf{The transition equation} specifies the evolution of latent factors and incorporates both autoregressive dependencies and volatility effects:

\begin{equation}
F_t = c + \sum_{j=1}^{P} \beta_j F_{t-j} + \sum_{k=1}^{K} b_k \tilde{h}_{t-k} + H_t^{1/2} e_t.   \label{eq:transition-factors-F}  
\end{equation}

The term $\sum_{k=1}^{K} b_k \tilde{h}_{t-k}$ captures the in-mean effect of past volatility on factor dynamics. This mechanism allows past uncertainty to directly affect current economic conditions. For instance, heightened financial volatility may depress real activity or amplify risk premia, consistent with uncertainty-driven business cycle models.

The stochastic volatility process $H_t$ evolves as a VAR:

\begin{equation}
    \tilde{h}_{t+1} = \alpha + \theta \tilde{h}_t + \sum_{j=1}^{Q} d_j F_{t-j} + S^{1/2} \eta_t, \label{eq:transition-vol-h}
\end{equation}

where $\tilde{h}_t=[h_{1t},h_{2t},\dots,h_{N,t}]$ is a vector of log stochastic volatilities, and $H_{t}=\text{diag}\left( \exp ( \tilde{h}_{t}) \right)$. Crucially, equation~\ref{eq:transition-vol-h} allows factor movements to directly influence volatility evolution, introducing endogeneity between the level and volatility of the factors. This channel operates with a lag: movements in the factors at $t-j$ shift volatility going forward. The contemporaneous link between factor levels and their volatilities instead operates through the correlated disturbances described below.

The disturbances $\varepsilon_t = [\eta_t, e_t]$ follow a normal distribution with a time-invariant covariance matrix:

\begin{equation}
\Sigma =\left( 
\begin{array}{cc}
\Sigma _{\eta } & \Sigma _{\eta e}^{\prime } \\ 
\Sigma _{\eta e} & \Sigma _{e}
\end{array}
\right),
\end{equation}
where we normalize the diagonal elements of $\Sigma$ to equal $1$ for identification. This normalization, combined with restrictions on the factor loading matrix $B$ described below, ensures the model is identified while allowing flexible correlation patterns between level and volatility shocks. The covariance matrix of the reduced form residuals in equations~\ref{eq:transition-factors-F} and~\ref{eq:transition-vol-h} is time-varying due to stochastic volatility: 

\begin{equation}
\label{eq:omega}
\Omega _{t}=\left( 
\begin{array}{cc}
S^{1/2} & 0 \\ 
0 & H_{t}^{1/2}%
\end{array}%
\right) \left( 
\begin{array}{cc}
\Sigma _{\eta } & \Sigma _{\eta e}^{\prime } \\ 
\Sigma _{\eta e} & \Sigma _{e}%
\end{array}%
\right) \left( 
\begin{array}{cc}
S^{1/2} & 0 \\ 
0 & H_{t}^{1/2}%
\end{array}%
\right) ^{\prime }.
\end{equation}

This formulation captures dynamic risk structures by allowing factor shocks and their volatilities to interact over time.

\subsection{\label{subsec:riskunderstand}How the Model Generates Asymmetric Risk Dynamics}

We begin by clarifying what we mean by uncertainty, risk, and risk asymmetry, as these concepts are central to our analysis. For any variable in our dataset, such as GDP growth or inflation, we construct the conditional distribution of future outcomes $h$ periods ahead, denoted $p(X_{i,t+h}|X^t)$. This distribution summarizes the full range of possible realizations given current information.

We define \textit{uncertainty} as the dispersion or width of this conditional distribution, measured by its standard deviation. We define \textit{downside risk} as the 5th percentile of the distribution and \textit{upside risk} as the 95th percentile. These tail quantiles capture the extreme adverse and favorable outcomes that might occur.

Crucially, we are interested in how these measures evolve over time. \textit{Risk asymmetry} occurs when the volatility of the lower tail differs from the volatility of the upper tail across time. For example, if the 5th percentile exhibits substantial variation during recessions while the 95th percentile remains relatively stable, we say the variable exhibits asymmetric downside risk, or ``growth-at-risk" behavior \citep{ABG19}. 

The key challenge is generating time-varying asymmetric risk dynamics in a model that spans many variables simultaneously. Standard approaches, as we discuss below, cannot capture this behavior.

\subsubsection*{How Our Model Works: An Illustrative Example}

Our model generates asymmetric tail risk through the \textit{correlation between factor levels and their volatilities}. When a factor declines and its volatility increases simultaneously, the lower tail of the conditional distribution shifts substantially leftward (both forces work in the same direction), while the upper tail shifts by less (the forces partially offset). This mechanism, introduced by \citet{ABG19} for univariate models and extended to VARs by \citet{CaldaraSVOL}, is what enables our model to capture rich risk dynamics across many variables.

\begin{figure}[h!]
\caption{Conditional Distributions with Two Factors}
\label{fig:twofactorcase}
\includegraphics[width=0.95\textwidth]{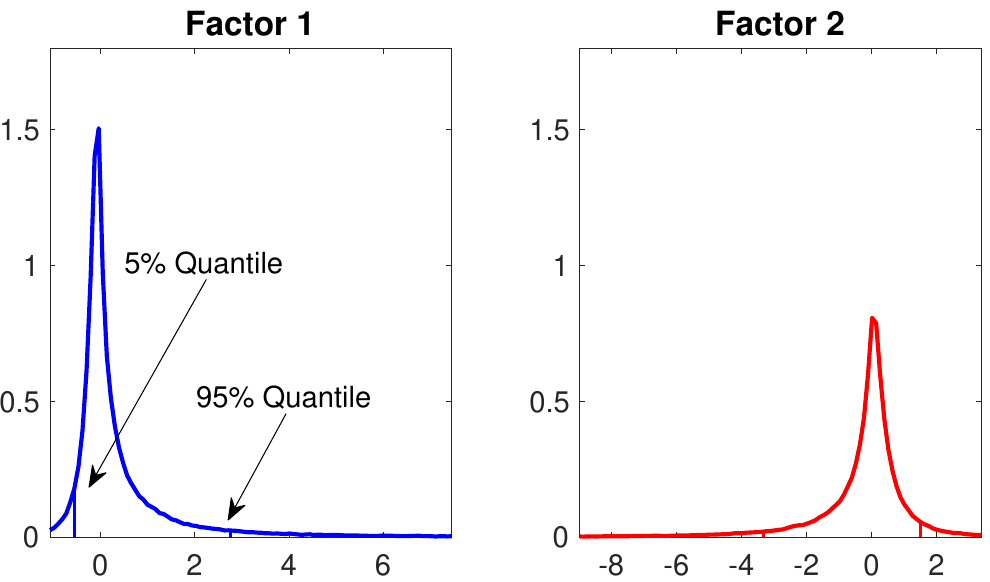}

\includegraphics[width=0.95\textwidth]{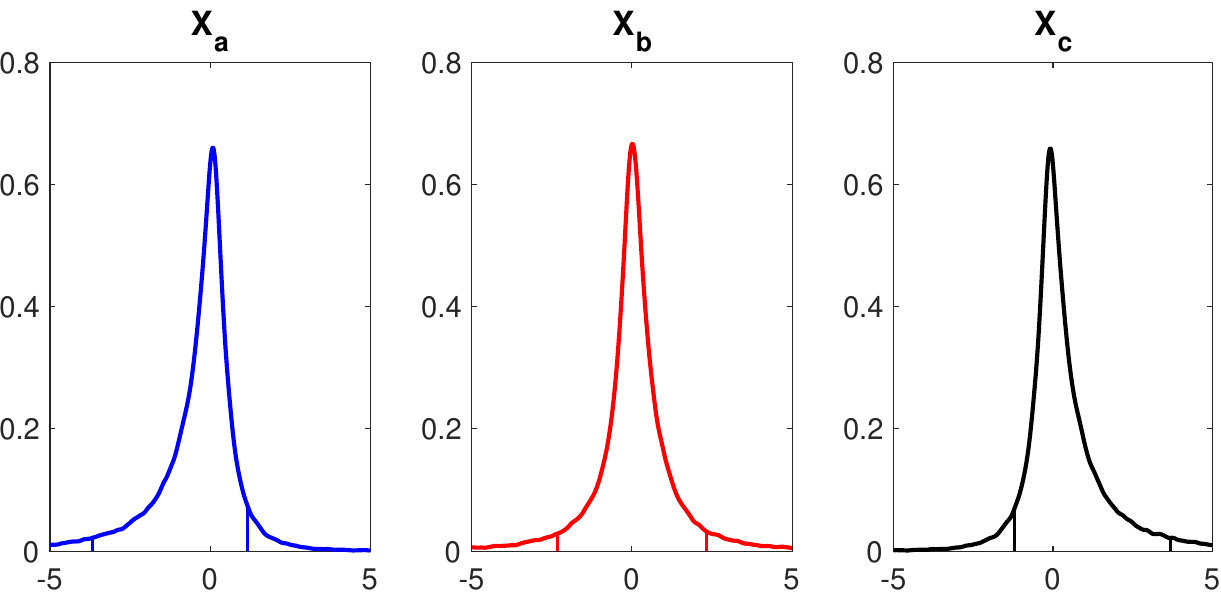}
\small{Note: This figure shows example conditional distributions for two factors (top row) and the implied distributions for three observable variables (bottom row). The conditional distributions in the bottom row are generated through linear combinations of the conditional distributions of the top row. We shut down the idiosyncratic errors when generating the conditional distributions of the observables. The vertical lines denote the $5$th and $95$th quantiles.}
\end{figure}

Figure~\ref{fig:twofactorcase} illustrates this mechanism using a simplified two-factor model.\footnote{For this illustration, we shut down the idiosyncratic shocks to focus on factor-driven dynamics.} The top row shows conditional distributions for two factors with opposite mean-volatility correlations. Factor 1 exhibits positive correlation: when it rises, uncertainty increases, creating a long right tail with elevated upside risk. Factor 2 exhibits negative correlation: when it falls, uncertainty increases, creating a long left tail. The vertical lines mark the 5\% and 95\% quantiles.

The bottom row demonstrates the key insight: observable variables inherit different risk patterns through the factor loading matrix $B$. Variable $X_a$ loads negatively on factors with positive mean-volatility correlation, producing growth-at-risk dynamics. Variable $X_b$ loads on both factors, yielding symmetric risk. Variable $X_c$ loads positively on factors with positive correlation, generating, for instance, inflation-at-risk dynamics. Thus, a small number of factors with distinct correlation patterns can generate diverse risk behaviors across many variables.

\subsubsection*{Implementation: Three Core Mechanisms}

We implement the mean-volatility correlation illustrated above through three interconnected mechanisms in equations~\ref{eq:transition-factors-F} and~\ref{eq:transition-vol-h}:

\begin{enumerate}
    \item \textbf{In-mean effects.} The term $\sum_{k=1}^{K} b_k \tilde{h}_{t-k}$ in equation~\ref{eq:transition-factors-F} allows past volatility to directly affect current factor dynamics, following \citet{MumtazTheodoridis2018}. Elevated past uncertainty can depress real activity or amplify risk premia.
    
    \item \textbf{Endogenous volatility.} Equation~\ref{eq:transition-vol-h} allows factor movements to influence volatility evolution through the term $\sum_{j=1}^{Q} d_j F_{t-j}$. Uncertainty responds to economic conditions rather than evolving independently.
    
    \item \textbf{Correlated shocks.} We allow non-zero correlation $\Sigma_{\eta e}$ between shocks to factors ($e_t$) and shocks to their volatilities ($\eta_t$). Combined with time-varying volatility (equation~\ref{eq:omega}), shocks that affect factor levels simultaneously influence volatility.
\end{enumerate}

Together, these channels generate both the \textit{leverage} and \textit{volatility-in-mean effects}. The leverage effect is the well-documented pattern whereby negative shocks raise subsequent volatility, first established for asset prices \citep{Black1976, Christie1982, Schwert1989}. The volatility-in-mean effect operates in the opposite direction, as heightened uncertainty feeds back into factor dynamics. \citet{LMN2020} show that both directions are active in macroeconomic data: adverse shocks to activity increase uncertainty, while uncertainty shocks reduce activity.

These correlations are not merely statistical regularities; they are consistent with several classes of macro models. In economies with financial frictions, adverse shocks that erode borrower net worth depress activity and amplify volatility as the economy moves toward constrained regions of the state space \citep{brunnermeier2014}; occasionally binding borrowing limits generate similarly asymmetric responses when constraints bind \citep{guerrieri-lorenzoni-2017}. Real-options effects make investment and hiring more sensitive to bad news when uncertainty is elevated \citep{Bloom2009}, while models with stochastic volatility assign real effects to uncertainty shocks themselves \citep{fernandez-villaverde-guerron-rubio-ramirez-2011}.

\subsubsection*{Why Standard DFMs Cannot Capture Risk Dynamics}

To appreciate the contribution of our modeling choices, consider how standard dynamic factor models and existing extensions handle uncertainty and risk.

\textbf{Standard linear DFMs.} In a linear dynamic factor model with constant volatility, a shock to a factor shifts the mean of the conditional distribution but leaves its width unchanged. The upper and lower tails move symmetrically, and there is no time variation in uncertainty. Such models can characterize average dynamics but cannot capture the risk fluctuations we document.

\textbf{DFMs with exogenous stochastic volatility.} \citet{delnegro-otrok-08} and \citet{ms2012} introduce time-varying volatility into factor models, allowing the width of conditional distributions to change over time. This captures uncertainty fluctuations. However, the tails still move symmetrically when volatility changes. A volatility shock widens the distribution uniformly, affecting upside and downside risks equally. These models cannot generate \textit{asymmetric} tail risk as a structural property of the model: the phenomenon where downside risk exhibits greater volatility over time than upside risk, or vice versa.\footnote{\citet{ccm2024} find that BVARs with exogenous volatility can nevertheless exhibit asymmetric tail risk when estimated, because the estimated historical level and volatility shocks turn out to be correlated even though the model assumes them independent. The asymmetry is inherited from the data rather than produced by the model: it does not arise endogenously in counterfactual exercises, which is where our specification and exogenous-volatility models diverge (Section~\ref{sec:shock-results}).}

\textbf{Comparison to \citet{JLN2013}.} The framework in \citet{JLN2013} is a related approach to measuring uncertainty and its effects. To illustrate the difference, consider a single-factor version of our model:
\begin{align} 
F_{t} &= c_{1} + \beta_{11} F_{t-1} + \beta_{12} h_{t-1} + \exp(h_{t}/2) \epsilon_{t} \\
h_{t+1} &= c_{2} + \beta_{21} F_{t-1} + \beta_{22} h_{t} + \tau \eta_{t} \notag
\end{align}
with $\text{Corr}(\epsilon_{t}, \eta_{t}) = \zeta$. \citet{JLN2013} effectively impose $\beta_{12} = \beta_{21} = \zeta = 0$, so that uncertainty evolves independently of factor movements. Time variation in uncertainty is driven primarily by past changes in volatility $h_t$. In our framework, shocks simultaneously affect both the level and uncertainty of factors ($\beta_{12} \neq 0$, $\beta_{21} \neq 0$, $\zeta \neq 0$), capturing aggregate disturbances with first- and second-moment effects—a channel absent in \citet{JLN2013}. Moreover, our specification allows predictor uncertainty to be endogenous, influenced by both current shocks and past movements in \textit{both} the factor level $F_t$ and its volatility $h_t$.\footnote{\citet{JLN2013} include squared factors and factors extracted from squared series as additional predictors, incorporating some stochastic volatility information through this alternative route.}

\textbf{Comparison to \citet{carriero2016}.} \citet{carriero2016} also estimate common macroeconomic and financial uncertainty jointly with mean effects in a factor volatility model. Relative to their work, our framework differs by specifying volatility as originating from the latent factors themselves rather than modeling a separate common volatility factor. This allows us to simultaneously model common fluctuations in levels and volatilities across many observable variables while maintaining, in principle, a clear economic interpretation through named factors, as each factor is associated with a specific observable variable. 

\subsection{\label{subsec:Gibbs}Estimation and Simulation Algorithms}

We estimate the model parameters using Gibbs sampling, cycling through conditional posterior distributions for factors, parameters, and volatilities. The key computational challenge is sampling stochastic volatilities given their correlation with factors. We address this using particle Gibbs with ancestor sampling \citep{RePEc:bla:jorssb:v:72:y:2010:i:3:p:269-342, JMLR:v15:lindsten14a}, which maintains good mixing properties even with few particles. The algorithm consists of seven main steps: (1) draw latent factors using the Carter-Kohn algorithm; (2) draw factor loadings and AR coefficients for idiosyncratic shocks; (3) draw idiosyncratic volatilities; (4) draw VAR coefficients; (5) draw variance of volatility shocks via Metropolis step; (6) draw the covariance matrix $\Sigma$ using the method of \citet{doi:10.1198/jcgs.2009.08095}; (7) draw stochastic volatilities using particle Gibbs. Complete technical details are provided in Appendix~\ref{appsec: model estimation}. 

We generate 50,000 draws, discard the first 30,000 as burn-in, and retain every 20th draw to minimize serial correlation. The resulting 1,000 draws provide reliable posterior inference for all quantities of interest. We verify convergence through trace plots and Geweke diagnostics.

A key output of our model is the conditional distribution of each variable at various horizons. These distributions summarize the full range of possible future outcomes, accounting for all sources of uncertainty. For any variable $X_i$ at horizon $h$, we generate the conditional distribution $p(X_{i,t+h}|X^t)$ by simulating forward from the current state. Each simulation draw proceeds as follows: (1) draw model parameters $\Theta$ and current states $(F_t, H_t, r_t)$ from their posterior distribution; (2) simulate factors, volatilities, and idiosyncratic shocks forward $h$ periods using equations~\ref{eq:measurement}--\ref{eq:transition-vol-h}; (3) construct $X_{i,t+h}$ from the simulated factors and idiosyncratic components. Repeating this process across many posterior draws yields the full conditional distribution, incorporating parameter, state, and innovation uncertainty.

From these distributions, we compute time-varying quantiles to measure downside and upside risk. The volatility of the lower tail (5th percentile) relative to the upper tail (95th percentile) captures the asymmetric risk dynamics we document throughout the paper.

\subsection{\label{subsec:data}Data and Model Specification}

We use 116 monthly U.S. macroeconomic and financial time series spanning January 1973 to December 2023, covering real activity, inflation, interest rates, asset prices, credit, and disaggregated sectoral indicators. The main data sources are \href{https://research.stlouisfed.org/econ/mccracken/fred-databases/}{FRED-MD}, the Bureau of Economic Analysis's National Income and Product Accounts, Kenneth French's website, and Global Financial Data. Variables are transformed to stationarity and standardized. A list of variable names is provided in Table~\ref{table:data}.\footnote{The degree of sectoral disaggregation reflects data availability at monthly frequency over the full 1973--2023 sample. Industrial production, personal consumption expenditures, and equity returns are available at a consistent sectoral breakdown throughout. The equity series are value-weighted industry portfolio returns rather than quantities, and therefore do not satisfy an adding-up constraint with respect to the aggregate index.}

We estimate seven latent factors (N=7), with six lags of factors and one lag of volatilities in the transition equation (equation~\ref{eq:transition-factors-F}), and one lag each of volatilities and factors in the volatility equation (equation~\ref{eq:transition-vol-h}). The number of factors balances model fit with computational tractability; additional factors provide only marginal improvements in explaining variation in the data.\footnote{We consider an extension of the model that allows for Markov regime switching on the intercept terms of the level and volatility equations to account for the pandemic. This extended model produces similar results for our risk estimates.} This number of factors is also consistent with \cite{mccracken2016}, who apply the information criteria of \citet{bai-ng-2002} to the FRED-MD panel, the core of our dataset, and select a similar number of factors.

To resolve rotational indeterminacy, we employ the named-factor normalization of \citet{sw2016}, restricting the first seven rows of the factor loading matrix $B$ to an identity matrix. This pins down each factor through its association with a specific observable variable. We select the following seven variables: Factor 1: Excess bond premium (financial conditions); Factor 2: 1-year Treasury rate (monetary policy); Factor 3: S\&P common stock price index (equity markets); Factor 4: Real personal consumption expenditures (PCE, aggregate consumption); Factor 5: Real PCE housing and utilities (housing sector); Factor 6: Total non-revolving credit (credit conditions); Factor 7: PCE inflation (price dynamics).

In a nonlinear model, different anchors define different factor and volatility dynamics rather than mere relabelings. The choice is not innocuous, and we discipline it with three criteria: the anchors span the growth, price, and financial dimensions of the dataset; each comoves with the series in its category; and each either exhibits substantial stochastic volatility or is closely tied to its sources, as with credit and financial vulnerabilities. As will be seen, the resulting estimates accord well with economic intuition and statistical assessments of model fit. Combined with the normalization of the diagonal elements of $\Sigma$ to equal one, these restrictions fully identify the model.

\section{Reading the Tails: Where Does Risk Live?\label{sec: reading-tails}}

In this section, we use the estimated model to ask: where does risk live in the economy? We begin by constructing three aggregate risk indices---growth, financial, and inflation---that distill tail risk information across many indicators into economy-wide measures. We then unpack heterogeneity across individual variables and sectors along three dimensions: cross-sectional patterns characterizing the dispersion and clustering of risk asymmetries across variables; the time series evolution of risk for consumption sectors, where quantity and price pairs trace how risk shifts across demand- and supply-driven episodes; and structural determinants of risk for industrial production, where Census Bureau establishment and labor flow data reveal why risk concentrates where it does.

\subsection{Aggregate Risk Indices: Growth, Financial, and Inflation \label{subsec: agg-indexes-construction}}

Unlike uncertainty, which is unsigned and reflects overall dispersion, risk is directional: it captures the potential for adverse or favorable outcomes. It is also asymmetric whenever uncertainty moves together with the conditional mean---the strength of that comovement determines how asymmetric risks are, and its sign determines which tail stretches. We construct three separate risk indices—growth, financial, and inflation—aligned with standard classifications of macroeconomic shocks. Adverse \textbf{demand shocks} generate downside risk to both growth and inflation. Adverse \textbf{supply shocks} produce divergent risks: lower growth, higher inflation. \textbf{Financial shocks} generate persistent downside risks to activity through credit spreads and funding conditions.

For each category, we construct two measures capturing downside risk (5th percentile) and upside risk (95th percentile). We use 12-month-ahead conditional distributions for real activity and inflation, reflecting the delayed and persistent transmission of shocks to these variables documented in the VAR literature, and 3-month-ahead distributions for financial variables, which respond immediately to economic stress.

A key advantage of this aggregation approach is robustness. By averaging across many variables, small idiosyncratic shocks cancel out in expectation, isolating the common tail risk signal shared across the economy. At the same time, unusually large idiosyncratic realizations, potentially harbingers of broader sectoral spillovers, remain visible in the index. This signal extraction logic suggests the indices provide a more comprehensive measure of economy-wide tail risk than any individual series, capturing dimensions of risk that standard aggregates may miss.

Each index aggregates the relevant percentile across many standardized indicators using equal weights. \textbf{Growth risk}: 60 real activity indicators (labor markets, consumption, housing, sales, inventories, industrial production). Variables are normalized so increases represent improving conditions. \textbf{Inflation risk}: 24 price indicators (aggregate and sectoral PCE inflation). \textbf{Financial risk}: 32 indicators (equity returns, credit volumes, spreads). Variables are normalized so increases indicate worsening conditions.

Figure~\ref{fig:agg_risk_indices} plots the indices from 1976 through 2023. Blue lines show the 5th and 95th percentiles; red horizontal lines mark standard normal reference values. All indices are in standard deviation units.\footnote{We generate the risk indices on data estimated through the end of 2019 and compare them to the benchmark indices in Figure \ref{fig:agg_risk_indices_through2019}. The two estimates are extremely close, thereby suggesting that including the pandemic observations does not appreciably change our results.}

Several patterns emerge. First, sharp spikes during major crises (2008–09 GFC, early 2020 pandemic) reflect synchronized macro-financial risk. Second, there are meaningful asymmetries: growth and financial risk tilt during adverse events; inflation risk moves upside in the 1970s and post-pandemic, but downside during the GFC. Third, the pandemic period differs from past episodes. Initially, all three indices show simultaneous upside and downside risk, indicating elevated uncertainty. From 2021, risk becomes asymmetric: growth and inflation tilt upside, though persistently so only for inflation. This contrasts with the 1970s, which showed sustained upside inflation risk, and the GFC, which showed downside growth risk. Finally, the 2010s show persistently subdued inflation risk, consistent with below-target dynamics. Empirically, the indices are substantially smoother than variable-specific tail risk measures outside major recessions, while remaining responsive during crises.

\subsection{Heterogeneity in Risk Across Variables}

The aggregate indices mask substantial variation in the nature and direction of risk across indicators and sectors of the economy. To unpack what drives these aggregate patterns, we characterize each variable's risk profile using two summary statistics computed from its conditional predictive distribution. The first is a measure of tail asymmetry: the log ratio of the standard deviation of the 95th percentile over time to the standard deviation of the 5th percentile over time. Values below zero indicate greater volatility in downside risk relative to upside risk, while values above zero suggest greater volatility in upside risk. The second is the mean-uncertainty correlation: the correlation over time between the conditional mean of a variable's predictive distribution and its uncertainty, which we measure as the standard deviation of that distribution. This statistic is the observable counterpart of the model's mean-volatility mechanism---the correlation between factor levels and factor volatilities, transmitted to each variable through its loadings. Together, these two statistics quantify both the direction of risk and the model mechanism generating it.

Figure~\ref{fig:scatter_tail_asymmetry} plots these two statistics for the 116 sign-normalized variables in our dataset from July 1976 to June 2019.\footnote{We exclude the pandemic for this analysis as it was a noneconomic shock that led to unprecedented moves in risk and uncertainty.} The aggregate risk indices, shown as pentagons, serve as anchors: the growth risk index sits firmly in the lower-left quadrant, reflecting more volatile downside risk and a strong negative mean-uncertainty correlation; the inflation and financial risk indices sit in the upper-right quadrant, reflecting more volatile upside risk and a positive mean-uncertainty correlation, where rising levels are accompanied by rising uncertainty.%\footnote{For the aggregate risk indices, mean and uncertainty are constructed by applying the same cross-sectional aggregation used to compute the index quantiles in Section~\ref{subsec: agg-indexes-construction}, averaging the relevant statistics across variables in each category.}

The most striking feature of the scatter is that points line up along a positive relationship between the two statistics, cutting across all variable categories. This pattern is informative about the source of tail risk: if idiosyncratic shocks dominated risk dynamics, there would be a cloud of points around the origin with much less range along both axes. The tight relationship over a wide range we observe thus reflects the dominance of common factors in shaping tail risk across the economy, with dispersion around the line capturing heterogeneity in factor exposures across variables.\footnote{In finite samples, realized idiosyncratic shocks can induce spurious mean-uncertainty correlation; Section~\ref{sec: robustness} confirms that shutting down idiosyncratic components preserves the structure of the scatter, validating the common factor interpretation.}

Within this common structure, there is meaningful heterogeneity across and within categories, reflecting differential factor exposures. Real activity variables cluster in the lower-left quadrant, exhibiting strong negative mean-uncertainty correlations and pronounced downside risk. Price variables cluster in the upper-right quadrant, exhibiting positive mean-uncertainty correlations and upside risk. Financial variables are the most dispersed. Interest rate and spread series sit in the upper-right quadrant, while sign-normalized equity variables cluster near the origin or in the lower-left quadrant, as volatility tends to dominate tail risk dynamics. Despite this within-category dispersion, the aggregate financial risk index sits in the upper-right quadrant, as systematic movements in rates and spreads dominate idiosyncratic equity fluctuations in driving the index.

The scatter reveals substantial heterogeneity across variables, both within and across groups. To better understand this variation, we identify the aggregate variables with the most pronounced tail asymmetries. Appendix Table~\ref{tab:top_variablesrisk} ranks variables with the largest average tail asymmetry in the three categories. On the growth side, the aggregate variables with the most asymmetric downside risk span multiple sectors, including manufacturing, labor markets, and industrial production. Among financial indicators, the strongest asymmetries appear in the VIX and various credit spreads. By contrast, equity returns have elevated uncertainty, and this uncertainty affects both tails.

We now examine sectoral patterns in consumption and production, where heterogeneity plays a central role in understanding risk transmission.

\subsection{Risk Dynamics Over History: Consumption Sectors}

The cross-sectional patterns in Figure~\ref{fig:scatter_tail_asymmetry} capture average asymmetry over the full sample, masking rich time variation across episodes with very different shock compositions. For consumption sectors, the availability of quantity and price pairs at a disaggregated level allows us to trace how tail risk shifts across demand- and supply-driven regimes. 

\begin{sidewaysfigure}
\caption{Heatmaps of Excessive Risk by Sector: Consumption}
\label{fig:heatmaps_asymmetry}
\includegraphics[width=\textwidth]{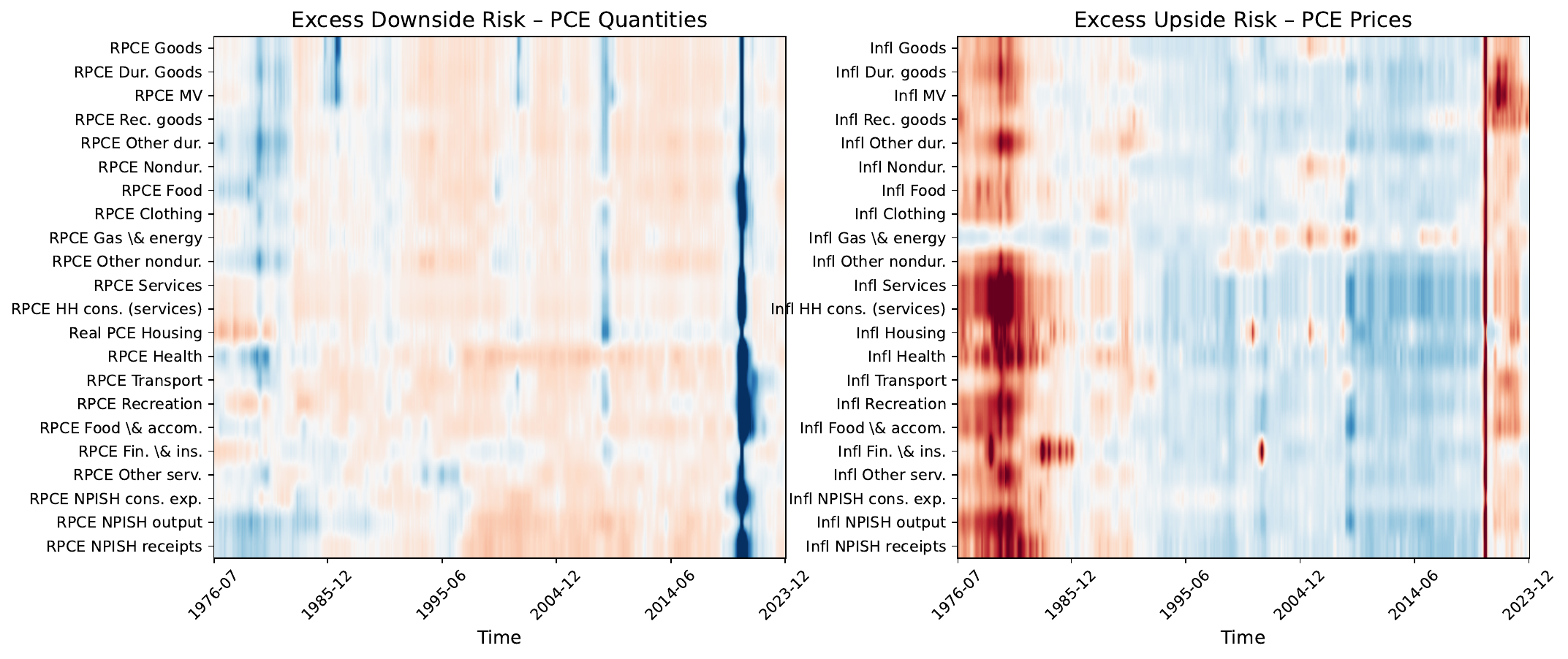}

\vspace{0.5cm}
\small{Note: The heatmaps display deviations of the conditional lower and upper quantiles of sectoral consumption expenditure  from their historical averages, both quantities (left panel) and prices (right panel). The left panel shows excess downside risk, computed as the deviation of the 5th percentile from its historical mean. The right panel shows excess upside risk, based on the deviation of the 95th percentile. For quantities, blue indicates periods of elevated downside risk relative to history, while red denotes periods of subdued risk; for prices, red denotes elevated upside risk, while blue denotes periods of subdued risk.}
\end{sidewaysfigure}

Figure~\ref{fig:heatmaps_asymmetry} examines the evolution of tail risk for sectoral personal consumption expenditures (PCE), both quantities and prices. The figure uses heatmaps to plot \textit{excess downside risk} for quantities and \textit{excess upside risk} for prices. Excess downside (upside) risk is computed as the deviation of the 5th (95th) percentile from its historical mean, expressed in standard deviations. Darker shading indicates more pronounced tail risk.

Several findings stand out. First, risk asymmetries are highly episodic and sector-specific. While crisis periods tend to see a generalized increase in risk across sectors, the magnitude and persistence vary considerably. Second, the 1970s were characterized by more acute, persistent, and broad-based upside risk to inflation compared to the post-pandemic period. During the 1970s, nearly all consumption categories exhibited elevated and sustained inflation-at-risk. In contrast, during late 2020 and early 2021, inflation risk spiked sharply but remained elevated only in a subset of sectors, particularly those related to goods consumption such as motor vehicles and furnishings.

Third, sector-specific risks are clearly visible. For instance, housing-related consumption shows pronounced downside quantity risk and upside price risk during the mid-2000s housing boom and subsequent bust. Energy consumption presents an interesting case: in the 1970s, price controls muted upside price risk relative to other sectors, whereas during the COVID period, energy prices exhibited sharp asymmetries reflecting the full impact of the demand collapse and subsequent recovery alongside supply constraints.

Fourth, the dynamics of risk in consumption quantities mirror, in opposite direction, those in prices during certain episodes. During supply-constrained periods such as the 1970s or specific phases of the pandemic recovery, downside risk in quantities coincided with upside risk in prices. By contrast, during the GFC, quantities exhibited pronounced downside risk while prices showed muted or even downside inflation risk, consistent with demand-driven dynamics.

\subsection{Industry Characteristics and Risk\label{subsec:industry-char}}

\begin{figure}
\caption{Tail Risk Asymmetry and Industry Characteristics}
\label{fig:scatter_tailvar}
{\centering
\includegraphics[width=0.57\textwidth]{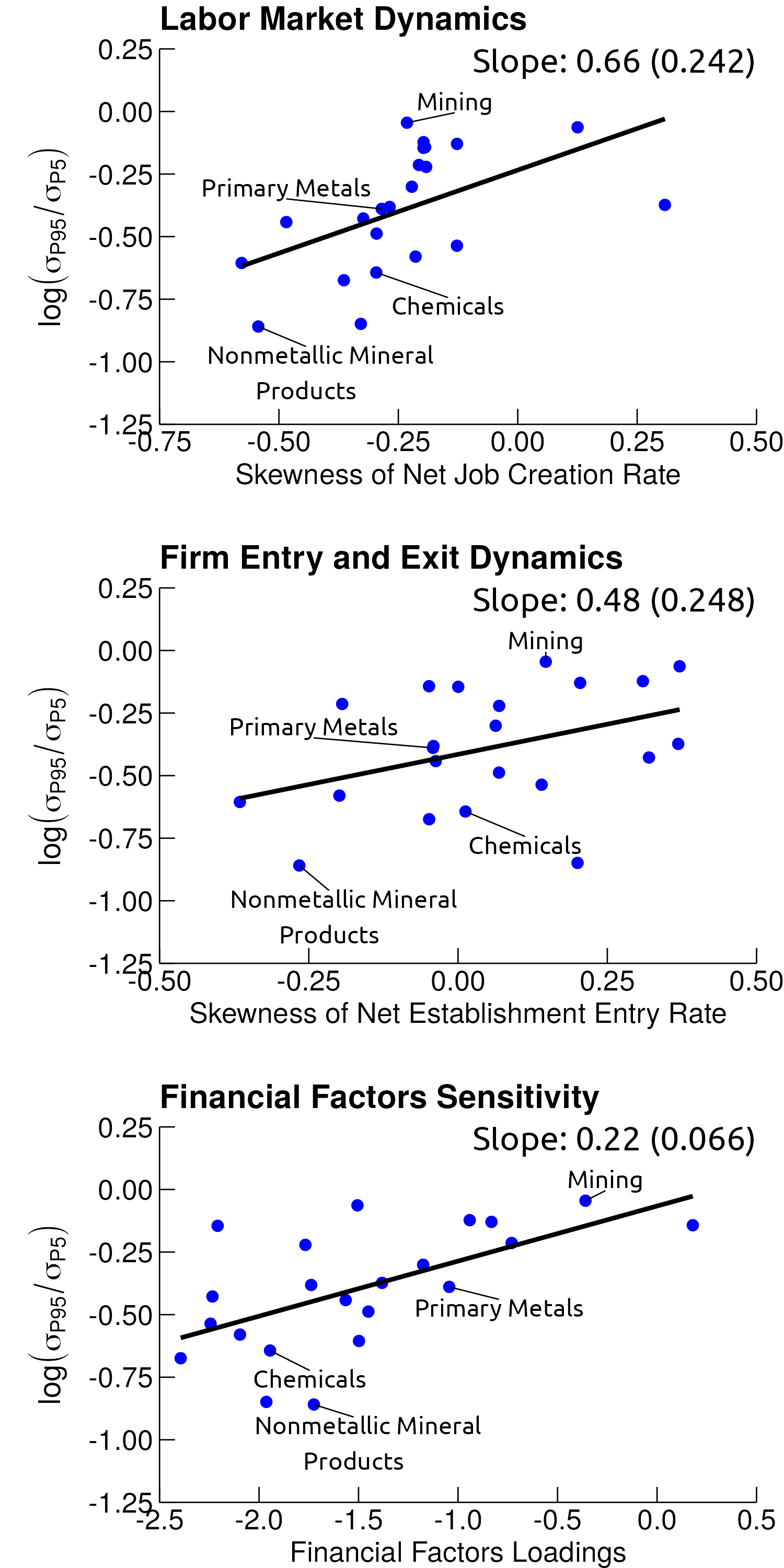}
\par}

\vspace{0.5cm}
\small{Note: This figure plots on the y-axis the log ratio of the conditional $95th$ and $5th$ percentile standard deviations for the IP sectoral variables from 1977 - 2019. On the x-axis, the panels plot the Kelley measure of skewness and the volatility for net job creation rate and net establishment entry rate by sector. The net establishment entry rate equals the establishment creation rate minus the establishment exit rate. The net job creation rate is the job creation rate minus the job destruction rate.}
\end{figure}

Finally, we ask: \textit{Why does risk live where it does?} We focus on industrial production, where Census Bureau microdata on establishment and labor market dynamics at the three-digit NAICS level allow us to link tail risk patterns to observable structural characteristics of industries. Figure~\ref{fig:scatter_tailvar} presents three panels that answer this question in two complementary ways: through external validation using industry dynamics Census Bureau data not used in estimation, and through the model mechanism linking tail asymmetry to differential exposure to financial conditions.

We leverage BDS microdata, which provides annual data on job flows and establishment dynamics at the three-digit NAICS level. We focus on two measures of industry dynamism: the net job creation rate (job creation minus destruction) and the net establishment entry rate (entry minus exit), calculated annually from 1978 to 2019.\footnote{The job destruction (creation) rate is calculated as the count of all employment losses (gains) from contracting (expanding) and closing (opening) establishments, divided by the average of employment for times $t$ and $t-1$. The establishment exit (entry) rate is defined as the count of establishments exiting (born) during the year, divided by the average of establishments for times $t$ and $t-1$. The results remain robust if we include data through the end of our sample but we exclude 2020 because of COVID.} For both variables, we compute the Kelley skewness by industry over this period.

The first two panels document that industries with more downside output risk also exhibit more left-skewed labor market and firm dynamics. Sectors such as nonmetallic mineral products and chemicals display pronounced left skewness in both net job creation and net establishment entry rates---they are more likely to shed jobs and close establishments during downturns than to expand during recoveries---and these same industries exhibit the most volatile downside risk in production. Since BDS data are not used in estimating the model, this correlation provides external validation of the risk patterns the model generates, and points to labor market adjustment and business formation characteristics as structural sources of asymmetry.

The third panel addresses the model mechanism. Industries with more negative loadings on the financial conditions factors---factors 1 and 3, the only factors whose loadings are individually and jointly statistically significant predictors of tail asymmetry across IP sectors---exhibit more pronounced downside risk. This association points to differential exposure to financial conditions as a key driver of heterogeneity in tail risk across industries. Nonmetallic mineral products anchors the bottom of the distribution, with the most negative financial factor loadings and the most volatile downside risk, while mining sits near the top with near-zero or positive loadings and relatively symmetric risk profiles. These results underscore that asymmetries in risk vary systematically across sectors and industries, and reflect deeper structural characteristics of production and industry dynamics.

\section{\label{sec: factorestimates}Factor Estimates and Model Mechanisms}

Having documented the behavior of risk across variables and over time, we now examine the model mechanisms that generate these patterns. The analysis proceeds in three parts. First, we examine the estimated paths of the latent factors and their volatilities over time. Second, we explore the joint dynamics of means and volatilities, focusing on the within-factor and cross-factor correlations that give rise to asymmetric risk and co-movement. Third, we examine how factors load onto variables, generating the heterogeneity in sectoral risk documented in Section~\ref{sec: reading-tails}.

\subsection{\label{subsec:factorvolestim}Latent Factors and Volatility Over Time}

\begin{figure}
\caption{Smoothed Estimates of the Factors and Volatilities}
\label{fig:factor_estim}
\includegraphics[width=\textwidth]{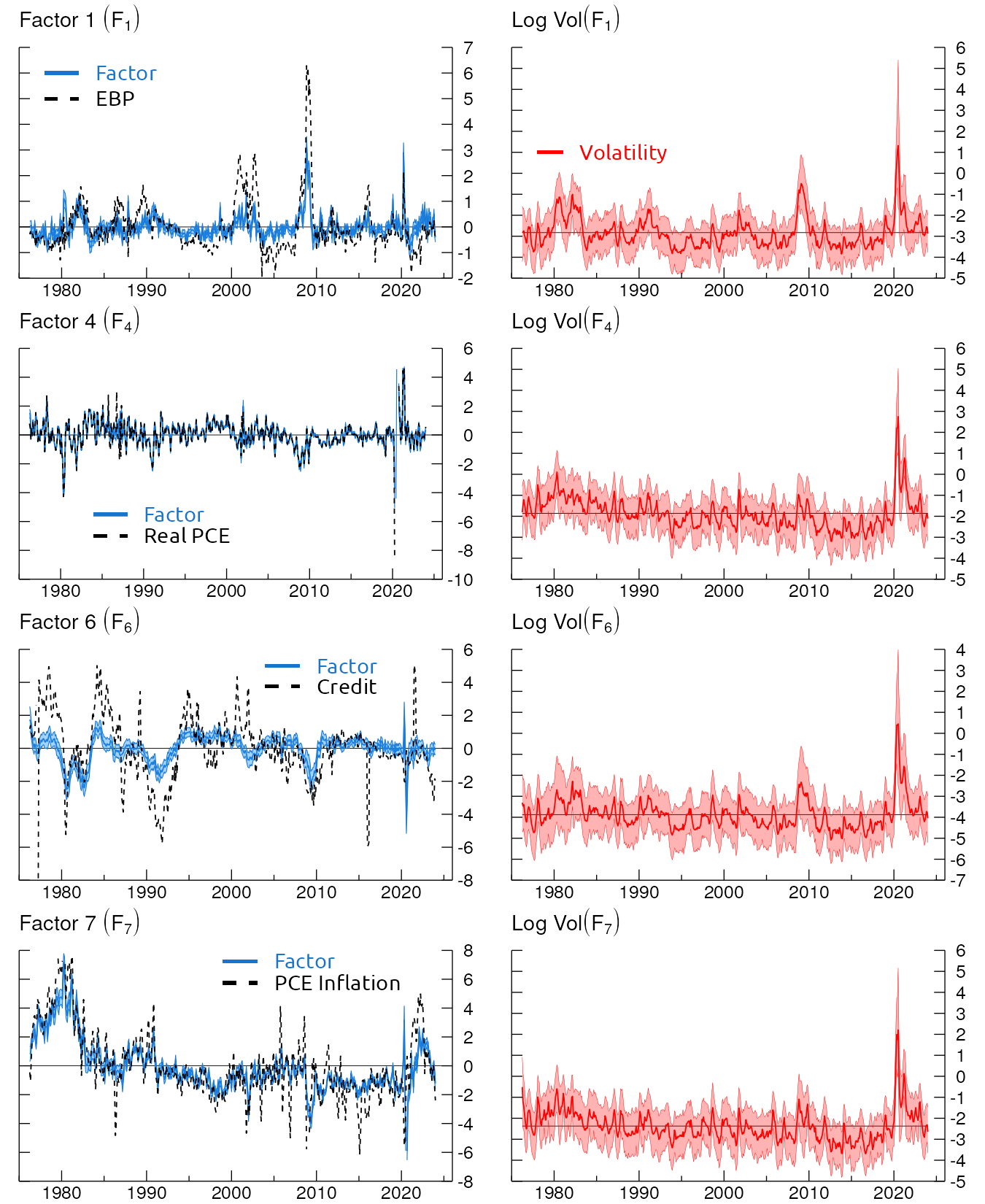}
\small{Note: The figure shows posterior median estimates of the level factors (blue lines) and of the log volatilities (red lines). Shaded areas denote the $5$th and $95$th of the posterior distributions. The dashed lines depict the observable variable associated to the factor for the normalization of matrix $B$. See Section~\ref{sec:model} for details on the normalization.}
\end{figure}

The left column of Figure~\ref{fig:factor_estim} plots the median estimates of four of the seven latent factors $F_t$ alongside the specific observable variable used in the normalization described in Section~\ref{sec:model}. The shaded regions indicate 90 percent posterior credible sets. The right column displays the estimated log volatilities of the same factors. The remaining factors and volatilities are reported in Appendix Figure~\ref{fig:factor_estim_appendix}.

The factors are estimated precisely and admit clear economic interpretations. The first factor captures financial conditions, closely tracking the excess bond premium. The fourth factor reflects aggregate consumption dynamics, mirroring real personal consumption expenditures. The sixth factor captures credit market conditions. The seventh factor is associated with PCE inflation dynamics. While these normalizations aid interpretation, each factor reflects information common across a broader set of variables beyond its anchoring observable. The dynamic factor model distinguishes between movements that are broadly shared, the common component captured by the factors, and those that are specific to individual series, the idiosyncratic component visualized as the gap between the factors and the observable series plotted in each panel.

Volatility varies markedly over time and is strongly countercyclical, rising during recessions and subsiding in expansions. Volatility spikes are particularly pronounced during well-known episodes of economic and financial stress: the early 1980s recessions, the GFC, and the onset of the pandemic. While some movements are factor-specific, there is substantial co-movement across volatilities, indicating systemic shifts in aggregate uncertainty.

\subsection{\label{subsec:distfactors}Mean–Volatility Interactions Across Factors}

\begin{figure}
\caption{Correlation across Factors}
\label{fig:corrmap-factors}
\includegraphics[width=\textwidth]{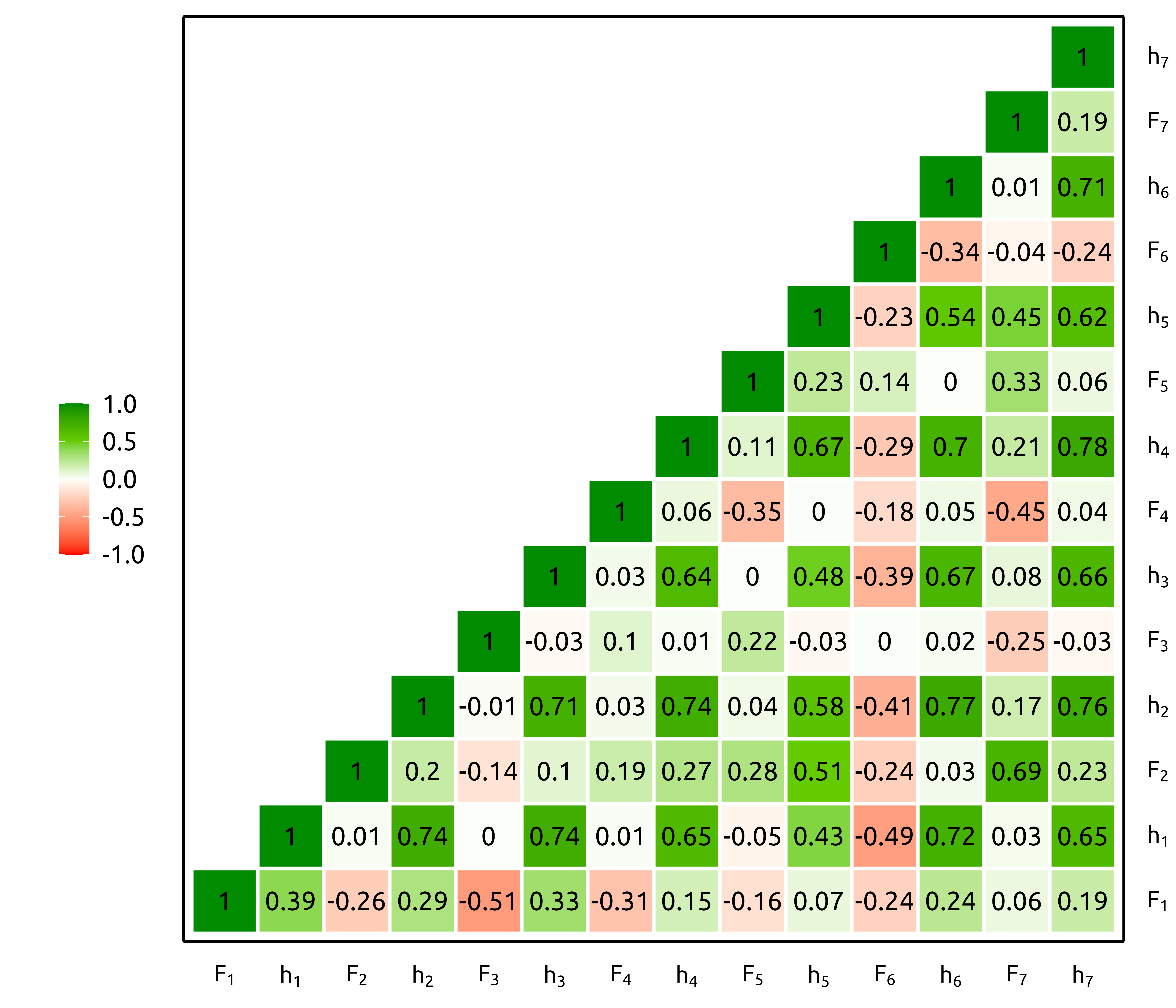}

\vspace{0.5cm}
\small{Note: This figure displays correlation coefficients among the estimated factors $F_x$ and the log volatilities $h_{x}$.}
\end{figure}

As discussed in Section~\ref{sec:model}, a central mechanism generating asymmetric risk is the correlation between each factor's level and its conditional variance. The leverage effect---adverse shocks raising volatility---is its classic form, and in our model the volatility-in-mean channel operates alongside it. Figure~\ref{fig:corrmap-factors} sheds light on this mechanism by visualizing the full correlation structure between the means and volatilities of the seven latent factors, computed from their smoothed values over the full sample. This heatmap reveals several economically meaningful patterns.

First, factors associated with financial conditions (factors 1 and 2) and inflation (factor 7) display strong positive mean–volatility correlations. When financial conditions tighten or inflation rises, uncertainty increases. This positive correlation generates more volatile upper tails for variables that load positively on these factors—consistent with the inflation-at-risk and spread-at-risk patterns documented in Section~\ref{sec: reading-tails}. Second, the factor related to credit (factor 6) exhibits negative mean–volatility correlations, consistent with a leverage effect. Negative shocks to credit raise volatility, amplifying downside risk. This mechanism, along with the fact that many real activity variables load negatively on the financial and inflation factors, explains the prevalence of growth-at-risk behavior—more volatile lower tails—observed across these indicators in Section~\ref{sec: reading-tails}.

Third, the heatmap reveals widespread co-movement in volatility across factors, indicated by the predominantly positive correlations in the volatility–volatility block (green shading). This co-movement of uncertainty drives the common risk patterns observed across many variables during crisis episodes. Finally, the means of certain factors, particularly the financial factor (factor 1) and the credit factor (factor 6), are strongly correlated with the volatilities of other factors. This highlights their central roles in transmitting and amplifying macroeconomic risk across the system.

Appendix Figure~\ref{fig:corr-structure-factors} provides additional granularity, showing that many of these mean volatility correlations are precisely estimated and vary across subsamples. For example, factors related to interest rates (factor 2) and inflation (factor 7) exhibit markedly more negative mean–volatility correlations during the GFC than in the pre-1983 period. This shift reflects the transition from an era dominated by upside inflation risk to one characterized by financial stress and disinflationary pressures. These time-varying correlations underscore the state-dependent nature of risk transmission in the model.

\subsection{\label{subsec:factorloadings}Factor Loadings and the Origins of Sectoral Risk}

The heterogeneity in risk across sectors documented in Section~\ref{sec: reading-tails} is explained by the mechanics of the model: it arises from the interaction between heterogeneous factor loadings and the mean-volatility correlations documented in Section~\ref{subsec:distfactors}. The factor loading matrix $B$ determines how the factors, and the volatility dynamics of their shocks, transmit to individual observables, shaping not just the level of risk exposure but its direction---whether a variable faces predominantly downside or upside risk---and its magnitude across sectors.

Table~\ref{tab:asym_loadings} quantifies this claim. We regress each variable's tail asymmetry on its seven factor loadings across the 116 variables in the panel.\footnote{The loadings are posterior medians and therefore generated regressors, which renders inference mildly conservative.} The loadings explain 55 percent of the cross-sectional variation in tail asymmetry, and the pattern of coefficients broadly accords with the mean-volatility correlations documented in Section~\ref{subsec:distfactors}.\footnote{For robustness, we also estimate this regression for our model estimated through the end of 2019. The loadings explain 64 percent of the cross-sectional variation in tail asymmetry, so the cross-sectional relationship is if anything tighter once the pandemic is excluded.} Loadings on the financial conditions factor carry the largest weight---explaining 38 percent of the variation on their own---followed by the inflation factor. Both enter the regression with positive coefficients, indicating that variables loading more positively on these factors tend to have more volatile upper than lower tails, consistent with the factors' positive mean-volatility correlations. The credit factor enters the regression negatively, consistent with its negative mean-volatility correlation. Notably, loadings on the consumption factor, which are among the largest in the panel, contribute nothing to asymmetry: exposure to common fluctuations determines how much a sector moves, but it is exposure to the factors with strong mean-volatility correlations that determines the direction and degree of its tail risk.\footnote{The equity factor is an exception to the broad correspondence. It enters the regression with a negative coefficient despite exhibiting little mean-volatility correlation. Its coefficient therefore captures cross-sectional variation in tail asymmetry that is not summarized by the factor's mean-volatility correlation alone.}

\begin{figure}
\caption{Sectoral Factor Loadings for PCE Quantities (top) and Prices (bottom)}
\label{fig:factorload_PCE}
\begin{center}
\includegraphics[width = \textwidth]{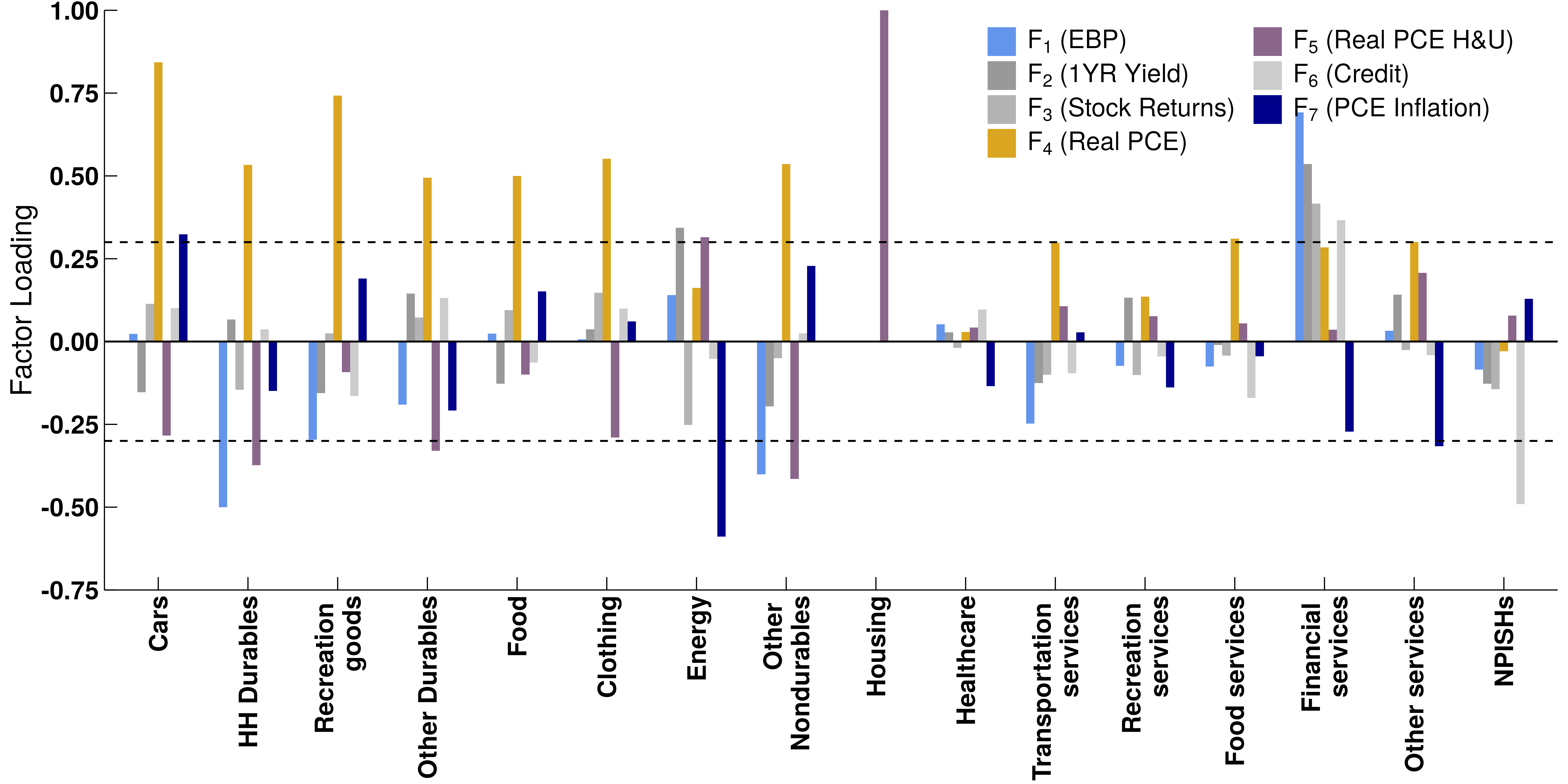}

\includegraphics[width = \textwidth]{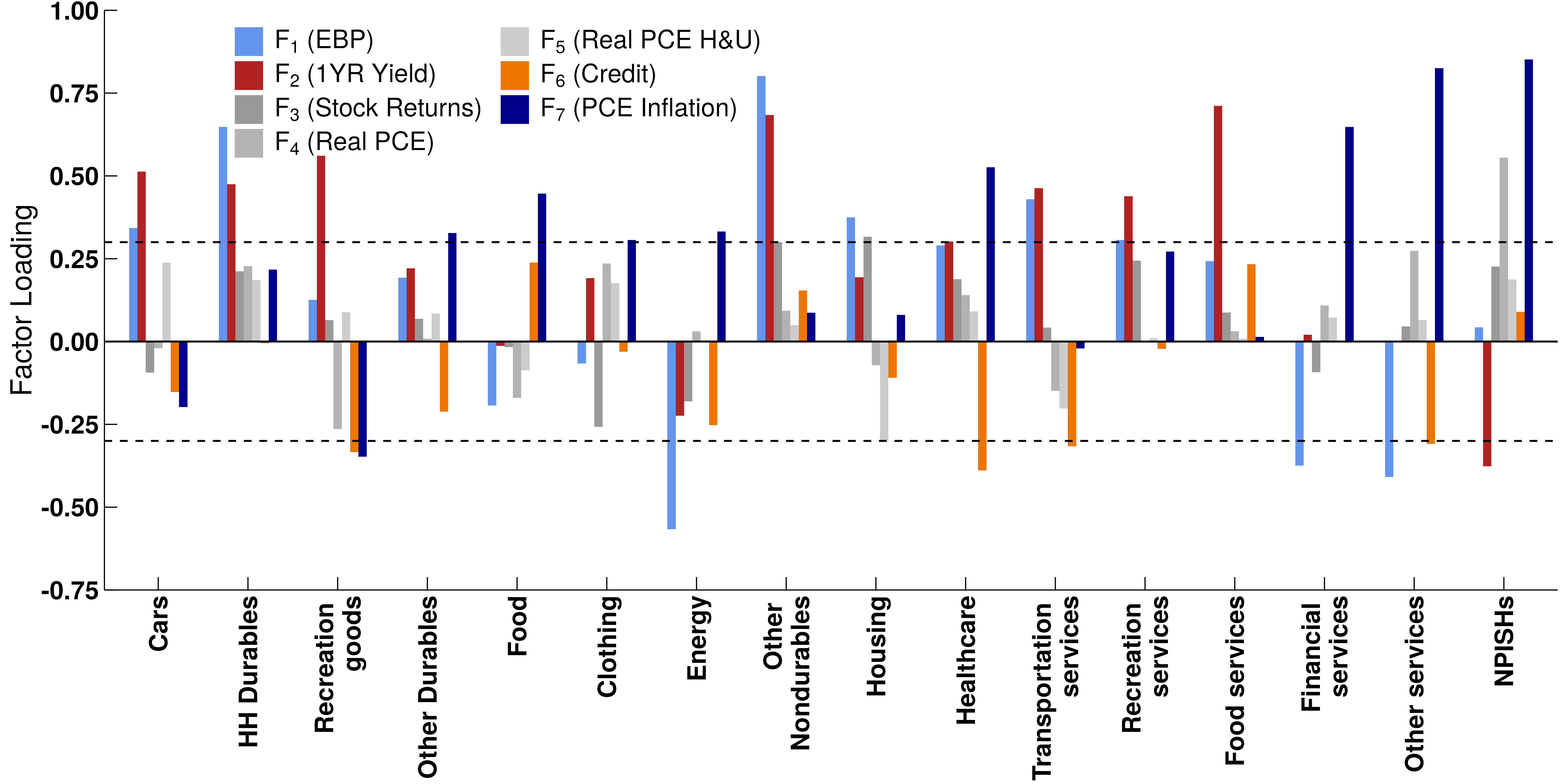}
\end{center}
\vspace{0.5cm}
\footnotesize{Note: This figure shows the median factor loading estimates for each PCE sector. The top panel shows the estimates for PCE Quantities and the bottom panel for PCE Prices. In the top panel, factors 1, 4, 5, and 7 are colored, representing the four most important factors for quantities. In the bottom panel, factors 1, 2, 6, and 7 are colored, representing the four most important for prices. The two dashed lines denote $0.3$ and $-0.3$.} 
\end{figure}

Figure~\ref{fig:factorload_PCE} displays the median posterior estimates of factor loadings for personal consumption expenditure (PCE) sectors. The top panel shows results for PCE quantities; the bottom panel for PCE prices. Colored bars highlight the four most influential factors on average in each case, determined by the magnitude of loadings and their contribution to cross-sectional variation.\footnote{Factor loadings for industrial production (IP) by industry are reported in Appendix Figure~\ref{fig:factorload_ip_appendix}. See Section~\ref{sec: reading-tails} for a discussion of the relationship between these loadings and tail asymmetry in IP.}

Several economically interpretable patterns emerge. For quantities, virtually all sectors load positively on factor 4, which captures aggregate real PCE dynamics. However, the magnitude varies substantially---explaining differences in volatility and cyclicality across consumption categories. Consistent with Table~\ref{tab:asym_loadings}, these loadings determine how strongly each sector comoves with aggregate consumption, while the asymmetry of sectoral risks is inherited through the loadings on the financial, credit, and inflation factors. Critically, some sectors load negatively on the financial conditions factor (factor 1). For example, durable goods consumption has a large negative loading, implying that tightening financial conditions translate directly into elevated downside risk for this sector. This mechanism explains the pronounced growth-at-risk behavior of durables documented in Section~\ref{sec: reading-tails}. Similarly, certain sectors load negatively on the inflation factor (factor 7), meaning that inflationary pressures depress real consumption in these categories.

For PCE prices, most sectors load positively on the inflation factor (factor 7), as expected. However, the magnitude varies considerably across categories. Many sectors also load positively on the financial factor (factor 1), suggesting that tighter financial conditions are associated with increased upside price risk, likely reflecting supply-side disruptions or demand inelasticity in certain goods. The combination of positive loadings on factors with positive mean-volatility correlations (factors 1 and 7) generates the inflation-at-risk patterns observed in Section~\ref{sec: reading-tails}.

Together, the regression in Table~\ref{tab:asym_loadings} and these loading patterns explain why certain sectors exhibit pronounced downside risk, others upside risk, and still others remain relatively symmetric. In particular, the negative loadings of real activity variables---both consumption and IP---on the financial conditions factor (factor 1), combined with that factor's positive mean-volatility correlation, provide a unified account of why financial conditions predict left-tail growth risk as documented by \citet{ABG19}, and show that this pattern extends pervasively across real activity variables.

\section{Risk Propagation: Mechanisms at Work\label{sec:shock-results}}

Sections~\ref{sec: reading-tails} and~\ref{sec: factorestimates} documented substantial heterogeneity in tail risk across sectors and showed that this heterogeneity correlates with factor loadings and mean-volatility correlations. We now assess whether these statistical relationships reflect genuine economic mechanisms by tracing how a financial shock propagates through the model. We do the same for an inflation shock in Appendix \ref{app:shocks}.

Beyond validation, this exercise demonstrates a practical application of the framework: \textit{granular risk analysis through counterfactual scenarios}. By conditioning on different economic states and varying shock magnitudes, the model can assess how tail risks evolve across sectors under alternative stress scenarios—a tool relevant for both policy analysis and risk management. We begin with the simplest such exercise: a financial shock during the GFC. Then, we examine counterfactual sectoral risk scenarios if the realized financial shocks had not occurred in the GFC and the realized inflation shocks had not occurred in the Great Inflation.

\subsection{\label{subsec:IRF}Factor and Volatility Responses}

\begin{figure}
\caption{Effects of a Shock to the First Factor on Factors and Volatilities in Oct $2008$}
\label{fig:shock_factor_vol}
\begin{center}
\includegraphics[scale = 0.45,keepaspectratio]{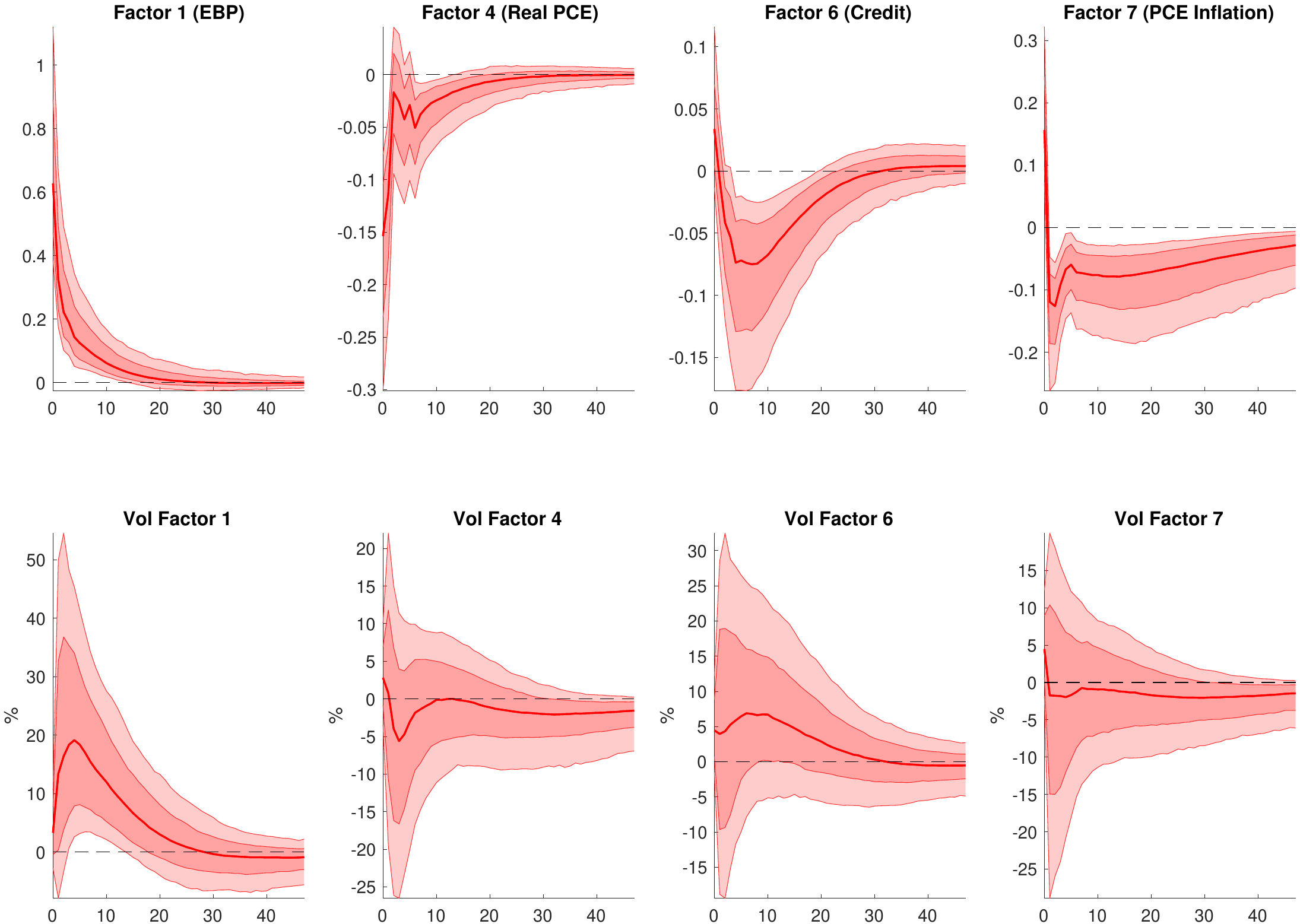}
\end{center}
\vspace{0.5cm}
\footnotesize{Note: This figure plots impulse response functions of a one standard deviation shock to the first factor in a Cholesky decomposition ordered first given October $2008$ conditions. The first column shows the level factor responses and the second column the log volatility responses. The line is the posterior median, the dark shaded area is the $68\%$ credible set and the light shaded area is the $90\%$ credible set.} 
\end{figure}

Figure~\ref{fig:shock_factor_vol} traces the propagation of a one standard deviation shock to financial conditions in October 2008. This shock is generated assuming a Cholesky decomposition and is ordered first. 

The shock generates rich co-movement across factors and volatilities, consistent with the systemic nature of financial stress. Financial conditions tighten persistently for approximately two years while credit conditions exhibit a hump-shaped decline, peaking around six months after the initial shock as financial stress transmits to lending standards and credit availability. Real activity declines sharply on impact and recovers only gradually, mirroring the sluggish pace of the post-crisis recovery. The inflation response is particularly revealing: an initial rise, likely reflecting cost-push pressures or temporary supply disruptions, followed by a persistent decline as weak demand dominates. This pattern aligns closely with observed inflation dynamics during the crisis, when commodity price spikes gave way to deflationary concerns.

Critically, log volatility increases substantially for financial and credit factors, amplifying uncertainty precisely when economic conditions deteriorate. This dual movement generates pronounced downside risk—the growth-at-risk patterns observed in Section~\ref{sec: reading-tails}. The inflation factor's volatility shows little response, consistent with its more muted mean-volatility correlation during this period.

\subsection{Sectoral Heterogeneity: Plausible or Surprising?\label{subsec:sectoral-dist}}

\begin{figure}
\caption{Distributional Effects of a Shock to the First Factor on IP and PCE Sectors in Oct $2008$}
\label{fig:shock_dist_sec}
\includegraphics[width=\textwidth]{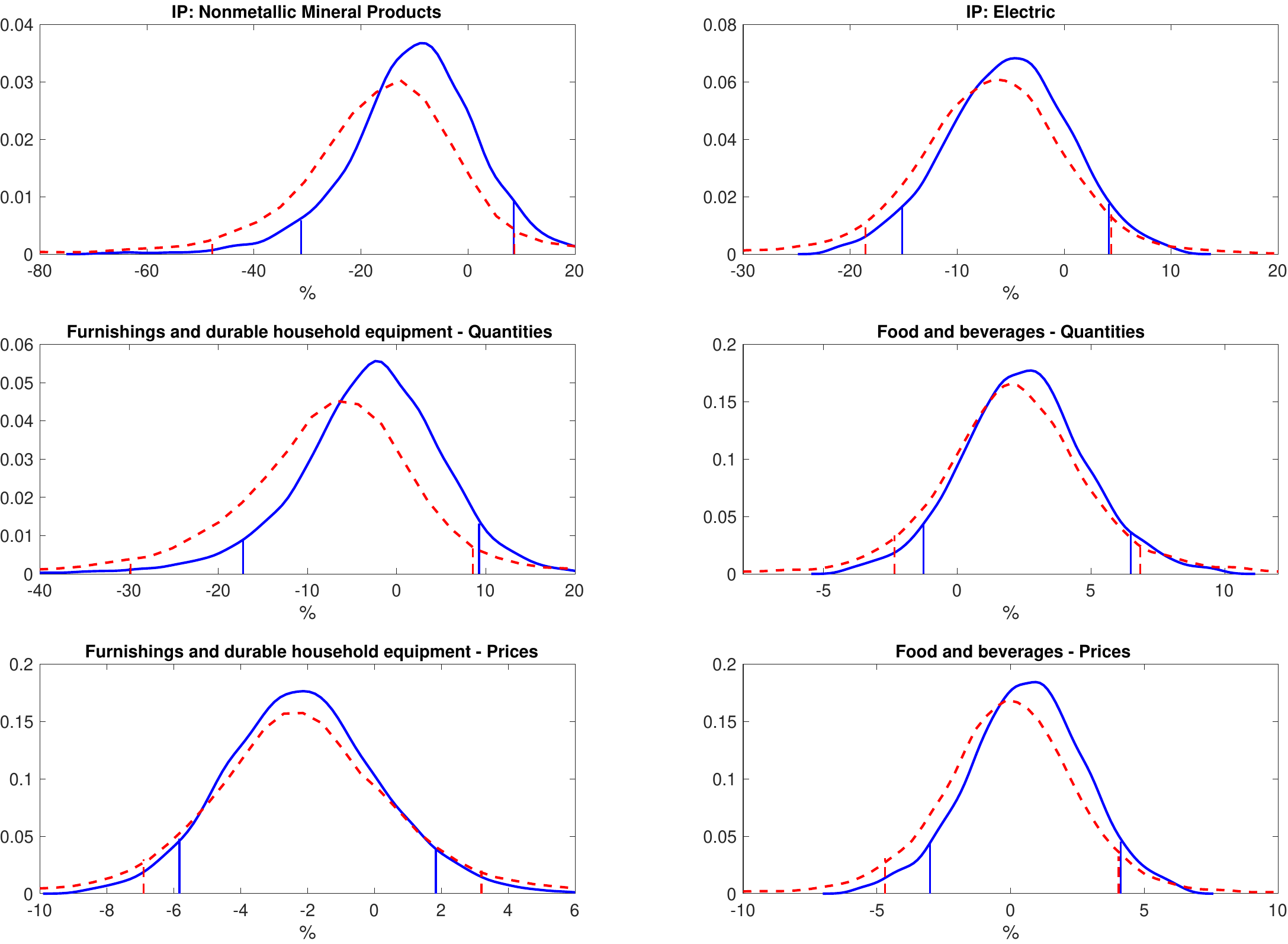}

\vspace{0.5cm}
\small{Note: This figure plots $12$ month ahead distributions given October $2008$ conditions. For variables in growth rates, we cumulative over the $12$ months. The blue distribution is a baseline without additional shocks and the red distribution is a counterfactual in which a three standard deviation shock to the first factor in a Cholesky decomposition ordered first realizes in November $2008$. The vertical lines denote the $5$th and $95$th quantiles.} 
\end{figure}

Figure~\ref{fig:shock_dist_sec} shows how the same financial shock generates strikingly different distributional responses across sectors. We present results for six sectors chosen to span the range of cyclicality and financial sensitivity in the economy. The heterogeneity is both economically plausible and quantitatively large.

\textbf{Industrial production.} Nonmetallic mineral products---which includes cement, concrete, and other construction materials---exhibits extreme downside risk sensitivity. This sector's production is closely tied to construction activity, which depends heavily on credit availability. When financial conditions tighten, construction projects are postponed or cancelled, causing severe contractions in demand for these materials. The sector's large negative loading on the financial factor captures this transmission channel. In contrast, electric power generation, while also showing downside risk, exhibits a much more muted response. As a regulated utility with stable demand and limited direct credit dependence, electricity production is partially insulated from financial shocks. 

Thus, both sectors experience downside risk in response to tighter financial conditions---consistent with the general growth-at-risk pattern---but the \textit{magnitude} differs dramatically, validating the model's ability to capture heterogeneity in cyclical sensitivity across industrial sectors.

\textbf{Personal consumption: quantities.} Furnishings and durable household equipment, big-ticket items financed through credit, exhibit sharp downside quantity risk. Food and beverages, by contrast, show minimal response. The model characterizes consumption heterogeneity that lines up with basic economic intuition about durables versus nondurables.

\textbf{Personal consumption: prices.} The price responses reveal additional nuance. Furnishings prices shift slightly to the \textit{right} with increased dispersion despite falling quantities, reflecting inflation-at-risk. This pattern is consistent with supply chain disruptions or cost-push pressures dominating demand effects for import-intensive durables during the crisis. Food prices shift left as weak demand dominates. These divergent price dynamics across sectors, inflationary for some goods and deflationary for others, align with the sectoral inflation heterogeneity observed during 2008-2009.

\textbf{Taking Stock.} The results pass a basic plausibility test: cyclical, credit-sensitive sectors exhibit large asymmetric responses; acyclical necessities remain stable. More importantly, the magnitudes are economically meaningful. They are not just statistically significant but also quantitatively large enough to matter for sectoral composition effects on aggregate risk. An economy more heavily weighted toward construction materials and durables would face substantially greater aggregate tail risk than one weighted toward utilities and food.

\subsection{Time-Varying Risk Dynamics: Supply versus Demand}
\label{subsec:scen_analysis}

The risk patterns documented in Section~\ref{sec: reading-tails} vary over time 
in economically interpretable ways. Figure~\ref{fig:price_ratios} 
illustrates this for sectoral inflation risk using two contrasting historical periods: the Great Inflation (1978-82) and the GFC (2007-11). For each period, we construct counterfactuals by shutting down realized shocks to the financial factor and the inflation factor separately over these two episodes. We assume a Cholesky decomposition with the shock to the financial factor ordered first and inflation factor ordered second. We emphasize that the goal of this exercise is not structural identification, which would require a dedicated framework beyond the scope of this paper, but rather to illustrate how the model's factor structure generates distinct risk patterns under different shock configurations. Appendix~\ref{app:historical} provides additional details about the implementation.

The contrast is striking. During the Great Inflation (left panel), sectoral 
price inflation exhibits pronounced upside risk—the upper tail is substantially 
more volatile than the lower tail across most categories. Shutting down inflation 
shocks (open triangles) eliminates this asymmetry, while shutting down financial 
shocks (open circles) has minimal effect. During the GFC (right panel), the 
pattern reverses: prices show downside risk alongside quantities, characteristic 
of demand-driven dynamics. Here, financial shocks (open circles) account for 
most tail asymmetries, while inflation shocks play a minor role.

This regime-dependent behavior demonstrates that the framework captures 
economically distinct shock types generating different risk patterns, which is 
the kind of state-dependence predicted by models with occasionally binding 
constraints or regime-switching dynamics. The same model structure produces both inflation-at-risk and growth-at-risk when inflation-factor shocks dominate---the supply-driven configuration---and primarily growth-at-risk when financial-factor shocks dominate---the demand-driven configuration.

\section{Model Fit and Robustness \label{sec: robustness}}

Having documented the pervasive heterogeneity in tail risk across variables and over time, we now assess how well the model fits the data and captures the dynamics of conditional distributions. We present three complementary pieces of evidence. First, we examine the share of variance in observable variables explained by the common factors. Second, we show that the heterogeneity and structure of risk that our model generates are driven by the loading structure on the common factors. Finally, we compare the model's implied conditional distributions to those from quantile regressions, a popular alternative in the tail risk literature, and assess whether the tail quantiles that underpin our results are well calibrated.

\subsection{Variance Decomposition: How Much Do Factors Explain?\label{subsec: sddecomp}}

The model's ability to generate time-varying and asymmetric risk across a large set of variables depends critically on the relevance of the common factors. Appendix Table~\ref{tab:sddecomps} decomposes the variance of each series into common and idiosyncratic components, for selected aggregates (Panel A) and sectoral series (Panel B).

At the aggregate level, the factors account for a large share of variation in real activity and prices: 90 percent for total industrial production, 62 percent for real PCE, and 74 percent for services inflation. Credit aggregates and some financial variables behave more idiosyncratically, with substantially lower shares. These rankings are consistent with earlier findings that dynamic factor models perform best for real activity and inflation \citep[e.g.,][]{sw2016}. At the sectoral level, explanatory power is lower on average but varies widely: the common component explains 34 percent of industrial production fluctuations across sectors, ranging from under 10 percent to over 70 percent, with cyclical, credit-intensive industries such as motor vehicles at the top of the distribution. The averages are about 20 percent for sectoral real PCE and PCE inflation, while equity returns are heavily common-driven, averaging 70 percent across industry portfolios.

The dispersion in explanatory power is itself informative, reflecting differences in how tightly variables are tied to common macroeconomic dynamics. It is also what generates the heterogeneity in tail risk documented in Section~\ref{sec: reading-tails}. Because idiosyncratic shocks in our specification are symmetric, they add dispersion to a series but no correlated movements in mean and volatility, tail asymmetry is inherited through the common component alone. A variable's tail behavior is therefore shaped by which factors it loads on and how strongly, even when those factors account for a minority of its total variance. The variance decomposition and the tail asymmetries we document are two manifestations of the same factor structure, which Section~\ref{subsec:heterogeneity} tests directly.

\subsection{Generating Heterogeneity and Structure in Risk\label{subsec:heterogeneity}}

The model generates the heterogeneity and structure in risk documented in Section \ref{sec: reading-tails} through the loadings on the common factors. Table~\ref{tab:asym_loadings} quantified this relationship in regression form; Figure \ref{fig:scatter_tail_asymmetry_combined} further drills down on it. The top panel reproduces Figure \ref{fig:scatter_tail_asymmetry}, which shows the relationship between mean-uncertainty correlation and tail risk across all variables. The second panel shuts off the idiosyncratic components of the variables. Comparing the first and second panels, the qualitative patterns continue to hold. Growth variables congregate in the bottom-left quadrant, inflation variables in the upper-right, and financial variables spread across both. The common component therefore plays a critical role in generating the risk patterns we see.

The bottom panel randomly reshuffles the factor loadings for each variable---reassigning each variable's loadings across factors in a single random draw---while keeping the idiosyncratic components off. The category structure dissolves: growth, inflation, and financial variables scatter across quadrants, no longer aligned with their economic type. The behavior of recreational goods and vehicles and transportation services inflation is instructive. Both inflation series are in the upper-right quadrant in the first two panels, showing the usual inflation-at-risk behavior. Upon reshuffling of the loadings, the two series switch to the bottom-left quadrant. This exercise shows that the factor loading structure, combined with the factors themselves, generates the structure of risk: destroying the loadings destroys the alignment between a variable's economic category and its risk profile that Table~\ref{tab:asym_loadings} quantifies.

The displayed reshuffling is one of many permutations that produce this result: across $10{,}000$ random reassignments of each variable's loadings across factors, the explanatory power of the loadings for tail asymmetry falls from $0.55$ to an average of $0.31$, and the actual regression exceeds every placebo draw. This permutation is deliberately conservative: reassigning a variable's loadings across factors preserves its overall exposure to the common component and scrambles only which factors it is exposed to. Severing the link entirely, by reshuffling loadings across variables within each factor, drives the average to $0.06$. The evidence from Census microdata in Section~\ref{subsec:industry-char} provides the complementary external check that these patterns reflect industry fundamentals rather than model artifacts.

\subsection{Tail Forecast Performance and Calibration \label{subsec:qrcomp}}

We evaluate performance along two dimensions. First, we ask whether our conditional densities are more accurate at the tails than those from quantile regressions. Second, we ask whether the tail quantiles are well calibrated, that is, whether realized outcomes fall below the model's 5th percentile or above the model's 95th percentile roughly 5 percent of the time. The first is a relative comparison against the standard tool in this literature; the second is an absolute property of our densities. Throughout, we report the risk object most relevant to each class of variables: the 12-month lower tail for growth variables, the 12-month upper tail for inflation variables, and the 3-month upper tail for financial variables.

Following \citet{ABG19}, we estimate quantile regressions of the form:
\begin{equation}
\label{eq:qr}
x^{(h)}_{t+h} = \beta_{0,\tau} + \beta_{1,\tau}x_t + \beta_{2,\tau}NFCI_t + \epsilon_{\tau,t+h},
\end{equation}

\noindent where $x^{(h)}_{t+h}$ denotes the variable of interest at horizon $h$, and $\tau \in \{5\%, 25\%, 75\%, 95\%\}$. The National Financial Conditions Index (NFCI) is included to capture time-varying financial conditions. We fit the \citet{ac2003} skewed-$t$ distribution to the resulting quantile forecasts to generate full conditional densities. Both models are evaluated in sample, with conditional distributions computed monthly from July 1976. We report an out-of-sample comparison for selected variables in the Appendix.

Panel (a) of Figure~\ref{fig:factor_qr_qwcrps_ratios} reports, for each of the 116 variables, the ratio of the factor model's quantile-weighted continuous ranked probability score (qwCRPS) of \citet{gr2011} to that of the quantile regression, with ratios below one favoring the factor model.\footnote{The qwCRPS is defined as $qwCRPS_t = \frac{2}{J-1} \sum^{J-1}_{j=1}v\left(\tau_j\right)QS_{\tau_j,t}$ where $QS_{\tau_j,t}$ is the quantile score of \cite{gk2005}. The left qwCRPS uses the weight function $v\left(\tau_j\right) = (1-\tau_j)^2$ and the right qwCRPS uses $v\left(\tau_j\right) = \tau_j^2$ where $\tau_j = 0.05, 0.1, ..., 0.95$ quantiles and $J = 20$. We follow \cite{ccm2024} in placing greater weight on the accuracy of tail predictions relative to the center of the distribution.} The factor model performs as well as or better than quantile regressions for most variables, with ratios at or below one for 80 of 116 series. For growth variables, the two approaches are essentially on par: ratios cluster tightly around one, and few differences are statistically significant in either direction. For inflation variables, the factor model is clearly more accurate, with 22 of 24 ratios below one and most differences significant. For financial variables, the factor model is modestly better, with 26 of 32 ratios below one. Appendix Table~\ref{tab:factor_qr_qwcrps_ratios} reports the underlying numbers by variable. Appendix Table~\ref{tab:qwcrps} reports ratios for four representative series---industrial production growth, real PCE growth, PCE inflation, and the excess bond premium---at both the 3- and 12-month horizons and for both tails. The rankings are similar in recessions and over the full sample, and, more importantly, are broadly robust in a recursive out-of-sample exercise running from 2007 through 2019.

We now turn to whether the tail quantiles are well calibrated.\footnote{A common alternative diagnostic is the probability integral transform (PIT), which is uniformly distributed under correct calibration. PIT histograms are impractical to display for 116 variables, and the two statistics we report capture the relevant information: empirical coverage at the 5th and 95th percentiles is the PIT evaluated at the tail quantiles of interest, while the qwCRPS evaluates the entire predictive density, weighting the tails most heavily.} Panel (b) of Figure~\ref{fig:factor_qr_qwcrps_ratios} reports empirical coverage rates with 95 percent confidence intervals. For growth variables, coverage of the 12-month lower tail---the growth-at-risk object---contains its nominal value for 51 of 60 series. For inflation variables, coverage of the 12-month upper tail contains nominal for 17 of 24 series. Where coverage departs from nominal, it does so in one direction only: realized outcomes fall in the tails \textit{less} often than the nominal rate implies. The model's predictive tails are therefore conservative: it over-signals risk rather than missing it.

Calibration is weaker for financial variables at the three-month horizon, where the upper tail is somewhat wider than realized outcomes require. This conservatism does not come at the cost of accuracy: as Panel (a) shows, the factor model remains competitive with quantile regressions for these same variables and horizon. Coverage also improves substantially at longer horizons, with 66 percent of financial series containing nominal coverage at twelve months.\footnote{Coverage in the subsample of NBER recessions is higher---above 90 percent of series across categories---but rests on 63 observations.}

Taken together, the factor model delivers tail forecasts for all 116 variables that are competitive with quantile regressions overall and clearly more accurate for inflation. It does so from a single estimated model---with an interpretable factor structure, sectoral heterogeneity, and counterfactual capability that variable-by-variable quantile regressions cannot provide jointly. Quantile regressions, which condition directly on financial conditions through the NFCI, retain an advantage for variables that respond immediately to financial stress. Importantly, both approaches produce qualitatively similar conditional distributions for key variables, validating the main risk patterns documented in this paper.

\section{Conclusions\label{sec: conclusions}}

This paper shows that asymmetric tail risk is a systematic, pervasive feature of macroeconomic dynamics. Across 116 U.S. time series spanning five decades, tail risk varies dramatically: real activity faces predominantly downside risk, prices exhibit regime-dependent asymmetries, and financial variables show upside risk during stress. Rather than requiring separate explanations, a unified framework with endogenous stochastic volatility in common factors generates growth-at-risk, inflation-at-risk, and sectoral heterogeneity simultaneously. Heterogeneity reflects differential factor exposures, not idiosyncratic noise: factor exposures explain over half of the cross-sectional variation in tail asymmetry. This structure is what makes the results useful in practice. Because sectoral vulnerabilities are inherited through observable exposures to common dynamics, tracking a small number of factors is enough to see where in the economy risk is concentrating, and how the balance of risks shifts across the business cycle.

The framework offers several directions for future work. First, incorporating time-varying parameters beyond volatility, for instance through a regime-switching structure, would allow the mechanisms themselves to change over time, testing whether the transmission we document is stable across history or differs systematically between expansions and recessions. Second, richer identification schemes would extend the counterfactuals to policy, asking how monetary and fiscal actions reshape the balance of risks across sectors. Third, extending to other economies could reveal whether these mechanisms operate similarly internationally or whether country-specific institutions alter propagation. Finally, the distributional measures we provide offer targets for calibrating and testing dynamic models with nonlinear features: models with occasionally binding constraints \citep{guerrieri-lorenzoni-2017}, financial frictions \citep{brunnermeier2014}, or regime-switching dynamics that predict asymmetric tail behavior but currently lack comprehensive empirical benchmarks.

\bibliographystyle{ecta}
\bibliography{references}

@article{gk2005,
author = {Raffaella Giacomini and Ivana Komunjer},
title = {Evaluation and Combination of Conditional Quantile Forecasts},
journal = {Journal of Business \& Economic Statistics},
volume = {23},
number = {4},
pages = {416--431},
year = {2005},
publisher = {Taylor \& Francis},
doi = {10.1198/073500105000000018},
URL = {
        https://doi.org/10.1198/073500105000000018
},
eprint = { 
        https://doi.org/10.1198/073500105000000018
}
}

@TechReport{am2025,
type={Working Papers},
institution={Federal Reserve Bank of St. Louis},
author={Aaron Amburgey and Michael W. McCracken},
title={Growth-at-Risk is Investment-at-Risk},
year={2025},
month={Aug},
number={2023-020},
doi={10.20955/wp.2023.020},
url={https://ideas.repec.org/p/fip/fedlwp/96594.html},
}

@article{dhk2025,
author = {Sharada Nia Davidson and Chenghan Hou and Gary Koop},
title = {Investigating Economic Uncertainty Using Stochastic Volatility in Mean VARs: The Importance of Model Size, Order-Invariance and Classification},
journal = {Journal of Business \& Economic Statistics},
volume = {43},
number = {4},
pages = {992--1007},
year = {2025},
publisher = {Taylor \& Francis},
doi = {10.1080/07350015.2025.2455064},
URL = { 
        https://doi.org/10.1080/07350015.2025.2455064
},
eprint = {
        https://doi.org/10.1080/07350015.2025.2455064
}
}

@article{gr2011,
author = {Tilmann Gneiting and Roopesh Ranjan},
title = {Comparing Density Forecasts Using Threshold- and Quantile-Weighted Scoring Rules},
journal = {Journal of Business \& Economic Statistics},
volume = {29},
number = {3},
pages = {411--422},
year = {2011},
publisher = {Taylor \& Francis},
doi = {10.1198/jbes.2010.08110},
URL = {
        https://doi.org/10.1198/jbes.2010.08110
},
eprint = {
        https://doi.org/10.1198/jbes.2010.08110
}
}

@article{odf2025,
author = {Opschoor, Daan and Dijk, Dick van and Franses, Philip Hans},
title = {Heterogeneity in Manufacturing Growth Risk},
year = {2025},
journal = {Journal of Money, Credit and Banking},
volume = {n/a},
number = {n/a},
pages = {},
doi = {https://doi.org/10.1111/jmcb.13256},
url = {https://onlinelibrary.wiley.com/doi/abs/10.1111/jmcb.13256},
eprint = {https://onlinelibrary.wiley.com/doi/pdf/10.1111/jmcb.13256}
}

@TechReport{fglz2016,
type={FEDS Notes},
institution={Board of Governors of the Federal Reserve System (U.S.)},
author={Giovanni Favara and Simon Gilchrist and Kurt F. Lewis and Egon Zakraj{\v{s}}ek},
title={Updating the Recession Risk and the Excess Bond Premium},
year={2016},
month={Oct},
number={2016-10-06},
doi={10.17016/2380-7172.1836},
url={https://ideas.repec.org/p/fip/fedgfn/2016-10-06.html},
}

@Article{sw2025,
journal={Brookings Papers on Economic Activity},
author={James H. Stock and Mark W. Watson},
title={Recovering from COVID},
year={2025},
month={None},
pages={297-374},
volume={56},
number={1 (Spring},
doi={None},
url={https://ideas.repec.org/a/bin/bpeajo/v56y2025i2025-01p297-374.html},
}

@Article{ABG19,
  author={Tobias Adrian and Nina Boyarchenko and Domenico Giannone},
  title={{Vulnerable Growth}},
  journal={American Economic Review},
  year=2019,
  volume={109},
  number={4},
  pages={1263-1289},
  month={April},
  doi={}
}

@article{guerrieri-lorenzoni-2017,
  title={Credit crises, precautionary savings, and the liquidity trap},
  author={Guerrieri, Veronica and Lorenzoni, Guido},
  journal={The Quarterly Journal of Economics},
  volume={132},
  number={3},
  pages={1427--1467},
  year={2017},
  publisher={Oxford University Press}
}

@article{fernandez-villaverde-guerron-rubio-ramirez-2011,
  title={Risk matters: The real effects of volatility shocks},
  author={Fern{\'a}ndez-Villaverde, Jes{\'u}s and Guerr{\'o}n-Quintana, Pablo and Rubio-Ram{\'i}rez, Juan F and Uribe, Mart{\'i}n},
  journal={American Economic Review},
  volume={101},
  number={6},
  pages={2530--2561},
  year={2011},
  publisher={American Economic Association}
}

@article{canova-ravn-2000,
  title={The macroeconomic effects of German unification: Real adjustments and the welfare state},
  author={Canova, Fabio and Ravn, Morten O},
  journal={Review of Economic Dynamics},
  volume={3},
  number={3},
  pages={423--460},
  year={2000},
  publisher={Elsevier}
}

@article{fernandez-villaverde-rubio-ramirez-2007,
  title={Estimating macroeconomic models: A likelihood approach},
  author={Fern{\'a}ndez-Villaverde, Jes{\'u}s and Rubio-Ram{\'i}rez, Juan F},
  journal={The Review of Economic Studies},
  volume={74},
  number={4},
  pages={1059--1087},
  year={2007},
  publisher={Wiley-Blackwell}
}

@article{canova-sala-2009,
  title={Back to square one: Identification issues in DSGE models},
  author={Canova, Fabio and Sala, Luca},
  journal={Journal of Monetary Economics},
  volume={56},
  number={4},
  pages={431--449},
  year={2009},
  publisher={Elsevier}
}

@article{canova-ciccarelli-2013,
  title={Panel vector autoregressive models: A survey},
  author={Canova, Fabio and Ciccarelli, Matteo},
  journal={VAR models in macroeconomics--New developments and applications: Essays in honor of Christopher A. Sims},
  volume={32},
  pages={205--246},
  year={2013},
  publisher={Emerald Group Publishing Limited}
}

@article{Black1976,
  author = {Black, Fischer},
  title = {Studies of Stock Market Volatility Changes},
  journal = {Proceedings of the American Statistical Association, Business and Economic Statistics Section},
  year = {1976},
  pages = {177--181}
}

@article{Christie1982,
  author = {Christie, Andrew A.},
  title = {The Stochastic Behavior of Common Stock Variances: Value, Leverage and Interest Rate Effects},
  journal = {Journal of Financial Economics},
  volume = {10},
  number = {4},
  year = {1982},
  pages = {407--432}
}

@article{Schwert1989,
  author = {Schwert, G. William},
  title = {Why Does Stock Market Volatility Change Over Time?},
  journal = {The Journal of Finance},
  volume = {44},
  number = {5},
  year = {1989},
  pages = {1115--1153}
}

@article{Bloom2009,
  author = {Bloom, Nicholas},
  title = {The Impact of Uncertainty Shocks},
  journal = {Econometrica},
  volume = {77},
  number = {3},
  year = {2009},
  pages = {623--685}
}

@article{lopezsalido-loria-2020,
title = {Inflation at risk},
journal = {Journal of Monetary Economics},
volume = {145},
pages = {103570},
year = {2024},
note = {Inflation: Expectations \& Dynamics October 14-15, 2022},
issn = {0304-3932},
doi = {https://doi.org/10.1016/j.jmoneco.2024.103570},
url = {https://www.sciencedirect.com/science/article/pii/S0304393224000230},
author = {David López-Salido and Francesca Loria}
}

@article{kiley-2022,
  title={Unemployment risk},
  author={Kiley, Michael T},
  journal={Journal of Money, Credit and Banking},
  volume={54},
  number={5},
  pages={1407--1424},
  year={2022},
  publisher={Wiley Online Library}
}

@TechReport{CaldaraSVOL,
  author={Dario Caldara and Chiara Scotti and Molin Zhong},
  title={{Macroeconomic and Financial Risks: A Tale of Mean and Volatility}},
  year=2021,
  month=Aug,
  institution={Board of Governors of the Federal Reserve System (U.S.)},
  type={International Finance Discussion Papers},
  url={https://ideas.repec.org/p/fip/fedgif/1326.html},
  number={1326},
  doi={10.17016/IFDP.2021.1326},
}

@article{doi:10.1198/073500102753410408,
author = {Eric Jacquier and Nicholas G Polson and Peter E Rossi},
title = {Bayesian Analysis of Stochastic Volatility Models},
journal = {Journal of Business \& Economic Statistics},
volume = {20},
number = {1},
pages = {69-87},
year  = {2002},
publisher = {Taylor & Francis},
doi = {10.1198/073500102753410408},

URL = { 
    
        https://doi.org/10.1198/073500102753410408
    
    

},
eprint = { 
    
        https://doi.org/10.1198/073500102753410408
    
    

}

}

@article{doi:10.1198/jcgs.2009.08095,
author = {Joshua Chi-Chun Chan and Ivan Jeliazkov},
title = {MCMC Estimation of Restricted Covariance Matrices},
journal = {Journal of Computational and Graphical Statistics},
volume = {18},
number = {2},
pages = {457-480},
year  = {2009},
publisher = {Taylor & Francis},
doi = {10.1198/jcgs.2009.08095},

URL = { 
        https://doi.org/10.1198/jcgs.2009.08095
    
},
eprint = { 
        https://doi.org/10.1198/jcgs.2009.08095
    
}

}

@Article{RePEc:bla:jorssb:v:72:y:2010:i:3:p:269-342,
  author={Christophe Andrieu and Arnaud Doucet and Roman Holenstein},
  title={{Particle Markov chain Monte Carlo methods}},
  journal={Journal of the Royal Statistical Society Series B},
  year=2010,
  volume={72},
  number={3},
  pages={269-342},
  month={},
  doi={},
  url={https://ideas.repec.org/a/bla/jorssb/v72y2010i3p269-342.html}
}

@article{JMLR:v15:lindsten14a,
  author  = {Fredrik Lindsten and Michael I. Jordan and Thomas B. Sch{\"o}n},
  title   = {Particle Gibbs with Ancestor Sampling},
  journal = {Journal of Machine Learning Research},
  year    = {2014},
  volume  = {15},
  pages   = {2145-2184},
  url     = {http://jmlr.org/papers/v15/lindsten14a.html}
}

@incollection{sw2016,
title = {Chapter 8 - Dynamic Factor Models, Factor-Augmented Vector Autoregressions, and Structural Vector Autoregressions in Macroeconomics},
editor = {John B. Taylor and Harald Uhlig},
series = {Handbook of Macroeconomics},
publisher = {Elsevier},
volume = {2},
pages = {415-525},
year = {2016},
issn = {1574-0048},
doi = {https://doi.org/10.1016/bs.hesmac.2016.04.002},
url = {https://www.sciencedirect.com/science/article/pii/S1574004816300027},
author = {J.H. Stock and M.W. Watson}
}

@article{ms2012,
    author = {Mumtaz, Haroon and Surico, Paolo},
    title = "{Evolving International Inflation Dynamics: World and Country-Specific Factors}",
    journal = {Journal of the European Economic Association},
    volume = {10},
    number = {4},
    pages = {716-734},
    year = {2012},
    month = {08},
    issn = {1542-4766},
    doi = {10.1111/j.1542-4774.2012.01068.x},
    url = {https://doi.org/10.1111/j.1542-4774.2012.01068.x},
    eprint = {https://academic.oup.com/jeea/article-pdf/10/4/716/10314602/jeea0716.pdf},
}

@article{MumtazTheodoridis2018,
author = {Haroon Mumtaz and Konstantinos Theodoridis},
title = {The Changing Transmission of Uncertainty Shocks in the U.S.},
journal = {Journal of Business \& Economic Statistics},
volume = {36},
number = {2},
pages = {239--252},
year = {2018},
publisher = {ASA Website},

}

@Article{carter-kohn-94,
  author={Carter, C K and Kohn, R},
  title={On Gibbs sampling for state space models},
  journal={Biometrika},
  year=1994,
  volume={81},
  number={3},
  pages={541--53}
}

@TechReport{delnegro-otrok-08,
  author={Marco {Del Negro} and Christopher Otrok},
  title={Dynamic factor models with time-varying parameters: measuring changes in international business cycles},
  year=2008,
  institution={Federal Reserve Bank of New York},
  type={Staff Reports},
  number={326}
}

@Article{bernanke-boivin-eliasz-05,
  author={Ben Bernanke and Jean Boivin and Piotr S. Eliasz},
  title={Measuring the Effects of Monetary Policy: A Factor-augmented Vector Autoregressive (FAVAR) Approach},
  journal={The Quarterly Journal of Economics},
  year=2005,
  volume={120},
  number={1},
  pages={387-422},
  month={January}
}

@ARTICLE{stock-watson-02,
AUTHOR={Stock, James H and Watson, Mark W},
TITLE={Macroeconomic Forecasting Using Diffusion Indexes},
JOURNAL={Journal of Business \& Economic Statistics},
YEAR=2002,
VOLUME={20},
NUMBER={2},
PAGES={147-62},
MONTH={April}
}

@Article{Banbura-Giannone-Reichlin-10Paper,
  author={Marta Banbura and Domenico Giannone and Lucrezia Reichlin},
  title={Large Bayesian vector auto regressions},
  journal={Journal of Applied Econometrics},
  year=2010,
  volume={25},
  number={1},
  pages={71-92},
}

@article{mccracken2016,
  title={{FRED-MD: A Monthly Database for Macroeconomic Research}},
  author={McCracken, Michael W and Ng, Serena},
  journal={Journal of Business \& Economic Statistics},
  volume={34},
  number={4},
  pages={574-89},
  year={2016},
  publisher={Taylor \& Francis}
}

@article{LMN2020,
  title = "Uncertainty and Business Cycles: Exogenous Impulse or Endogenous Response?",
  journal = "American Economic Journal: Macroeconomics",
  volume = "13",
  pages = "369-410",
  year = "2021",
  Number = {10},
  author = "Sydney C. Ludvigson and Sai Ma and Serena Ng"
}

@article{JLN2013,
  title = {{Measuring Uncertainty}},
  journal = "American Economic Review",
  volume = "105",
  number = "3",
  year = "2015",
  pages= "1177-1216",
  author= "Jurado, K. and Ludvigson, S. and Ng, S."
}

@Article{carriero2016,
	author = {Carriero, A. and Clark, T.E. and Marcellino, M},
	title = {{Measuring Uncertainty and Its Impact on the Economy}},
	journal = {Review of Economics and Statistics},
	year = 2018,
    volume = 100,
	pages = {799-815},
	number = 5,
}

@Article{brunnermeier2014,
  author={Markus K. Brunnermeier and Yuliy Sannikov},
  title={{A Macroeconomic Model with a Financial Sector}},
  journal={American Economic Review},
  year=2014,
  volume={104},
  number={2},
  pages={379-421},
  month={February},
  doi={},
  url={https://ideas.repec.org/a/aea/aecrev/v104y2014i2p379-421.html}
}

@article{GORODNICHENKO201752,
Author = {Gorodnichenko, Yuriy and Ng, Serena},
title = {{Level and Volatility Factors in Macroeconomic Data}},
journal = "Journal of Monetary Economics",
volume = "91",
pages = "52 - 68",
year = "2017",
issn = "0304-3932",
doi = "https://doi.org/10.1016/j.jmoneco.2017.09.004",
url = "http://www.sciencedirect.com/science/article/pii/S0304393217300934"
}

@article{ac2003,
    author = {Azzalini, Adelchi and Capitanio, Antonella},
    title = "{Distributions Generated by Perturbation of Symmetry with Emphasis on a Multivariate Skew t-Distribution}",
    journal = {Journal of the Royal Statistical Society Series B: Statistical Methodology},
    volume = {65},
    number = {2},
    pages = {367-389},
    year = {2003},
    month = {04},
    issn = {1369-7412},
    doi = {10.1111/1467-9868.00391},
    url = {https://doi.org/10.1111/1467-9868.00391},
    eprint = {https://academic.oup.com/jrsssb/article-pdf/65/2/367/49684466/jrsssb\_65\_2\_367.pdf},
}

@Article{ccm2024,
  author={Andrea Carriero and Todd E. Clark and Massimiliano Marcellino},
  title={{Capturing Macro‐Economic Tail Risks with Bayesian Vector Autoregressions}},
  journal={Journal of Money, Credit and Banking},
  year=2024,
  volume={56},
  number={5},
  pages={1099-1127},
  month={August},
  doi={10.1111/jmcb.13121},
  url={https://ideas.repec.org/a/wly/jmoncb/v56y2024i5p1099-1127.html}
}

@article{DieboldMariano1995,
  author  = {Diebold, Francis X. and Mariano, Roberto S.},
  title   = {Comparing Predictive Accuracy},
  journal = {Journal of Business \& Economic Statistics},
  year    = {1995},
  volume  = {13},
  number  = {3},
  pages   = {253--263},
  doi     = {10.1080/07350015.1995.10524599}
}

@article{Andrews1991,
  author  = {Andrews, Donald W. K.},
  title   = {Heteroskedasticity and Autocorrelation Consistent Covariance Matrix Estimation},
  journal = {Econometrica},
  year    = {1991},
  volume  = {59},
  number  = {3},
  pages   = {817--858},
  doi     = {10.2307/2938229}
}

@article{AndrewsMonahan1992,
  author  = {Andrews, Donald W. K. and Monahan, J. Christopher},
  title   = {An Improved Heteroskedasticity and Autocorrelation Consistent Covariance Matrix Estimator},
  journal = {Econometrica},
  year    = {1992},
  volume  = {60},
  number  = {4},
  pages   = {953--966},
  doi     = {10.2307/2951574}
}

@article{bai-ng-2002,
  title = {Determining the Number of Factors in Approximate Factor Models},
  author = {Bai, Jushan and Ng, Serena},
  journal = {Econometrica},
  volume = {70},
  number = {1},
  pages = {191--221},
  year = {2002}
}

@article{gabaix2011,
  title = {The Granular Origins of Aggregate Fluctuations},
  author = {Gabaix, Xavier},
  journal = {Econometrica},
  volume = {79},
  number = {3},
  pages = {733--772},
  year = {2011}
}

\clearpage

\begin{table}[htbp]
\begin{center}
\caption{Cross-Sectional Determinants of Tail Asymmetry\label{tab:asym_loadings}}
\footnotesize
\begin{tabular}{llS[table-format=-1.3]S[table-format=1.3]S[table-format=1.3]}
\toprule
 & & \multicolumn{2}{c}{Joint regression} & {Univariate} \\
\cmidrule(lr){3-4}\cmidrule(lr){5-5}
Factor & Anchor variable & {Coefficient} & {Std.\ error} & {$R^2$} \\
\midrule
Factor 1 & Excess bond premium      & 0.389  & 0.073 & 0.385 \\
Factor 2 & 1-Year Treasury rate     & 0.113  & 0.082 & 0.074 \\
Factor 3 & S\&P 500                 & -0.104 & 0.045 & 0.041 \\
Factor 4 & Real PCE                 & 0.004  & 0.074 & 0.052 \\
Factor 5 & PCE housing              & -0.048 & 0.080 & 0.055 \\
Factor 6 & Non-revolving credit     & -0.187 & 0.061 & 0.023 \\
Factor 7 & PCE inflation            & 0.132  & 0.051 & 0.245 \\
\addlinespace
\multicolumn{2}{l}{Intercept}       & -0.055 & 0.038 & {} \\
\midrule
\multicolumn{2}{l}{Observations}          & \multicolumn{3}{c}{$116$} \\
\multicolumn{2}{l}{Adjusted $R^2$}        & \multicolumn{3}{c}{$0.522$} \\
\bottomrule
\end{tabular}
\end{center}

{\small Note: The table reports cross-sectional regressions of tail asymmetry on
estimated factor loadings across the 116 variables in the panel. Tail asymmetry is
the log ratio of the time-series standard deviation of the 95th percentile of each
variable's conditional distribution to that of the 5th percentile, as defined in
Section~\ref{sec: reading-tails}, using the 12-month horizon for growth and
inflation variables and the 3-month horizon for financial variables. The joint
regression includes all seven loadings simultaneously; the final column reports the
$R^2$ from regressing tail asymmetry on each loading separately. Standard errors
are heteroskedasticity-robust (HC1). Anchor variables identify each factor under
the named-factor normalization described in Section~\ref{sec:model}.}
\end{table}

\clearpage

\clearpage
\begin{figure}[h!]
\caption{Aggregate Risk Indices: Growth, Inflation, and Financial Conditions}
\label{fig:agg_risk_indices}
\includegraphics[width=\textwidth]{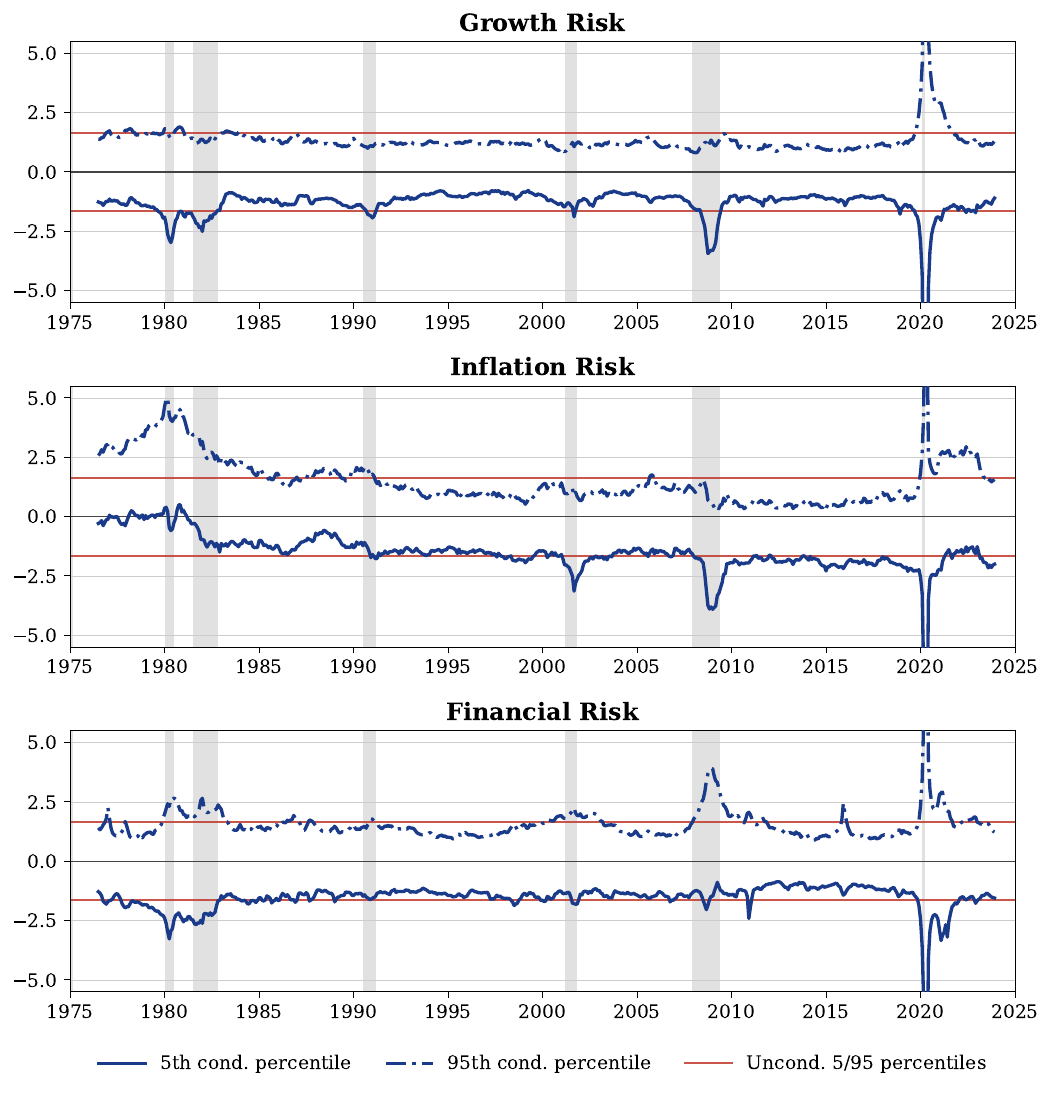}

\small{Note: This figure shows aggregate risk indices for economic growth, inflation, and financial conditions. Each series is constructed as the cross-sectional average of the $5th$ and $95th$ percentiles of the conditional distributions implied by the model for a group of related indicators. The blue solid and dashed lines report the averages of the $5th$ and $95th$ percentiles, respectively. The red solid lines report as reference the $5th$ and $95th$ percentiles of a standard normal distribution. The y-axis is truncated at ±5.5 standard deviations to preserve readability of pre-pandemic dynamics. During the COVID-19 pandemic, peak values exceed this range: the growth risk index reaches -21.7 (5th percentile) and 16.9 (95th percentile), the financial risk index reaches -11.0 and 10.3, and the inflation risk index reaches -13.7 and 10.5. See Section~\ref{subsec: agg-indexes-construction} for details about the construction of the indices. Shaded areas denote NBER recessions.}
\end{figure}

\clearpage

\begin{figure}
\caption{Tail Risk Asymmetry and the Mean-Uncertainty Correlation Across the Macroeconomy}
\label{fig:scatter_tail_asymmetry}
\includegraphics[width=\textwidth]{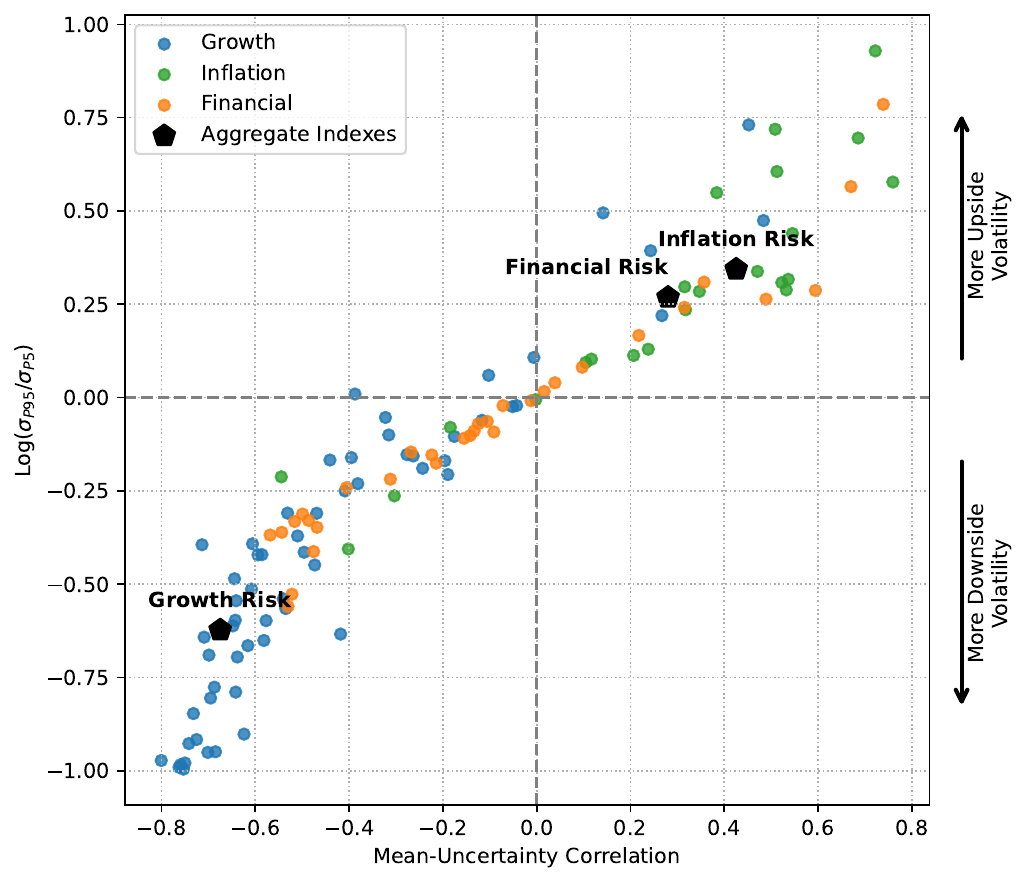}
\small{Note: For each variable, the x-axis shows the correlation 
between the conditional mean and uncertainty of its predictive 
distribution over time---a measure of the leverage and volatility-in-mean effects---and 
the y-axis shows tail asymmetry, defined as the log ratio of the 
standard deviation of the 95th to the 5th percentile of the 
conditional distribution over time. Positive values indicate more 
volatile upper tails (upside risk); negative values indicate more 
volatile lower tails (downside risk). Colors denote variable 
categories. Pentagons denote the aggregate risk indices, whose
mean and uncertainty are computed by the same cross-sectional
averaging used for the index quantiles
(Section~\ref{subsec: agg-indexes-construction}). Sample:
July 1976--June 2019.}

\end{figure}

\clearpage

\clearpage

\clearpage

\clearpage

\clearpage

\clearpage 

\clearpage

\clearpage

\begin{figure}[htbp]
\caption{Tail Variability of Price Ratios Across Expenditure Categories in the GFC (left) and Great Inflation (right) \label{fig:price_ratios}}
    \begin{center}
    \includegraphics[width=\textwidth]{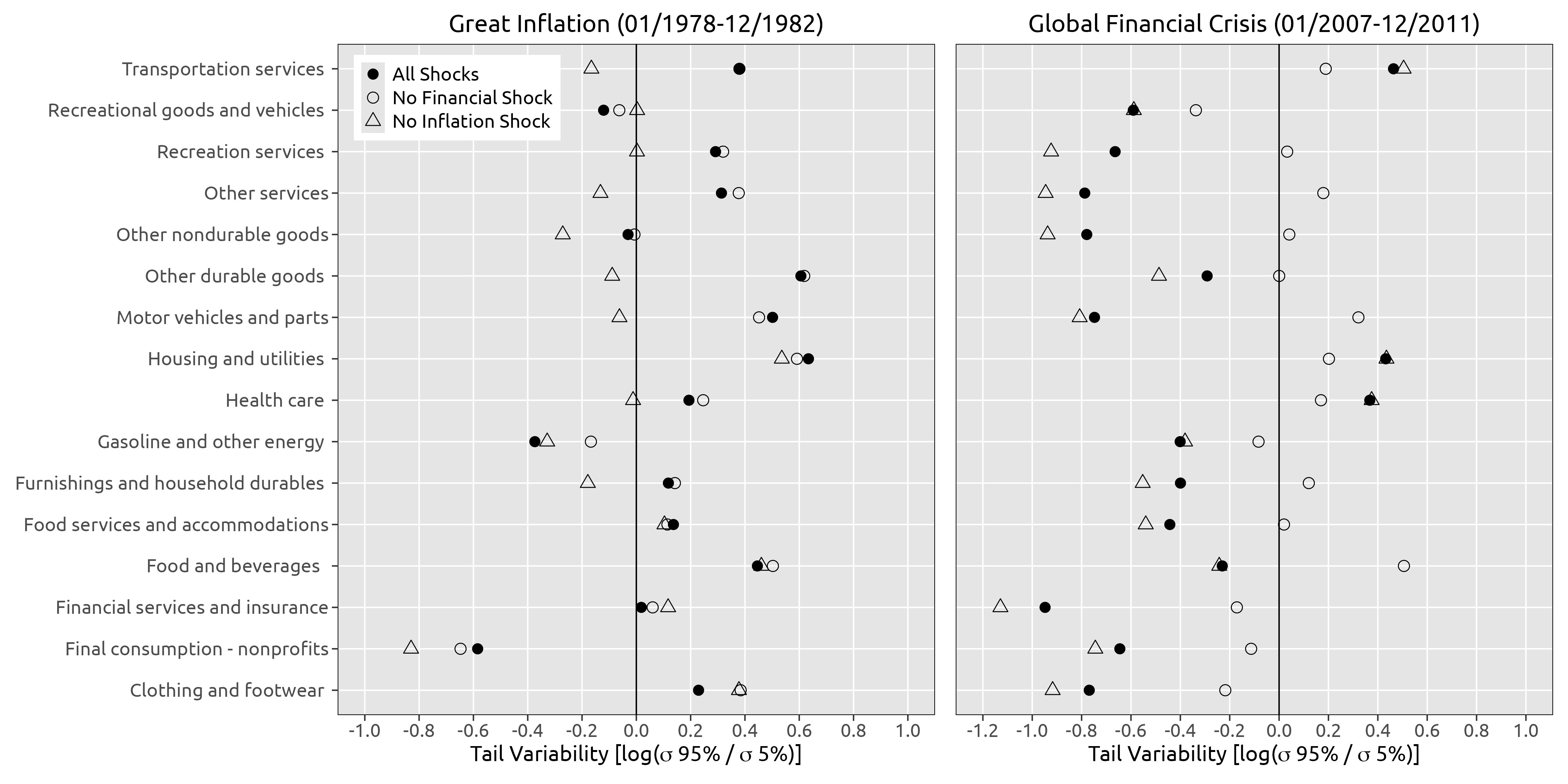}
    \end{center}
    
    {\small Note: Tail variability of price ratios across different expenditure categories during two crisis periods: Great Inflation (January 1978 - December 1982) and Global Financial Crisis (January 2007 - December 2011). The x-axis shows the log ratio of the 95th to 5th percentiles. Different markers represent scenarios with all shocks (filled circles), no financial shock (open circles), and no inflation shock (triangles).}

\end{figure}

\clearpage

\begin{figure}[htbp]
    \caption{Mean-Uncertainty Correlation versus Tail Risk: Robustness}
    \begin{center}
        
    \includegraphics[width=0.62\linewidth]{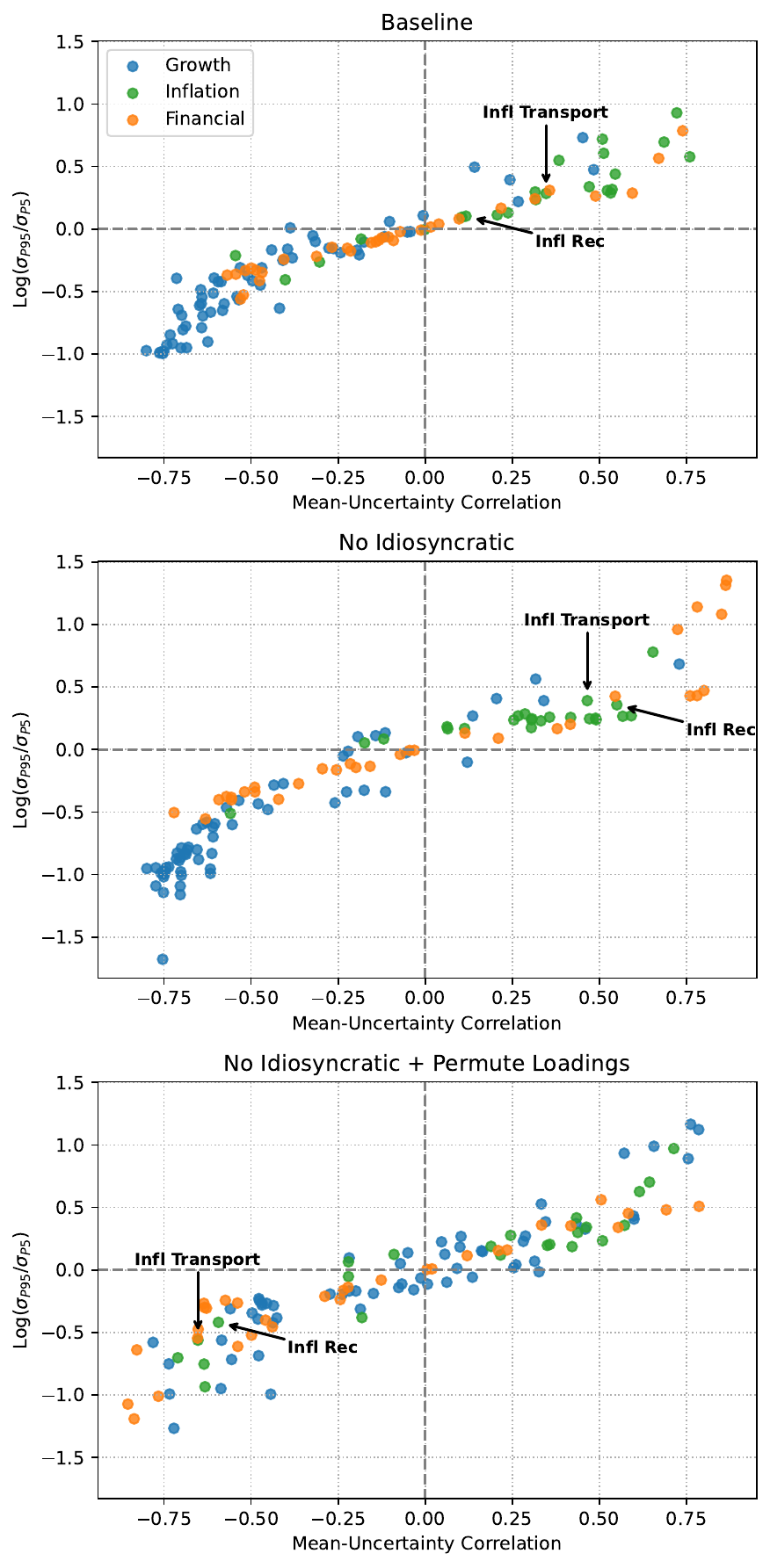}
    \end{center}
    \vspace*{-0.5cm}
    {\small Note: The top figure reproduces the scatter in Figure \ref{fig:scatter_tail_asymmetry}. The second figure generates the scatter shutting off the idiosyncratic components of all variables. The third figure generates the scatter shutting off the idiosyncratic components of all variables and randomly reshuffling the factor loadings for each variable.}
    \label{fig:scatter_tail_asymmetry_combined}
\end{figure}

\begin{figure}[!htbp]
\caption{Tail Forecast Performance and Calibration}
\label{fig:factor_qr_qwcrps_ratios}
\vspace{-0.2cm}
\begin{center}
\begin{subfigure}{\textwidth}
    \caption{Relative accuracy: factor model vs.\ quantile regression}
    \label{fig:panel_qwcrps}
    \includegraphics[width=\textwidth]{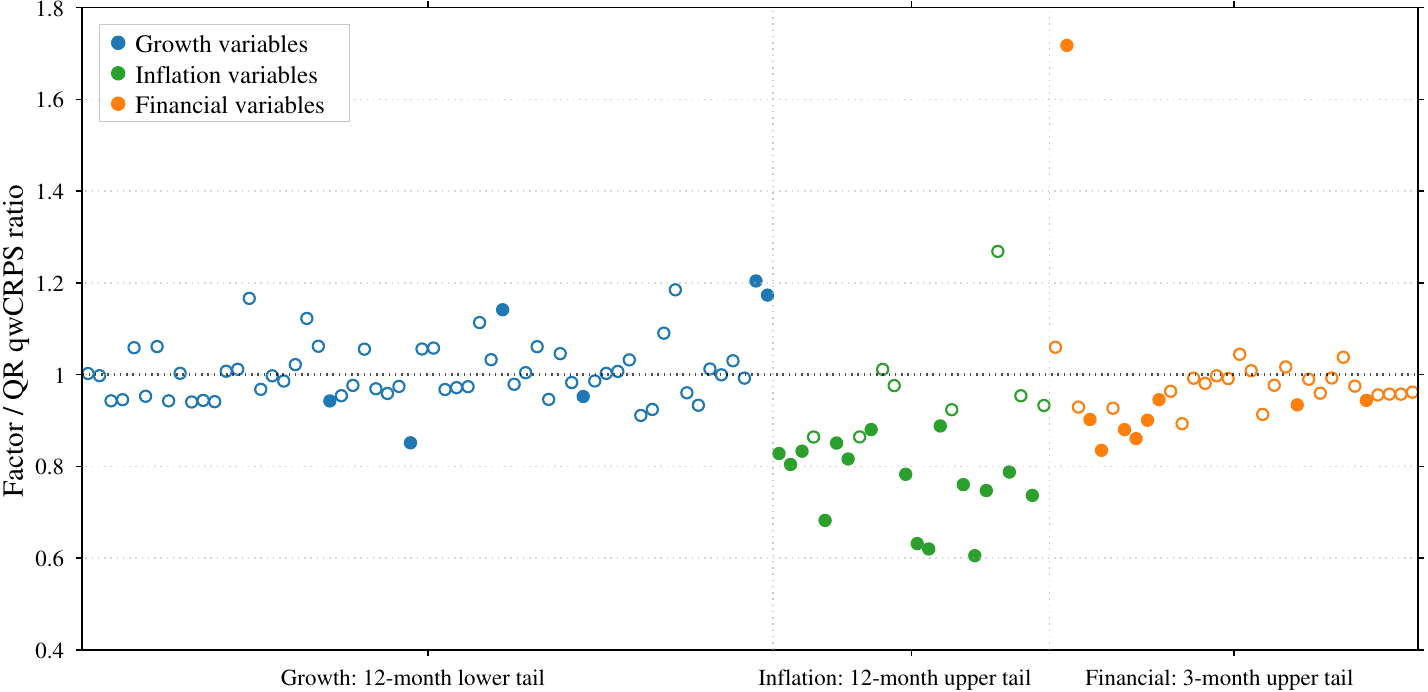}
\end{subfigure}

\vspace{0.4cm}

\begin{subfigure}{\textwidth}
    \caption{Tail calibration: empirical coverage}
    \label{fig:panel_coverage}
    % PLACEHOLDER: replace with the coverage figure formatted as in Panel (a)
    \includegraphics[width=\textwidth]{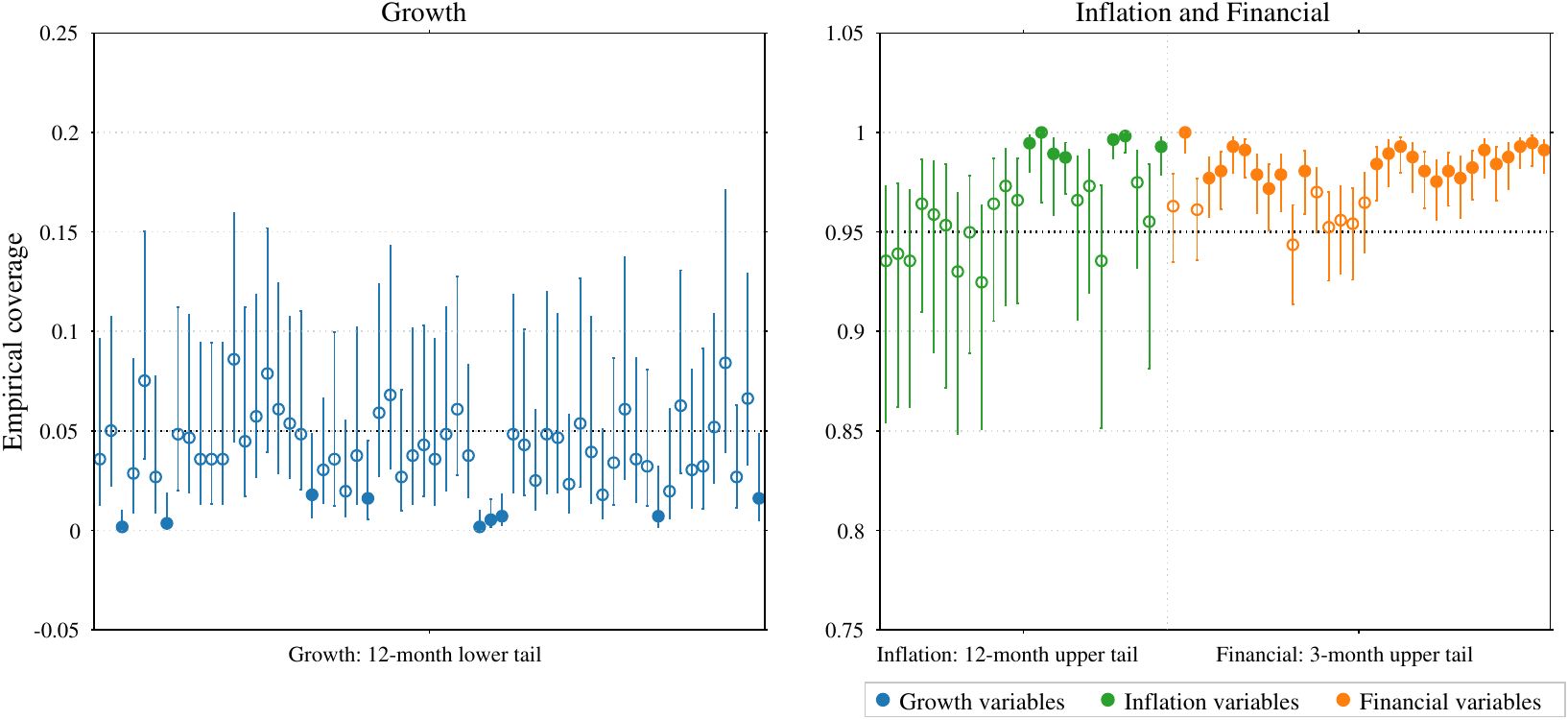}
\end{subfigure}
\end{center}

\small{\noindent Note: Each dot is one of 116 variables, grouped and colored by
category and ordered as in Table~\ref{table:data}. Growth uses the 12-month lower tail, inflation uses the 12-month upper tail, and financial variables use the 3-month upper tail. Both panels are based on in-sample conditional distributions computed monthly from July 1976 through September 2023 (3 months ahead) and December 2022 (12 months ahead). Panel (a) reports the ratio of the
mean quantile-weighted continuous ranked probability score for the factor
model to that for the quantile regression; ratios below one (dotted line) favor the
factor model, and filled circles denote rejection of equal predictive accuracy at
the 5 percent level using a two-sided Diebold--Mariano test. Panel (b) reports
empirical quantile coverage with 95 percent confidence intervals; dotted lines mark
nominal coverage.}

\end{figure}

\clearpage
\appendix
\renewcommand{\thefigure}{A.\arabic{figure}}
\renewcommand{\thetable}{A.\arabic{table}}
\setcounter{page}{1}
\renewcommand{\thepage}{A-\arabic{page}}

\setcounter{figure}{0}
\setcounter{table}{0}
\setcounter{equation}{0}

\begin{center}
    \textbf{\Large{Online Appendix}}
\end{center}
\vspace{0.5cm}

\section{\label{appsec: data}Data}

Our dataset includes 116 monthly time series from January 1973 to December 2023, which Table \ref{table:data} summarizes. We use FRED for the aggregate real activity, financial, and inflation series, as well as the sectoral industrial production series. We use NIPA Tables 2.8.3 and 2.8.4 for the sectoral real consumption and price series. For the sectoral equity series, we use Kenneth French's database. For the excess bond premium, we use the updates provided by \cite{fglz2016} and for the corporate spread series not found in FRED, we use Global Financial Database. 

We consider four modifications to the data. First, we induce stationarity to most series by either converting them to growth rates with a log difference transformation or computing a first difference. For the series that are already stationary, we either do not perform any transformation or take their logs. Second, we standardize the data using full-sample standard deviations. Third, when computing the risk indices (Figure \ref{fig:agg_risk_indices}) and scatters (Figures \ref{fig:scatter_tail_asymmetry} and \ref{fig:scatter_tail_asymmetry_combined}), we normalize each series to be in line with the discussion in Section \ref{subsec: agg-indexes-construction}. Finally, for robustness, we check real activity series for outliers pre-COVID and replace them with the median of the preceding five observations, following \cite{sw2025}. Doing so does not make a difference for estimation compared to our benchmark results that do not use outlier detection.

The normalization column specifies the sign in which the series enters into the growth, financial, or inflation index. For aggregate growth variables, we classify the sign based off the sum of coefficients from a regression of the series on six leads and lags of industrial production growth, following \cite{sw2025}. The sectoral industrial production and real PCE series enter the index with a positive sign. For the financial series, we specify that rates, spreads, and the VIX enter the index with a positive sign, while credit quantities and stock returns enter with a negative sign. Prices all enter the inflation index with a positive sign.

\scriptsize
\setlength{\tabcolsep}{3pt}
\begin{longtable}{@{}l>{\raggedright\arraybackslash}p{0.42\textwidth}llc@{}}
\caption{List of Data Series Used in Estimation.} \label{table:data}\\
\hline
Mnemonic & Full Name & Category & Transf. & Norm. \\ 
\hline
\endfirsthead
\multicolumn{5}{l}{Table \ref{table:data}: Data description, continued}\\[1ex]
\hline
Mnemonic & Full Name & Category & Transf. & Norm. \\ 
\hline
\endhead
\hline
\endfoot

\multicolumn{5}{l}{\textbf{Aggregate Variables}} \\[1ex]
\hline
EBP & Excess Bond Premium & Financial & No transform & 1 \\
Y1\_TSY & 1-Year Treasury Rate & Financial & No transform & 1 \\
REAL\_PERS\_X\_TR & Real personal income ex transfer receipts & Growth & Log diff & 1 \\
REAL\_MANU & Real Manu. and Trade Industries Sales & Growth & Log diff & 1 \\
INIT\_CLAI & Initial Claims & Growth & Log diff & -1 \\
ALL\_EMPL & All Employees: Total nonfarm & Growth & Log diff & 1 \\
HOUS\_STAR & Housing Starts: Total New Privately Owned & Growth & Log & 1 \\
NEW\_ORDE & New Orders for Durable Goods & Growth & Log diff & 1 \\
TOTA\_BUSI & Total Business: Inventories to Sales Ratio & Growth & First diff & -1 \\
COMM\_LOANS & Commercial and Industrial Loans & Financial & Log diff & -1 \\
REAL\_ESTA & Real Estate Loans at All Commercial Banks & Financial & Log diff & -1 \\
TOTA\_NONR & Total Nonrevolving Credit & Financial & Log diff & -1 \\
NONR\_CONS & Nonrevolving consumer credit to Personal Income & Financial & First diff & -1 \\
CONS\_MOTO & Consumer Motor Vehicle Loans Outstanding & Financial & Log diff & -1 \\
TOTA\_CONS & Total Consumer Loans and Leases Outstanding & Financial & Log diff & -1 \\
SECU\_IN & Securities in Bank Credit at All Commercial Banks & Financial & Log diff & -1 \\
EQ\_SP500\_COMP & S\&P's Common Stock Price Index: Composite & Financial & Log diff & -1 \\
VIX & VIX & Financial & No transform & 1 \\
MON3\_TREA & 3-Month Treasury Bill & Financial & No transform & 1 \\
Y10\_TSY & 10-Year Treasury C Minus FEDFUNDS & Financial & No transform & 1 \\
BAA\_MINU & Baa - 10-Year Treasury & Financial & No transform & 1 \\
DJ\_YIELD & Dow Jones Corporate Bond - 10-Year Treasury & Financial & No transform & 1 \\
HIGH10\_Y & Federal Reserve 10-year High Quality Corporate Bond - 10-Year Treasury & Financial & No transform & 1 \\
BARRON\_Y & Barron's Best Grade Bond Yield - 10-Year Treasury & Financial & No transform & 1 \\
IP\_INDE & IP Index & Growth & Log diff & 1 \\
REAL\_PERS & Real personal consumption expenditures & Growth & Log diff & 1 \\
PERS\_CONS & PCE: Chain Index Prices & Inflation & Log diff & 1 \\[2ex]

\multicolumn{5}{l}{\textbf{Equity}} \\[1ex]
\hline
EQ\_NODUR & Equities: Consumer Nondurables & Financial & Log diff & -1 \\
EQ\_DURBL & Equities: Consumer Durables & Financial & Log diff & -1 \\
EQ\_MANUF & Equities: Manufacturing & Financial & Log diff & -1 \\
EQ\_ENRGY & Equities: Energy & Financial & Log diff & -1 \\
EQ\_HITEC & Equities: Business Equipment & Financial & Log diff & -1 \\
EQ\_TELCM & Equities: Telecommunications & Financial & Log diff & -1 \\
EQ\_SHOPS & Equities: Shops & Financial & Log diff & -1 \\
EQ\_HLTH & Equities: Healthcare & Financial & Log diff & -1 \\
EQ\_UTILS & Equities: Utilities & Financial & Log diff & -1 \\
EQ\_OTHER & Equities: Other & Financial & Log diff & -1 \\
EQ\_LO20 & Equities: Size, Bottom 20\% & Financial & Log diff & -1 \\
EQ\_QNT2 & Equities: Size, 21\%-40\% & Financial & Log diff & -1 \\
EQ\_QNT3 & Equities: Size, 41\%-60\% & Financial & Log diff & -1 \\
EQ\_QNT4 & Equities: Size, 61\%-80\% & Financial & Log diff & -1 \\
EQ\_HI20 & Equities: Size, Top 20\% & Financial & Log diff & -1 \\[2ex]

\multicolumn{5}{l}{\textbf{Industrial Production}} \\[1ex]
\hline
IP\_MANU\_SIC & IP: Manufacturing (SIC) & Growth & Log diff & 1 \\
IP\_MANU\_NAICS & IP: Manufacturing (NAICS) & Growth & Log diff & 1 \\
IP\_DURA\_MAN & IP: Durable Manufacturing & Growth & Log diff & 1 \\
IP\_WOOD & IP: Wood Products (NAICS = 321) & Growth & Log diff & 1 \\
IP\_NONM & IP: Nonmetallic Mineral Products (NAICS = 327) & Growth & Log diff & 1 \\
IP\_PRIM & IP: Primary Metals (NAICS = 331) & Growth & Log diff & 1 \\
IP\_FABR & IP: Fabricated Metal Products (NAICS = 332) & Growth & Log diff & 1 \\
IP\_MACH & IP: Machinery (NAICS = 333) & Growth & Log diff & 1 \\
IP\_COMP & IP: Computer and Electronic Products (NAICS = 334) & Growth & Log diff & 1 \\
IP\_ELEC\_EQUIP & IP: Electrical Equip, Appliances, and Components (NAICS = 335) & Growth & Log diff & 1 \\
IP\_MOTO & IP: Motor Vehicles and Parts (NAICS = 3361, 3362, 3363) & Growth & Log diff & 1 \\
IP\_AERO & IP: Aerospace and Misc Transportation Equipment (NAICS = 3364-3369) & Growth & Log diff & 1 \\
IP\_FURN & IP: Furniture and Related Products (NAICS = 337) & Growth & Log diff & 1 \\
IP\_DURA\_MISC & IP: Durable Misc (NAICS = 339) & Growth & Log diff & 1 \\
IP\_NOND & IP: Nondurable Manufacturing & Growth & Log diff & 1 \\
IP\_FOOD & IP: Food, Beverage, and Tobacco Products (NAICS = 311, 312) & Growth & Log diff & 1 \\
IP\_TEXT & IP: Textile and Product Mills (NAICS = 313, 314) & Growth & Log diff & 1 \\
IP\_APPA & IP: Apparel and Leather (NAICS = 315, 316) & Growth & Log diff & 1 \\
IP\_PAPE & IP: Paper (NAICS = 322) & Growth & Log diff & 1 \\
IP\_PRIN & IP: Printing and Support (NAICS = 323) & Growth & Log diff & 1 \\
IP\_PETR & IP: Petroleum and Coal Products (NAICS = 324) & Growth & Log diff & 1 \\
IP\_CHEM & IP: Chemicals (NAICS = 325) & Growth & Log diff & 1 \\
IP\_PLAS & IP: Plastics and Rubber Products (NAICS = 326) & Growth & Log diff & 1 \\
IP\_OTHE & IP: Other Manufacturing (Non-NAICS) (NAICS = 1133, 5111) & Growth & Log diff & 1 \\
IP\_MINI & IP: Mining (NAICS = 21) & Growth & Log diff & 1 \\
IP\_UTIL & IP: Utilities (NAICS = 2211, 2212) & Growth & Log diff & 1 \\
IP\_ELEC & IP: Electric (NAICS = 2211) & Growth & Log diff & 1 \\
IP\_NATU & IP: Natural Gas (NAICS = 2212) & Growth & Log diff & 1 \\[2ex]

\multicolumn{5}{l}{\textbf{Real Personal Consumption Expenditures (Quantities)}} \\[1ex]
\hline
RPCE\_GOOD\_QUAN & Goods & Growth & Log diff & 1 \\
RPCE\_DURA\_GOOD & Durable goods & Growth & Log diff & 1 \\
RPCE\_MOTO\_VEHI & Motor vehicles and parts & Growth & Log diff & 1 \\
RPCE\_FURN & Furnishings and durable household equipment  & Growth & Log diff & 1 \\
RPCE\_RECR\_GOOD & Recreational goods and vehicles  & Growth & Log diff & 1 \\
RPCE\_OTHE\_DURA & Other durable goods  & Growth & Log diff & 1 \\
RPCE\_NOND\_GOOD & Nondurable goods  & Growth & Log diff & 1 \\
RPCE\_FOOD & Food and beverages purchased for off-premises consumption  & Growth & Log diff & 1 \\
RPCE\_CLOT & Clothing and footwear  & Growth & Log diff & 1 \\
RPCE\_GASO & Gasoline and other energy goods  & Growth & Log diff & 1 \\
RPCE\_OTHE\_NOND & Other nondurable goods  & Growth & Log diff & 1 \\
RPCE\_SERV\_QUAN & Services  & Growth & Log diff & 1 \\
RPCE\_HOUS\_CONS & Household consumption expenditures (for services)  & Growth & Log diff & 1 \\
RPCE\_HOUS & Housing and utilities  & Growth & Log diff & 1 \\
RPCE\_HEAL\_CARE & Health care  & Growth & Log diff & 1 \\
RPCE\_TRAN\_SERV & Transportation services  & Growth & Log diff & 1 \\
RPCE\_RECR\_SERV & Recreation services  & Growth & Log diff & 1 \\
RPCE\_FOOD\_SERV & Food services and accommodations  & Growth & Log diff & 1 \\
RPCE\_FINA\_SERV & Financial services and insurance  & Growth & Log diff & 1 \\
RPCE\_OTHE\_SERV & Other services  & Growth & Log diff & 1 \\
RPCE\_FINA\_CONS & Final consumption expenditures of NPISHs  & Growth & Log diff & 1 \\
RPCE\_GROS\_OUTP & Gross output of nonprofit institutions  & Growth & Log diff & 1 \\
RPCE\_LESS\_RECE & Less: Receipts from sales of goods and services by nonprofit inst  & Growth & Log diff & 1 \\[2ex]

\multicolumn{5}{l}{\textbf{Personal Consumption Expenditure Price Indices}} \\[1ex]
\hline
P\_GOOD\_PRIC & Goods  & Inflation & Log diff & 1 \\
P\_DURA\_GOOD & Durable goods  & Inflation & Log diff & 1 \\
P\_MOTO\_VEHI & Motor vehicles and parts  & Inflation & Log diff & 1 \\
P\_FURN\_ & Furnishings and durable household equipment  & Inflation & Log diff & 1 \\
P\_RECR\_GOOD & Recreational goods and vehicles  & Inflation & Log diff & 1 \\
P\_OTHE\_DURA & Other durable goods  & Inflation & Log diff & 1 \\
P\_NOND\_GOOD & Nondurable goods  & Inflation & Log diff & 1 \\
P\_FOOD & Food and beverages purchased for off-premises consumption  & Inflation & Log diff & 1 \\
P\_CLOT & Clothing and footwear  & Inflation & Log diff & 1 \\
P\_GASO & Gasoline and other energy goods  & Inflation & Log diff & 1 \\
P\_OTHE\_NOND & Other nondurable goods  & Inflation & Log diff & 1 \\
P\_SERV\_PRIC & Services  & Inflation & Log diff & 1 \\
P\_HOUS\_CONS & Household consumption expenditures (for services)  & Inflation & Log diff & 1 \\
P\_HOUS & Housing and utilities  & Inflation & Log diff & 1 \\
P\_HEAL\_CARE & Health care  & Inflation & Log diff & 1 \\
P\_TRAN\_SERV & Transportation services  & Inflation & Log diff & 1 \\
P\_RECR\_SERV & Recreation services  & Inflation & Log diff & 1 \\
P\_FOOD\_SERV & Food services and accommodations  & Inflation & Log diff & 1 \\
P\_FINA\_SERV & Financial services and insurance  & Inflation & Log diff & 1 \\
P\_OTHE\_SERV & Other services  & Inflation & Log diff & 1 \\
P\_FINA\_CONS & Final consumption expenditures of NPISHs  & Inflation & Log diff & 1 \\
P\_GROS\_OUTP & Gross output of nonprofit institutions  & Inflation & Log diff & 1 \\
P\_LESS\_RECE & Less: Receipts from sales of goods and services by nonprofit inst  & Inflation & Log diff & 1 \\
\hline
\end{longtable}

\small{Note: Log diff (diff) refers to the first difference of the log (first difference). Normalization indicates the sign of the loading on the aggregate risk indices.}

\setlength{\tabcolsep}{6pt}
\normalsize

\section{\label{appsec: model estimation} Model Estimation} 

This appendix provides technical details on the estimation of the dynamic factor model with endogenous stochastic volatility presented in Section~\ref{sec:model}. The model is specified by equations~\ref{eq:measurement}--\ref{eq:transition-vol-h} in the main text. We describe the prior distributions, the Gibbs sampling algorithm, and provide Monte Carlo evidence that the algorithm accurately recovers model parameters.

\subsection{Prior Distributions}

We employ standard conjugate priors where possible and non-informative priors otherwise. Table~\ref{tab:priors} summarizes the prior specifications. Prior hyperparameters are calibrated using the first 30 observations (training sample) when needed.

\begin{table}[h]
\centering
\caption{Prior Specifications\label{tab:priors}}
\begin{tabular}{lcc}
\toprule
Parameter & Prior Distribution & Hyperparameters \\
\midrule
\multicolumn{3}{l}{\textit{Observation Equation}} \\
Factor loadings $B_i$ (row $i$) & $N(B_0, V_{B0})$ & $B_0$: OLS, $V_{B0} = 10 \times \text{OLS var}$ \\
AR coefficients $\tilde{\rho}_i$ & $N(0, 0.1 \times I_L)$ & $L$: number of lags \\
Idiosyncratic vol. variance $g_i$ & $IG(0.001, 1)$ & Scale $= 0.001$, df $= 1$ \\
\midrule
\multicolumn{3}{l}{\textit{Transition Equation}} \\
VAR coefficients $\Gamma$, $\tilde{\Gamma}$ & $N(\Gamma_0, P_0)$ & Via dummy observations (see below) \\
Volatility shock variance $S$ & $IG(0.001, 1)$ & Scale $= 0.001$, df $= 1$ \\
Correlation matrix $\Sigma$ & Via Huang-Wand (2013) & $a_{kj} \sim N(0,1)$ \\
\midrule
\multicolumn{3}{l}{\textit{Initial Conditions}} \\
Factors $F_0$ & $N(\hat{F}_1, I_N)$ & $\hat{F}_1$: first principal component \\
Log volatilities $h_{i0}$ & $N(\mu_{0,i}, 1)$ & $\mu_{0,i}$: from training sample VAR \\
\bottomrule
\end{tabular}
\end{table}

\subsubsection*{VAR Coefficient Priors via Dummy Observations}

The priors for VAR coefficients $\Gamma = \text{vec}([c; \beta_j; b_k])$ in equation~\ref{eq:transition-factors-F} and $\tilde{\Gamma} = \text{vec}([\alpha; \theta; d_j])$ in equation~\ref{eq:transition-vol-h} are implemented through dummy observations, following \cite{Banbura-Giannone-Reichlin-10Paper}. This approach allows us to specify Minnesota-style priors in a straightforward manner.

The dummy observations are:
\begin{equation}
y_D = \begin{bmatrix}
\frac{\text{diag}(\gamma_1 s_1, \ldots, \gamma_N s_N)}{\tau} \\
0_{N(P-1) \times N} \\
0_{EX \times N}
\end{bmatrix}, \quad
x_D = \begin{bmatrix}
\frac{J_P \otimes \text{diag}(s_1, \ldots, s_N)}{\tau} & 0_{NP \times EX} \\
0_{N \times NP} & 0_{N \times EX} \\
0_{EX \times NP} & I_{EX} / c
\end{bmatrix}
\end{equation}
where $\gamma_i$ denotes the prior mean for the first own-lag coefficient (obtained from univariate AR(1) regressions), $s_i$ is the residual standard deviation from the AR(1) regression, $\tau$ controls overall tightness, $c$ controls tightness on exogenous regressors, $EX$ is the number of exogenous variables, and $J_P = \text{diag}(1, 2, \ldots, P)$ implements lag decay.

We set $\tau = 0.1$ (tight prior on VAR dynamics). For coefficients on lagged volatilities, we use $c = 0.1$; for intercepts, we use $c = 1000$ (flat prior). The prior moments are then:
\begin{align}
\Gamma_0 &= (x_D' x_D)^{-1}(x_D' y_D), \\
P_0 &= S \otimes (x_D' x_D)^{-1},
\end{align}
where $S$ is a diagonal matrix with training sample estimates of factor variances on the diagonal.

\subsubsection*{Initial Conditions}

We initialize the factors at $F_0 \sim N(\hat{F}_1, I_N)$ where $\hat{F}_1$ denotes the first observation of the principal component estimate of the factors. For the log volatilities, we estimate a VAR on the training sample, compute the Cholesky decomposition of the residual covariance matrix, and set $h_{i0} \sim N(\mu_{0,i}, 1)$ where $\mu_{0,i}$ are the log of the Cholesky diagonal elements.

\subsection{Gibbs Sampling Algorithm}

We draw from the posterior distribution using a Gibbs sampler that cycles through seven blocks. Let $\Theta$ denote all model parameters. In each step below, we draw from the conditional posterior given the most recent draws of all other parameters.

\subsubsection*{Step 1: Latent Factors, $G(F_t | \Theta)$}

Conditional on the stochastic volatilities $\tilde{h}_t$, $r_t$ and other parameters, the model has a linear Gaussian state-space representation. We stack the system as:
\begin{align}
Z_t &= \Lambda + \Phi Z_{t-1} + E_t, \quad \text{var}(E_t) = Q_{F,t}, \\
\tilde{X}_t &= \tilde{B} Z_t + V_t, \quad \text{var}(V_t) = R_{F,t},
\end{align}
where the state vector is $Z_t = [\tilde{h}_{t+1}', F_t', \tilde{h}_t', F_{t-1}', \ldots]'$. The matrices $\Phi$ and $\Lambda$ capture the VAR dynamics with appropriate lags, $\tilde{X}_t$ contains the AR-filtered observables, and $\tilde{B}$ adjusts for the AR structure in idiosyncratic shocks. The covariance matrix $Q_{F,t}$ contains $\Omega_t$ (equation~\ref{eq:omega}) in the upper-left block with zeros elsewhere, and $R_{F,t} = \text{diag}(0_N, R_t)$ reflects time-varying idiosyncratic volatility.

We apply the \cite{carter-kohn-94} algorithm to draw the state vector from its conditional posterior distribution, which is Gaussian given the linear structure and known volatilities.

\subsubsection*{Step 2: Factor Loadings $B$ and AR Coefficients $\rho_i$}

Given factors $F_t$ and volatilities $R_t$, the observation equation becomes $n$ independent regressions with known heteroscedasticity. For row $i$ of $B$:
\begin{equation}
X_{it}^* = B_i F_t^* + \tilde{v}_{it}, \quad \tilde{v}_{it} \sim N(0, 1),
\end{equation}
where $X_{it}^* = (X_{it} - \sum_{l=1}^L \rho_{i,l} X_{it-l}) / \sqrt{\exp(r_{it})}$ and $F_t^*$ is similarly transformed. The conditional posterior is $N(m_i, v_i)$ with:
\begin{equation}
v_i = (V_{B0}^{-1} + F_t^{*'} F_t^*)^{-1}, \quad m_i = v_i(V_{B0}^{-1} B_0 + F_t^{*'} X_{it}^*).
\end{equation}

The AR coefficients $\tilde{\rho}_i = [\rho_{i,1}, \ldots, \rho_{i,L}]$ are drawn similarly after computing residuals $v_{it} = X_{it} - B_i F_t$ and GLS-transforming by $\sqrt{\exp(r_{it})}$.

\subsubsection*{Step 3: Idiosyncratic Volatilities, $G(r_{it} | \Theta)$}

Conditional on factors and factor loadings, we obtain $n$ independent univariate stochastic volatility models:
\begin{align}
u_{it} &= \sqrt{\exp(r_{it})} \bar{u}_{it}, \quad \bar{u}_{it} \sim N(0,1), \\
r_{it} &= r_{it-1} + g_i^{1/2} \epsilon_{it}, \quad \epsilon_{it} \sim N(0,1).
\end{align}
We draw $r_{it}$ using the algorithm of \cite{doi:10.1198/073500102753410408}. Given $r_{it}$, the variance parameter $g_i$ has an inverse Gamma posterior: $IG((r_{it} - r_{it-1})' (r_{it} - r_{it-1}) + 0.001, T + 1)$.

\subsubsection*{Step 4: VAR Coefficients, $G(\Gamma, \tilde{\Gamma} | \Theta)$}

Conditional on $F_t$, $S$, $H_t$, and $\Sigma$, equations~\ref{eq:transition-factors-F} and~\ref{eq:transition-vol-h} form a SUR system with time-varying heteroscedasticity:
\begin{equation}
Y_t = X_t \Pi_t + E_t, \quad \text{var}(E_t) = G_t \Sigma G_t',
\end{equation}
where $Y_t = [\tilde{h}_{t+1}', F_t']'$, $G_t = \text{diag}([\tilde{s}^{1/2}, \exp(\tilde{h}_t)^{1/2}])$ and $\Pi_t = [\Gamma', \tilde{\Gamma}']'$ stacks the coefficients. Given the time-varying covariance structure, we use the Kalman filter as a computational device to accumulate the posterior recursively. The transition equation $\Pi_t = \Pi_{t-1}$ carries no state innovation, so the coefficients are constant over time; the filter serves only to handle the time-varying error covariance. We draw the coefficients from the terminal distribution $N(\Pi_{T|T}, P_{T|T})$, which is the posterior for the fixed coefficient vector. The Kalman recursions are:
\begin{align}
\Pi_{t|t-1} &= \Pi_{t-1|t-1}, \quad P_{t|t-1} = P_{t-1|t-1}, \\
\eta_{t|t-1} &= Y_t - X_t \Pi_{t|t-1}, \quad f_{t|t-1} = X_t P_{t|t-1} X_t' + G_t \Sigma G_t', \\
K_t &= P_{t|t-1} X_t' f_{t|t-1}^{-1}, \\
\Pi_{t|t} &= \Pi_{t|t-1} + K_t \eta_{t|t-1}, \quad P_{t|t} = P_{t|t-1} - K_t X_t P_{t|t-1}.
\end{align}
The filter is initialized at $\Pi_0$ and $P_0$ from the prior.

\subsubsection*{Step 5: Variance of Volatility Shocks, $G(S | \Theta)$}

Given the residuals $\eta_t$ from equation~\ref{eq:transition-vol-h} and $\Sigma$, the transition equation can be written as:
\begin{equation}
\tilde{h}_{t+1} - S^{1/2} \mu_{\eta_t | e_t} = \alpha + \theta \tilde{h}_t + \sum_{j=1}^Q d_j F_{t-j} + \eta_t^*, \quad \text{var}(\eta_t^*) = S^{1/2} \Sigma_{\eta | e} S^{1/2'},
\end{equation}
where $\mu_{\eta_t | e_t} = e_t \Sigma_e^{-1} \Sigma_{\eta e}$ and $\Sigma_{\eta | e} = \Sigma_\eta - \Sigma_{\eta e}' \Sigma_e^{-1} \Sigma_{\eta e}$.

The correlation between volatility shocks makes the conditional posterior non-standard, requiring a Metropolis step. We use a mixture proposal:
\begin{equation}
q(S_j) = \varkappa \cdot IG(v_1, T_1) + (1 - \varkappa) \cdot IG(v(S_{j-1}), T(\bar{V})),
\end{equation}
where $v_1 = \tilde{\eta}_{it}' \tilde{\eta}_{it} + 0.001$ and $T_1 = T + 1$ (based on residuals $\tilde{\eta}_{it}$ from the $i$th equation). The second component is centered at the previous draw with standard deviation $\bar{V}$:
\begin{equation}
v(S_{j-1}) = 2 S_{j-1} \left(1 + \frac{S_{j-1}^2}{\bar{V}^2}\right), \quad
T(\bar{V}) = 2\left(2 + \frac{S_{j-1}^2}{\bar{V}^2}\right).
\end{equation}
We set $\varkappa = 0.5$ and tune $\bar{V}$ for an acceptance rate near 25-40\%. The draw is accepted with probability:
\begin{equation}
\alpha = \min\left\{1, \frac{g(E_t | S_j) q(S_{j-1})}{g(E_t | S_{j-1}) q(S_j)}\right\},
\end{equation}
where $g(E_t | S_j)$ is the likelihood of the transition equation residuals given $S_j$.

\subsubsection*{Step 6: Covariance Matrix, $G(\Sigma | \Theta)$}

We draw the restricted covariance matrix $\Sigma$ (with unit diagonal) using the independence Metropolis algorithm of \cite{doi:10.1198/jcgs.2009.08095}. They decompose $\Sigma = L^{-1} D L^{-1'}$ where $L$ is lower triangular with ones on the diagonal and $D$ is diagonal. The unit diagonal restriction on $\Sigma$ implies:
\begin{align}
\lambda_1 &= 1, \\
\lambda_k &= 1 - \sum_{j=1}^{k-1} (a_{kj})^2 \lambda_j, \quad k = 2, \ldots, N+n,
\end{align}
where $a_{kj}$ are the lower-triangular elements of $L^{-1}$ and $\lambda_k$ are the diagonal elements of $D$.

The proposal density for the vector $\mathbf{a} = [a_{21}, a_{31}, a_{32}, \ldots]'$ is:
\begin{equation}
f(\mathbf{a} | \varepsilon_t) = N(\mu, \tau V),
\end{equation}
where $V = (A_0^{-1} + \sum_{t=1}^T U_t \hat{D}^{-1} U_t)^{-1}$, $\mu = V(A_0^{-1} \mathbf{a}_0 + \sum_{t=1}^T U_t \hat{D}^{-1} \varepsilon_t)$, and $U_t$ is a matrix constructed from the residuals $\varepsilon_t$ as described in \cite{doi:10.1198/jcgs.2009.08095}. The diagonal matrix $\hat{D}$ is obtained by iterating between the formula for $\mu$ and the restriction equations for $\lambda_k$. The draw is accepted with probability:
\begin{equation}
\alpha = \min\left\{1, \frac{g(\varepsilon_t | \Sigma_{\text{new}}) f(\mathbf{a}_{\text{old}} | \varepsilon_t)}{g(\varepsilon_t | \Sigma_{\text{old}}) f(\mathbf{a}_{\text{new}} | \varepsilon_t)}\right\},
\end{equation}
with the restriction that $\lambda_k > 0$ for all $k$ to ensure positive definiteness.

\subsubsection*{Step 7: Stochastic Volatilities, $G(\tilde{h}_t | \Theta)$}

This is the most computationally intensive step. Conditional on factors and other parameters, we have a multivariate non-linear state-space system. We rewrite it as:
\begin{align}
\digamma_t &= C + \Psi \digamma_{t-1} + N_t, \\
F_t - H_t^{1/2} \mu_{e_t | \eta_t} &= c + \sum_{j=1}^P \beta_j F_{t-j} + \sum_{k=1}^K b_k \tilde{h}_{t-k} + \tilde{e}_t,
\end{align}
where $\digamma_t = [\eta_{t+1}', \eta_t', \tilde{h}_{t+1}', \tilde{h}_t', \tilde{h}_{t-1}', \ldots]'$ includes the volatility shocks as state variables, $\mu_{e_t | \eta_t} = \eta_t \Sigma_\eta^{-1} \Sigma_{\eta e}'$ is the conditional mean, $\Sigma_{e_t | \eta_t} = \Sigma_e - \Sigma_{\eta e} \Sigma_\eta^{-1} \Sigma_{\eta e}'$ is the conditional variance, and $\text{var}(\tilde{e}_t) = H_t^{1/2} \Sigma_{e | \eta} H_t^{1/2'}$. Importantly, $\tilde{e}_t$ is uncorrelated with $N_t$.

We use particle Gibbs with ancestor sampling \citep{RePEc:bla:jorssb:v:72:y:2010:i:3:p:269-342, JMLR:v15:lindsten14a} to sample from the conditional posterior. Let $\digamma_t^{(i-1)}$ denote the trajectory from the previous Gibbs iteration. The algorithm with $\tilde{M}$ particles proceeds as:

\textbf{For $t=1$:}
\begin{enumerate}
\item Draw $\digamma_1^{(j)}$ for $j=1, \ldots, \tilde{M}-1$. Fix $\digamma_1^{(\tilde{M})} = \digamma_1^{(i-1)}$.
\item Compute normalized weights $p_1^{(j)} = w_1^{(j)} / \sum_{j=1}^{\tilde{M}} w_1^{(j)}$ where $w_1^{(j)}$ is the likelihood:
\begin{equation}
w_1^{(j)} = |\Omega_1^{(j)}|^{-1/2} \exp\left(-\frac{1}{2} \tilde{e}_1' (\Omega_1^{(j)})^{-1} \tilde{e}_1\right).
\end{equation}
\end{enumerate}

\textbf{For $t=2$ to $T$:}
\begin{enumerate}
\item Resample indices $a_t^{(j)}$ for $j=1, \ldots, \tilde{M}-1$ with $P(a_t^{(j)} = k) \propto p_{t-1}^{(k)}$.
\item Draw $\digamma_t^{(j)} | \digamma_{t-1}^{(a_t^{(j)})}$ from the transition equation for $j=1, \ldots, \tilde{M}-1$.
\item Fix $\digamma_t^{(\tilde{M})} = \digamma_t^{(i-1)}$.
\item \textbf{Ancestor sampling:} Draw $a_t^{(\tilde{M})}$ with probability proportional to:
\begin{equation}
w_{t-1}^{(j)} \prod_{s=t}^{t-1+\mathcal{L}} g(F_s | \digamma_{1:t-1}^{(j)}, \digamma_{t:s}^{(i-1)}) f(\digamma_s^{(i-1)} | \digamma_{1:t-1}^{(j)}, \digamma_{t:s-1}^{(i-1)}),
\end{equation}
where $\mathcal{L}$ is a look-ahead parameter (we set $\mathcal{L} = 5$). This step breaks the reference trajectory into pieces, preventing particle degeneracy.
\item Update weights as in step (2) above.
\end{enumerate}

\textbf{Final step:} Draw $\digamma_t^{(i)}$ by sampling index $j$ with probability $p_T^{(j)}$.

We use $\tilde{M} = 20$ particles. The ancestor sampling step is critical for good mixing; without it, the particle system collapses to the reference trajectory.

\subsection{Monte Carlo Validation}

To verify the algorithm's performance, we conduct a Monte Carlo experiment with a simplified two-factor version of the model. We generate 600 observations, discarding the first 100 as burn-in and retaining 500 for analysis, from the following data generating process:
\begin{align}
\begin{pmatrix} \ln h_{1,t+1} \\ \ln h_{2,t+1} \end{pmatrix} &= 
\begin{pmatrix} 0.97 & -0.01 \\ 0.01 & 0.97 \end{pmatrix}
\begin{pmatrix} \ln h_{1t} \\ \ln h_{2t} \end{pmatrix} +
\begin{pmatrix} -0.05 & 0.01 \\ -0.05 & 0.01 \end{pmatrix}
\begin{pmatrix} F_{1,t-1} \\ F_{2,t-1} \end{pmatrix} +
\begin{pmatrix} 0.1^{1/2} e_{1t} \\ 0.1^{1/2} e_{2t} \end{pmatrix}, \\
\begin{pmatrix} F_{1t} \\ F_{2t} \end{pmatrix} &= 
\begin{pmatrix} 0.3 \\ -0.3 \end{pmatrix} +
\begin{pmatrix} 0.75 & -0.1 \\ 0.1 & 0.75 \end{pmatrix}
\begin{pmatrix} F_{1,t-1} \\ F_{2,t-1} \end{pmatrix} +
\begin{pmatrix} -0.05 & 0.01 \\ -0.05 & 0.01 \end{pmatrix}
\begin{pmatrix} \ln h_{1,t-1} \\ \ln h_{2,t-1} \end{pmatrix} +
\begin{pmatrix} h_{1t}^{1/2} e_{3t} \\ h_{2t}^{1/2} e_{4t} \end{pmatrix},
\end{align}
with $\text{Corr}(\mathbf{e}_t) = \begin{pmatrix} 1 & 0.2 & 0.3 & -0.4 \\ 0.2 & 1 & 0.6 & 0.2 \\ 0.3 & 0.6 & 1 & -0.2 \\ -0.4 & 0.2 & -0.2 & 1 \end{pmatrix}$.

We generate $n=100$ observable series as $X_t = BF_t + v_t$ where $B$ is drawn from $N(0,1)$, the persistence parameters $\rho_i \sim U(0.1, 0.7)$, and idiosyncratic volatilities evolve as $r_{it} = r_{it-1} + 0.01^{1/2} \epsilon_{it}$.

\begin{figure}
\caption{Estimation on Simulated Data \label{fig:estim-simul-data}}
\includegraphics[width=\textwidth,clip=true,trim = 2cm 6cm 2cm 5.5cm]{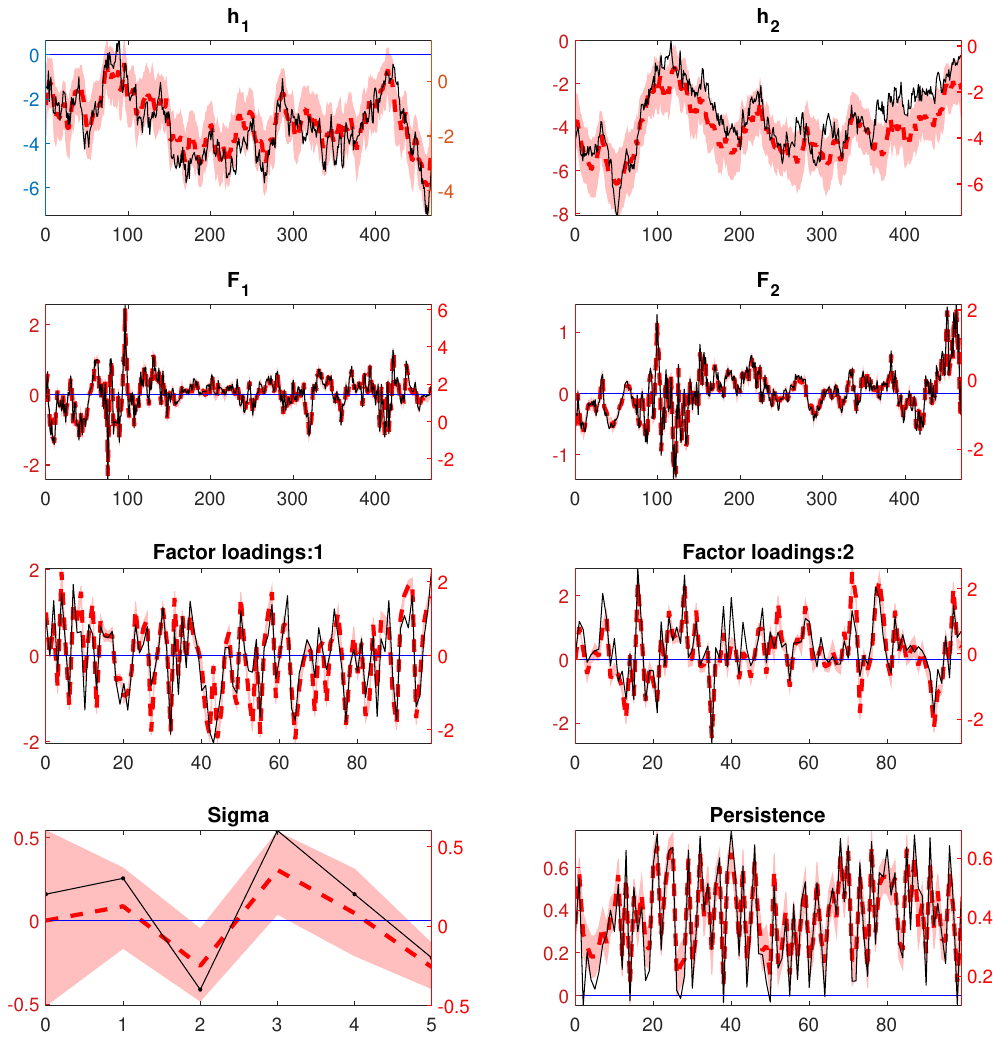}
\small{Note: Estimated Parameters and true values. The two factors are denoted by $F_{1}$ and $F_{2}$, the log stochastic volatility by $h_{1}$ and $h_{2}$. The covariance of the error terms of the transition equation for the factors and stochastic volatilities is denoted by \textit{Sigma}, the autoregressive coefficients for the idiosyncratic errors by \textit{persistence} and two columns of the factor loadings matrix by \textit{Factor loadings:1} and \textit{Factor loadings:2}}
\end{figure}

The model is estimated using 21,000 MCMC draws, discarding the first 1,000 as burn-in and retaining every 10th draw. Figure~\ref{fig:estim-simul-data} compares true parameter values (black) with posterior medians (red dots) and 95\% credible intervals (red shaded). The algorithm recovers VAR coefficients, volatility persistence, in-mean effects, and correlation structure accurately. Some scale differences arise because we standardize the data before estimation, but the structural relationships are correctly identified.

\clearpage

\clearpage

\section{Effects of Shocks to Financial and Inflation Factors \label{app:shocks}}

This section presents the effects of a shock to the first factor (financial) in October 2008 and to the seventh factor (inflation) in December 1980. We assume a Cholesky decomposition with a shock to the financial factor ordered first and shock to the inflation factor ordered second.

\begin{figure}[b]
\caption{Effects of a Shock to the First Factor on Factors and Volatilities in Oct $2008$}
\label{fig:shock_factor_vol_appendix}
\begin{center}
\includegraphics[scale = 0.55,keepaspectratio]{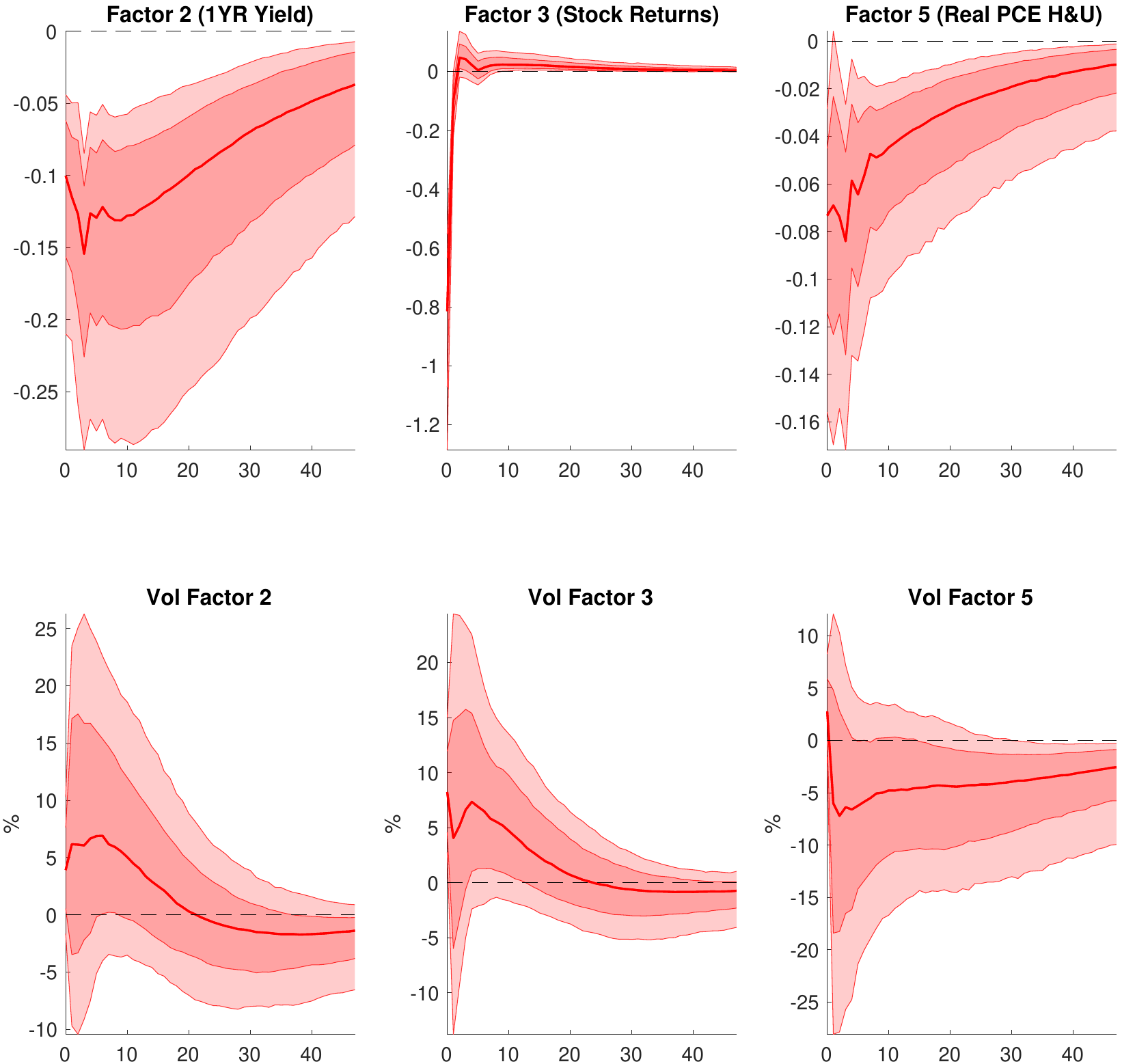}
\end{center}
\vspace{0.5cm}
\footnotesize{Note: This figure plots impulse response functions of a one standard deviation shock to the first factor in a Cholesky decomposition ordered first given October $2008$ conditions. The first column shows the level factor responses and the second column the log volatility responses. The line is the posterior median, the dark shaded area is the $68\%$ credible set and the light shaded area is the $90\%$ credible set.} 
\end{figure}

Figure \ref{fig:shock_factor_vol_appendix} shows the effects of the shock to the financial factor on the second (1 year Treasury yield), third (stock returns), and fifth (housing consumption) factors, which are the responses not shown in the main draft. The shock lowers the Treasury and housing consumption factors persistently, consistent with the effects of a financial shock. The stock return factor sharply declines. The volatilities of the Treasury and stock return factor innovations increase, while the volatility of the housing consumption factor innovations slightly decline.

\begin{figure}
\caption{Effects of a Shock to the Seventh Factor on Factors and Volatilities in Dec $1980$}
\label{fig:factor7shock_factor_vol}
\begin{center}
\includegraphics[scale = 0.45,keepaspectratio]{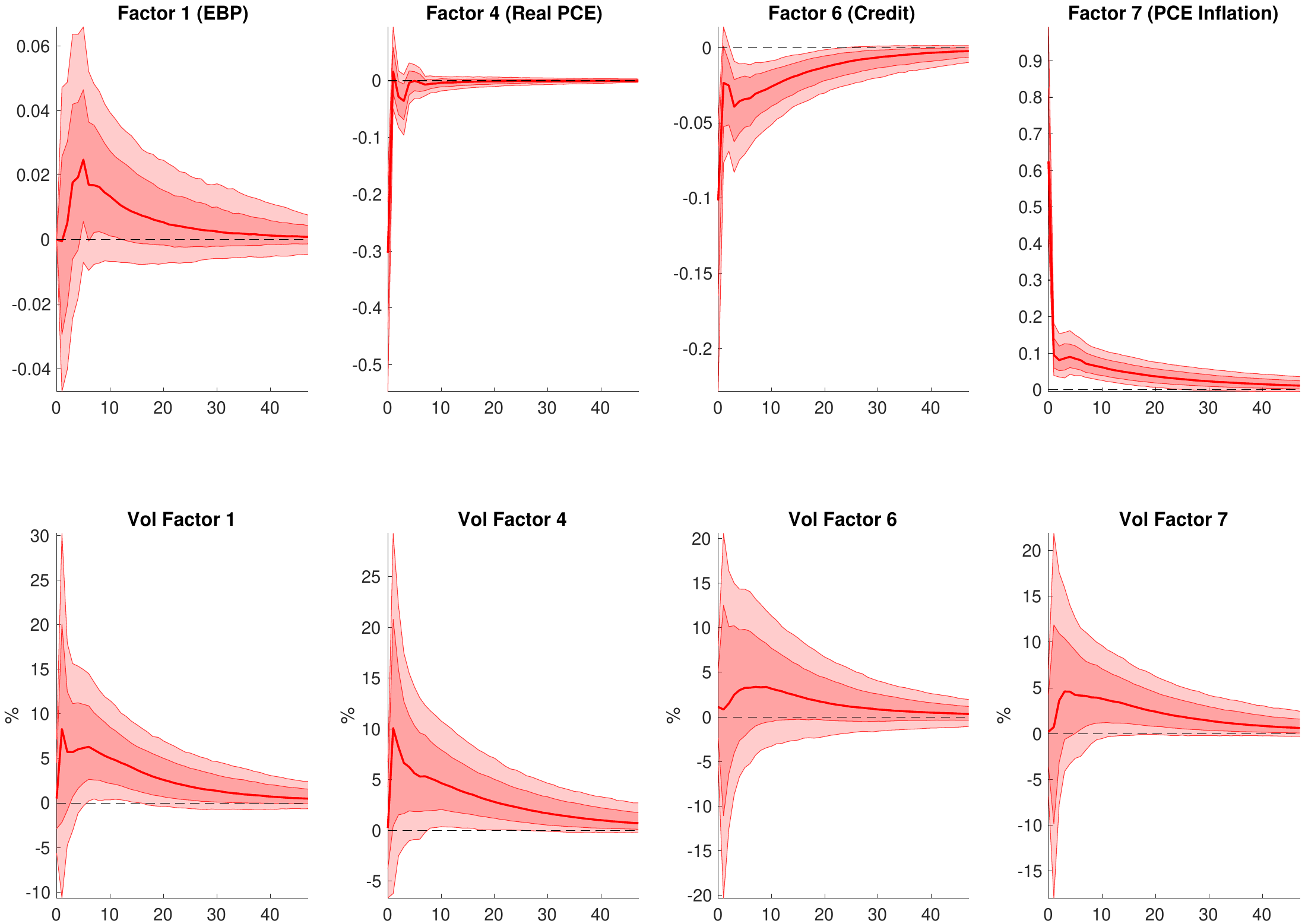}
\end{center}
\vspace{0.5cm}
\footnotesize{Note: This figure plots impulse response functions of a one standard deviation shock to the seventh factor in a Cholesky decomposition ordered second given December $1980$ conditions. The first column shows the level factor responses and the second column the log volatility responses. The line is the posterior median, the dark shaded area is the $68\%$ credible set and the light shaded area is the $90\%$ credible set.} 
\end{figure}

\begin{figure}[b]
\caption{Effects of a Shock to the Seventh Factor on Factors and Volatilities in Dec $1980$}
\label{fig:factor7shock_factor_vol_appendix}
\begin{center}
\includegraphics[scale = 0.55,keepaspectratio]{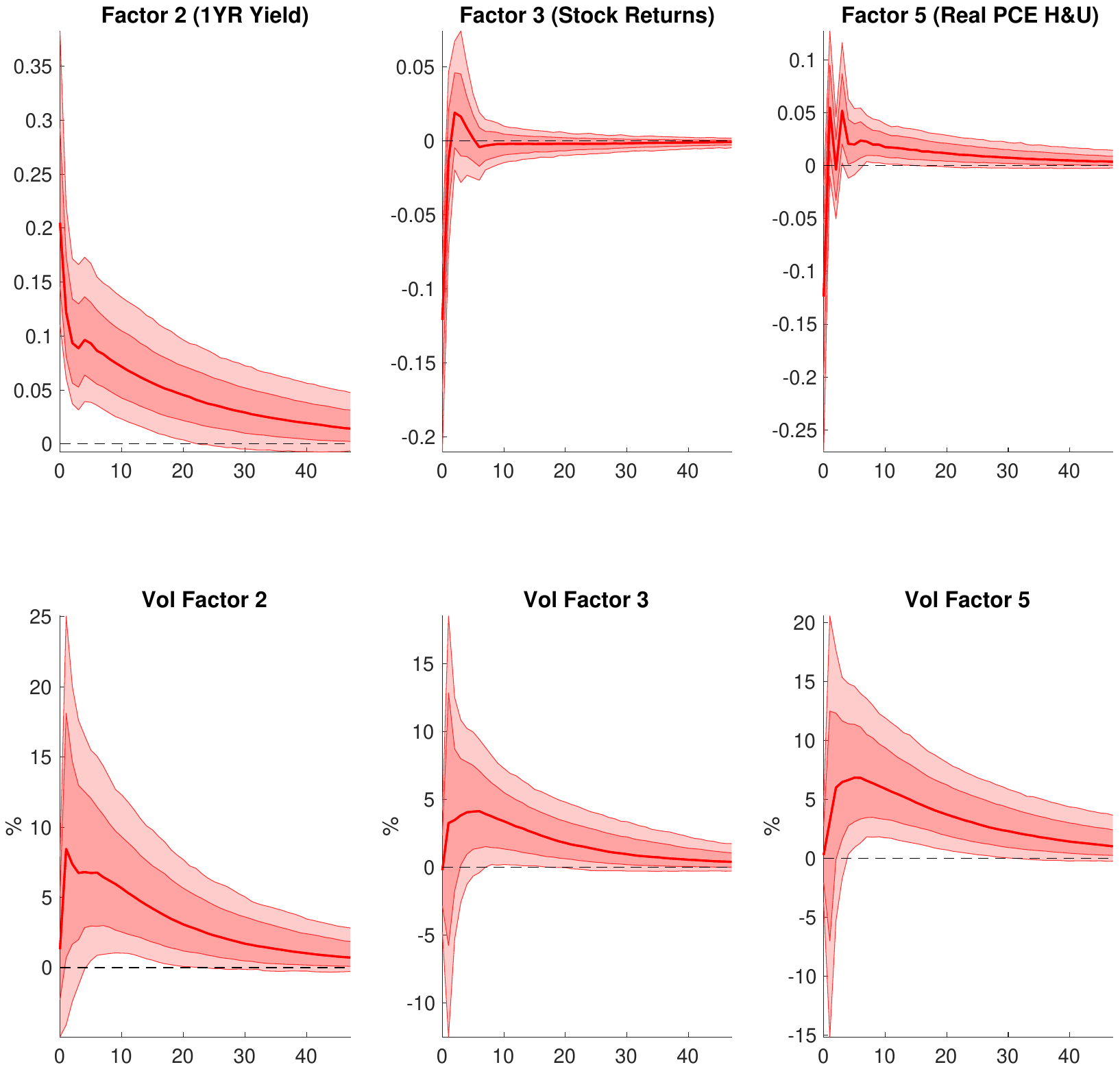}
\end{center}
\vspace{0.5cm}
\footnotesize{Note: This figure plots impulse response functions of a one standard deviation shock to the seventh factor in a Cholesky decomposition ordered second given December $1980$ conditions. The first column shows the level factor responses and the second column the log volatility responses. The line is the posterior median, the dark shaded area is the $68\%$ credible set and the light shaded area is the $90\%$ credible set.} 
\end{figure}

Figures \ref{fig:factor7shock_factor_vol} and \ref{fig:factor7shock_factor_vol_appendix} show the responses of the shock to the inflation factor. The inflation factor itself sharply increases before unwinding most of its initial impact and persistently returning to baseline. The consumption and credit factors immediately decline, while the financial factor modestly increases. The Treasury factor increases in response to the increase in inflation, while the stock return and housing consumption factors decline on impact. Volatilities of the innovations across all factors increase following an inflation factor shock, reflecting the destabilizing effects of high inflation.

\clearpage

\section{Details of the Scenario Analysis \label{app:historical}}

\begin{figure}[b]
\caption{Great Inflation Scenario: Shutting Off Inflation Shocks}
\label{fig:scen_GI_noinfl_appendix}
\begin{center}
   
\begin{tabular}{@{}c@{}}
\includegraphics[width=0.85\textwidth]{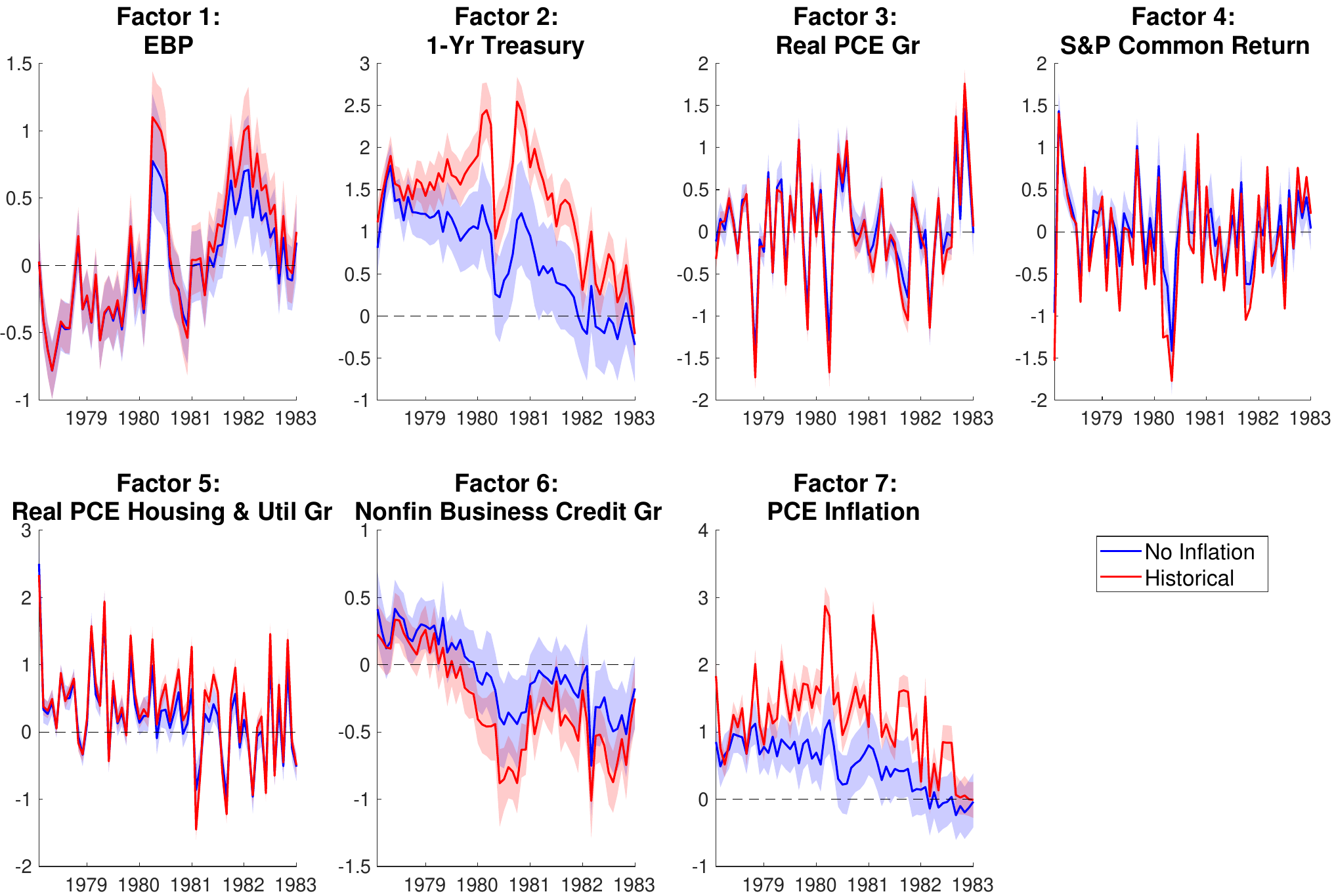} \\[0.3cm]
\includegraphics[width=0.85\textwidth]{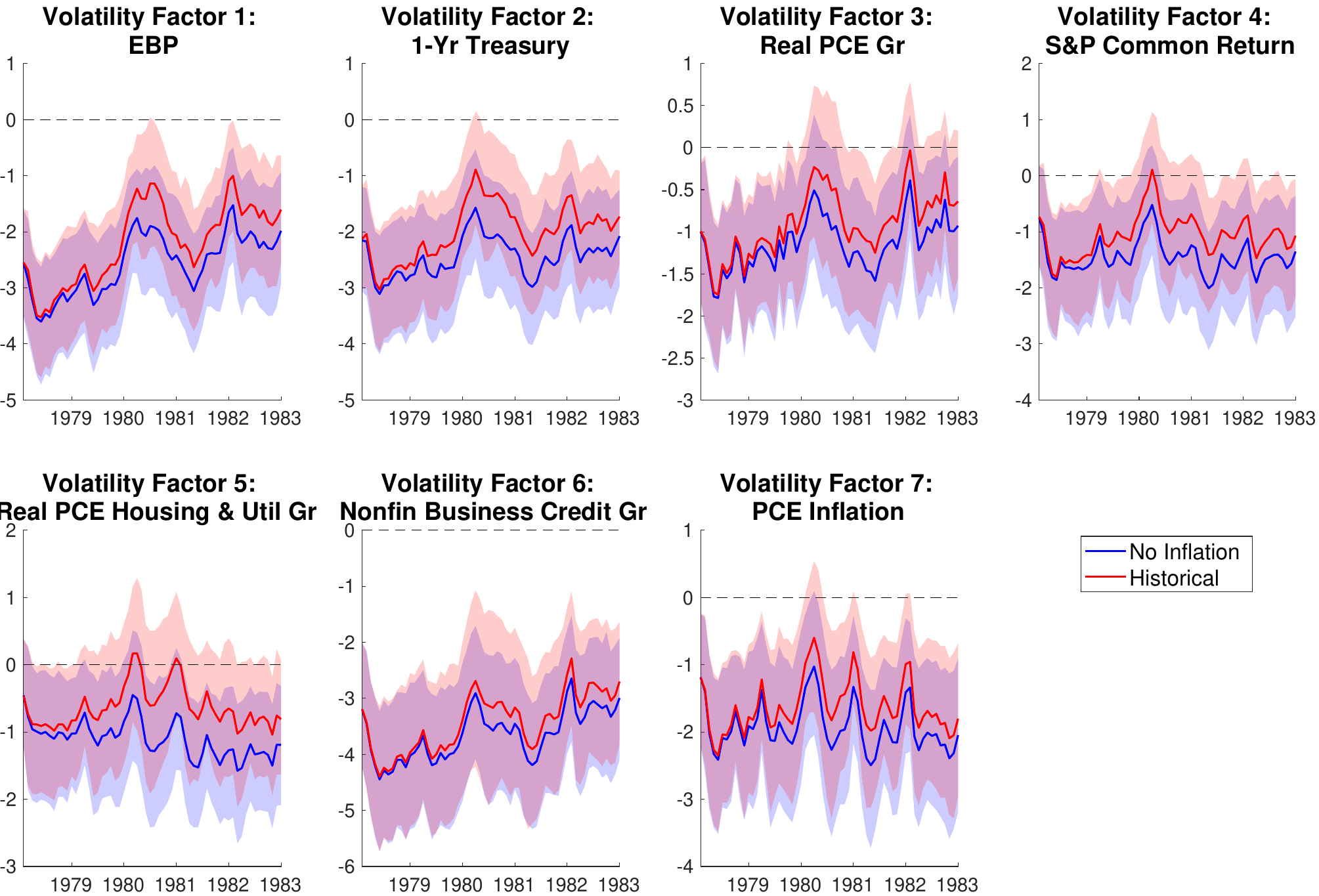}
\end{tabular}
\end{center}

\small{Note: This figure shows the historical evolution of the factors (top) and volatilities (bottom) from January 1978--December 1982. The red lines are the historical estimated evolution of the factors and volatilities along with 90\% credible sets. The blue lines are counterfactuals assuming no shocks to the seventh factor ordered second in a Cholesky decomposition occurred over the period.} 
\end{figure}

\begin{figure}[b]
\caption{GFC Scenario: Shutting Off Financial Shocks}
\label{fig:scen_GFC_nofin_appendix}
\begin{center}
  
\begin{tabular}{@{}c@{}}
\includegraphics[width=0.85\textwidth]{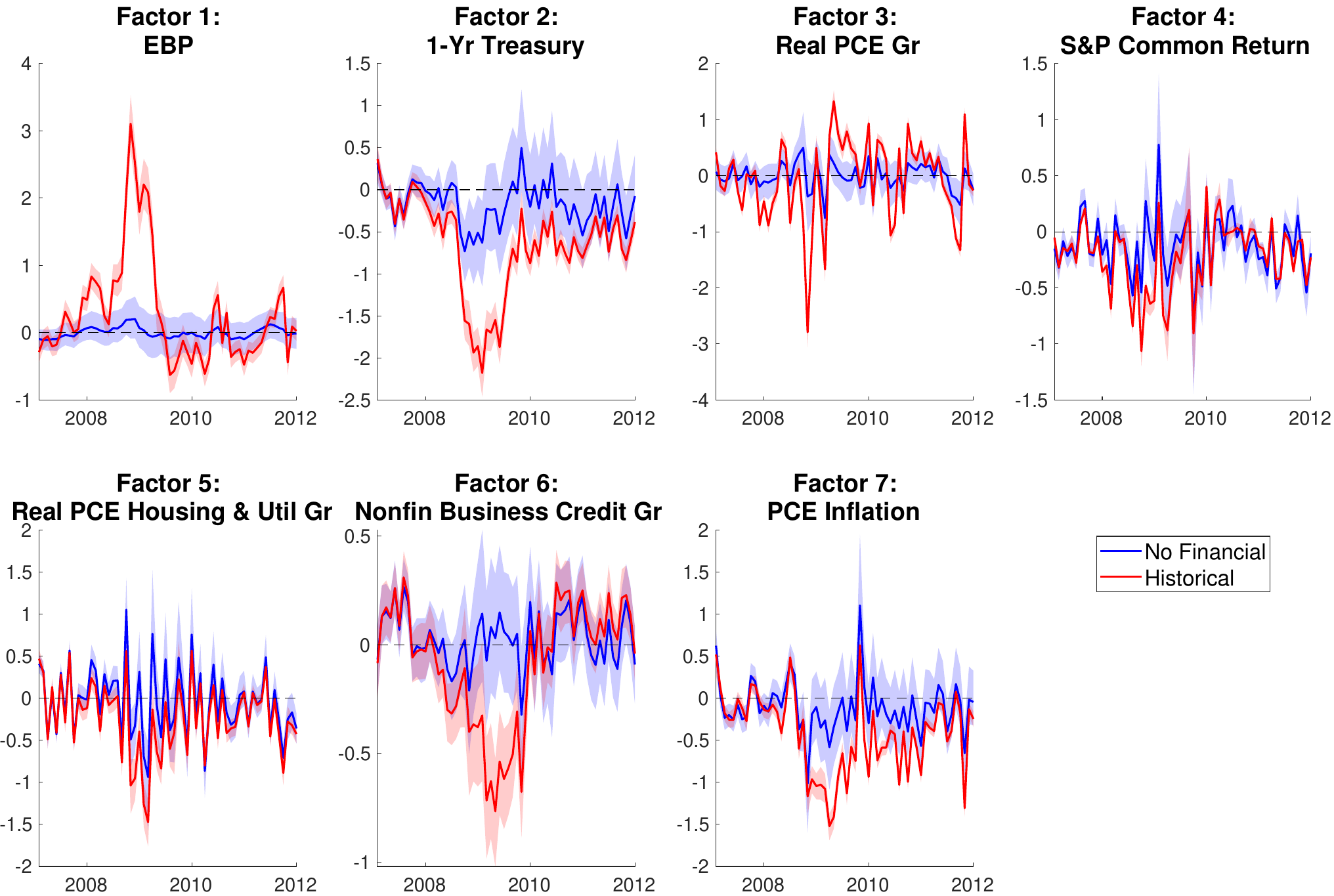} \\[0.3cm]
\includegraphics[width=0.85\textwidth]{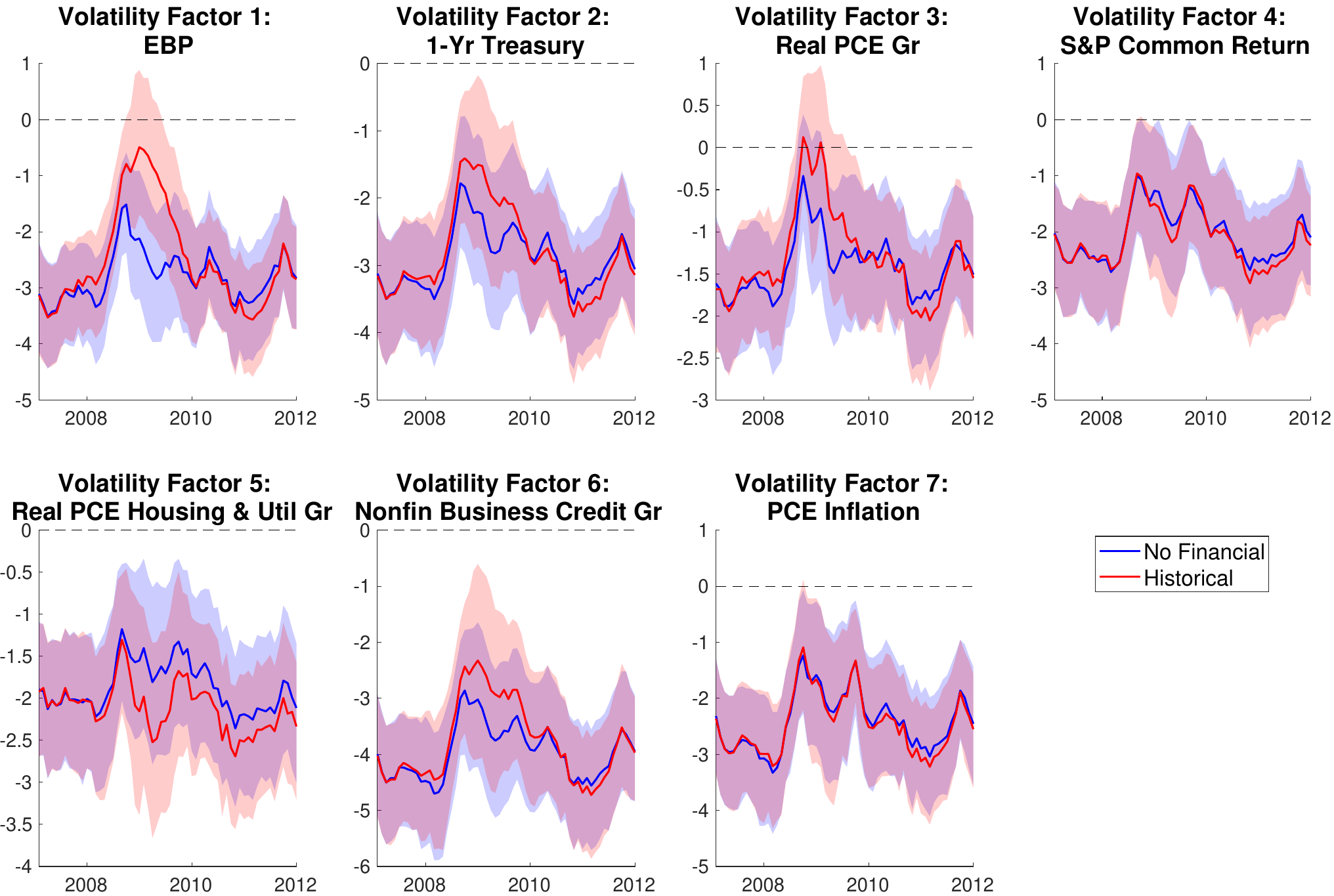}
\end{tabular}
\end{center}

\small{Note: This figure shows the historical evolution of the factors (top) and volatilities (bottom) from January 2007--December 2011. The red lines are the historical estimated evolution of the factors and volatilities along with 90\% credible sets. The blue lines are counterfactuals assuming no shocks to the first factor ordered first in a Cholesky decomposition occurred over the period.} 
\end{figure}

In this section, we discuss how we implement the scenario analysis exercise in Section \ref{subsec:scen_analysis}. The scenario runs from January 1978 - December 1982 for the Great Inflation and from January 2007 - December 2011 for the GFC. We begin with the historical smoothed estimates of the factors and volatilities produced by our estimation, which are shown in the red lines in Figures \ref{fig:scen_GI_noinfl_appendix} and \ref{fig:scen_GFC_nofin_appendix}. Then, we form counterfactual paths (blue lines in the figures) by setting either the realized shocks to the financial factor (factor 1) to 0 or shocks to the inflation factor (factor 7) to 0, keeping the other shocks as before. We use a Cholesky ordering with shocks to the financial factor ordered first and shocks to the inflation factor ordered second.

We generate $12-$month ahead conditional distributions of the sectoral prices given either the historical or counterfactual states assuming that all shocks are active. We then compute the tail variability statistics shown in Figure \ref{fig:price_ratios}.

\clearpage

\section{\label{appsec: additional figures}Additional Tables and Figures}

\begin{figure}[h!]
\caption{Aggregate Risk Indices to 2019: Growth, Inflation, and Financial Conditions}
\label{fig:agg_risk_indices_through2019}
\includegraphics[width=\textwidth]{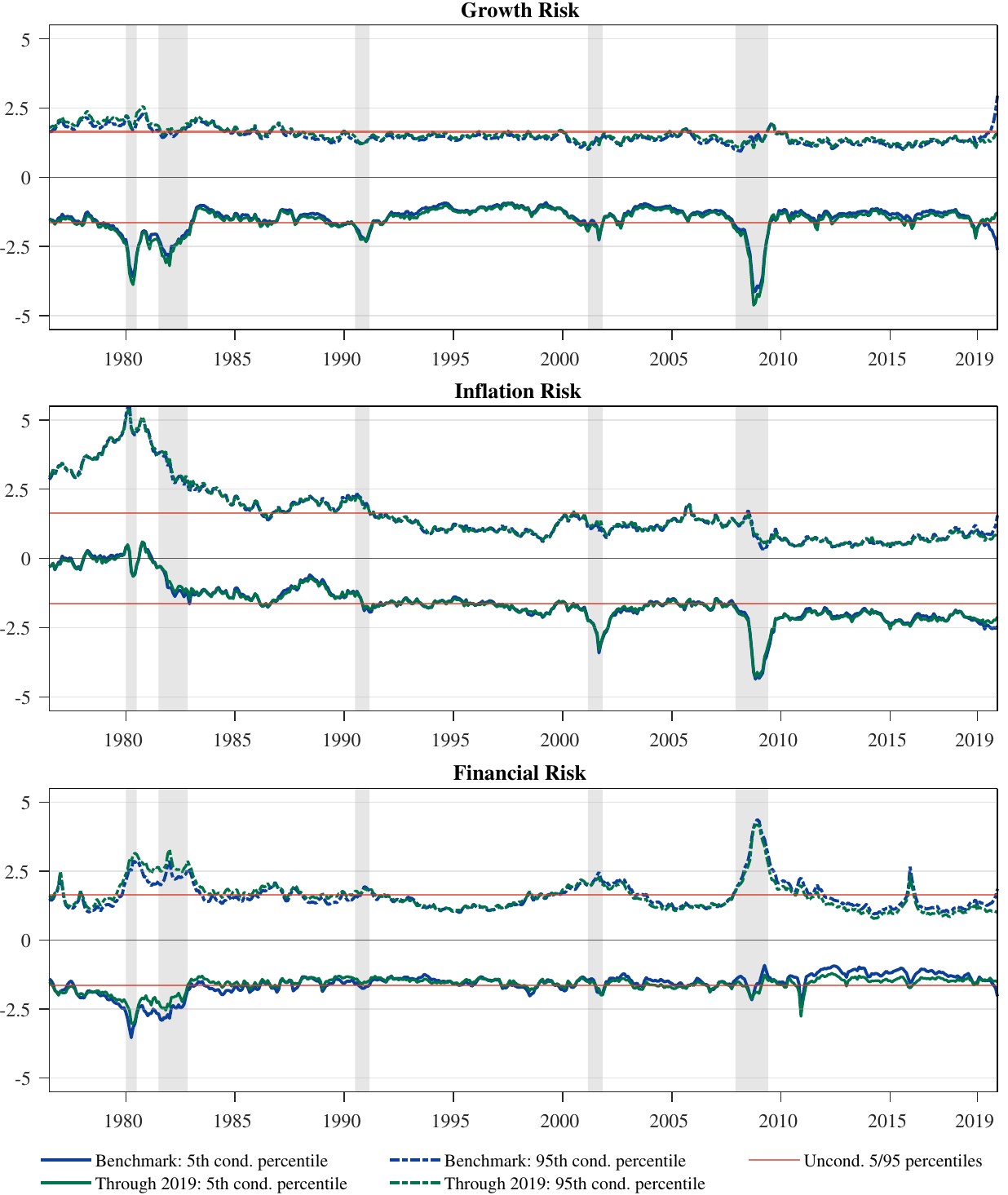}

\footnotesize{Note: This figure shows aggregate risk indices for economic growth, inflation, and financial conditions. Each series is constructed as the cross-sectional average of the $5th$ and $95th$ percentiles of the conditional distributions implied by the model for a group of related indicators. The blue solid and dashed lines report the averages of the $5th$ and $95th$ percentiles, respectively, for the model estimated on the full-sample and normalized with data through 2019. The green solid and dashed lines report the same for the model estimated on data through 2019. The red solid lines report as reference the $5th$ and $95th$ percentiles of a standard normal distribution. See Section~\ref{subsec: agg-indexes-construction} for details about the construction of the indices. Shaded areas denote NBER recessions.}
\end{figure}

\begin{figure}[b]
\caption{Smoothed Estimates of the Factors and Volatilities}
\label{fig:factor_estim_appendix}
\includegraphics[width=\textwidth,trim=0cm 6.2cm 0cm 0cm, clip,]{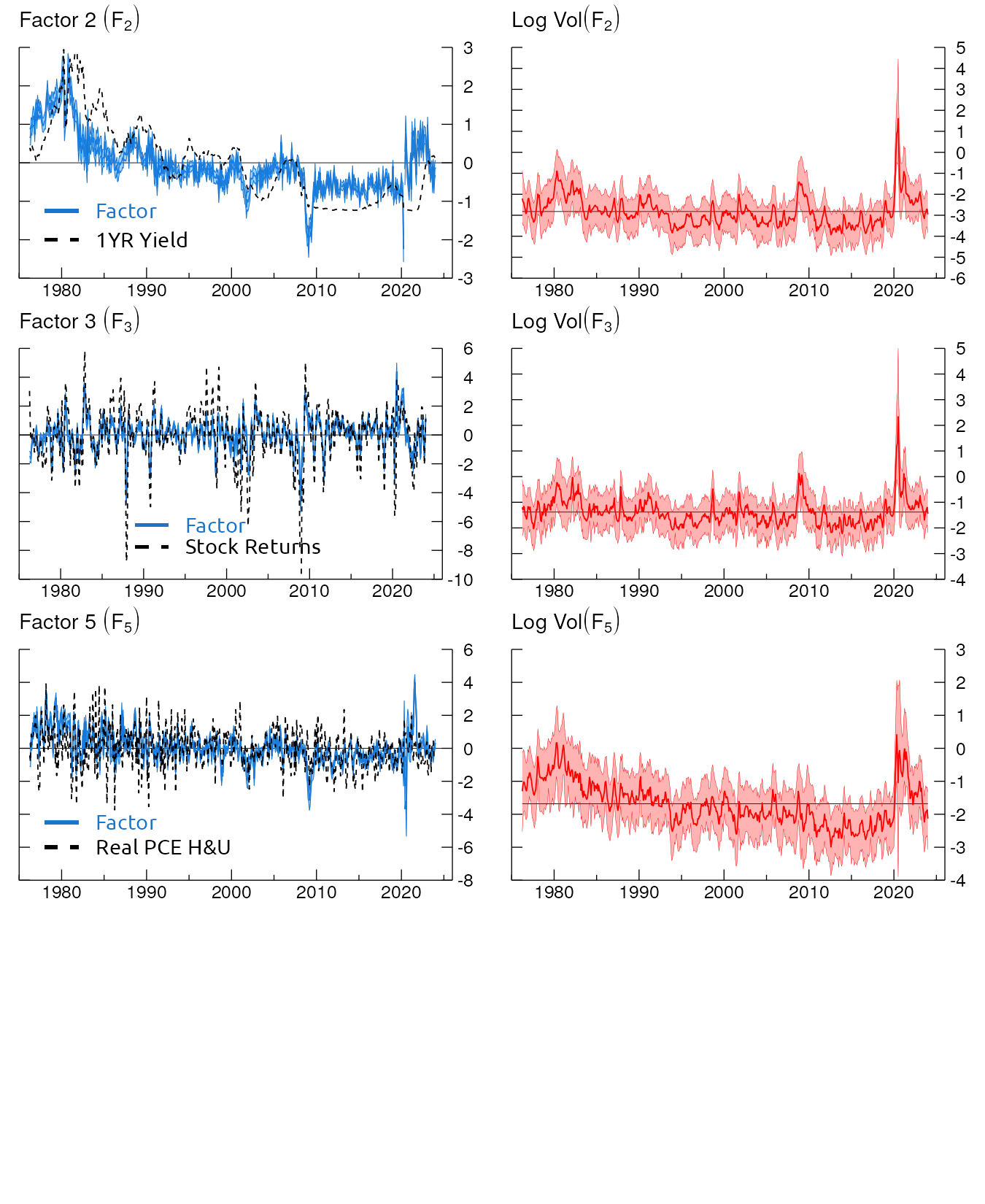}
\small{Note: The figure shows posterior median estimates of the level factors (blue lines) and of the log volatilities (red lines). Shaded areas denote the $5$th and $95$th of the posterior distributions. The dashed lines depict the observable variable associated to the factor for the normalization of matrix $B$. See Section~\ref{sec:model} for details on the normalization.}
\end{figure}

\begin{figure}[h!]
    
    \begin{center}        
        \caption{Correlation of Factors and Log Volatilities.\label{fig:corr-structure-factors}}
    % First subplot
    \begin{subfigure}[b]{\textwidth}
        \centering
        \includegraphics[width=\textwidth]{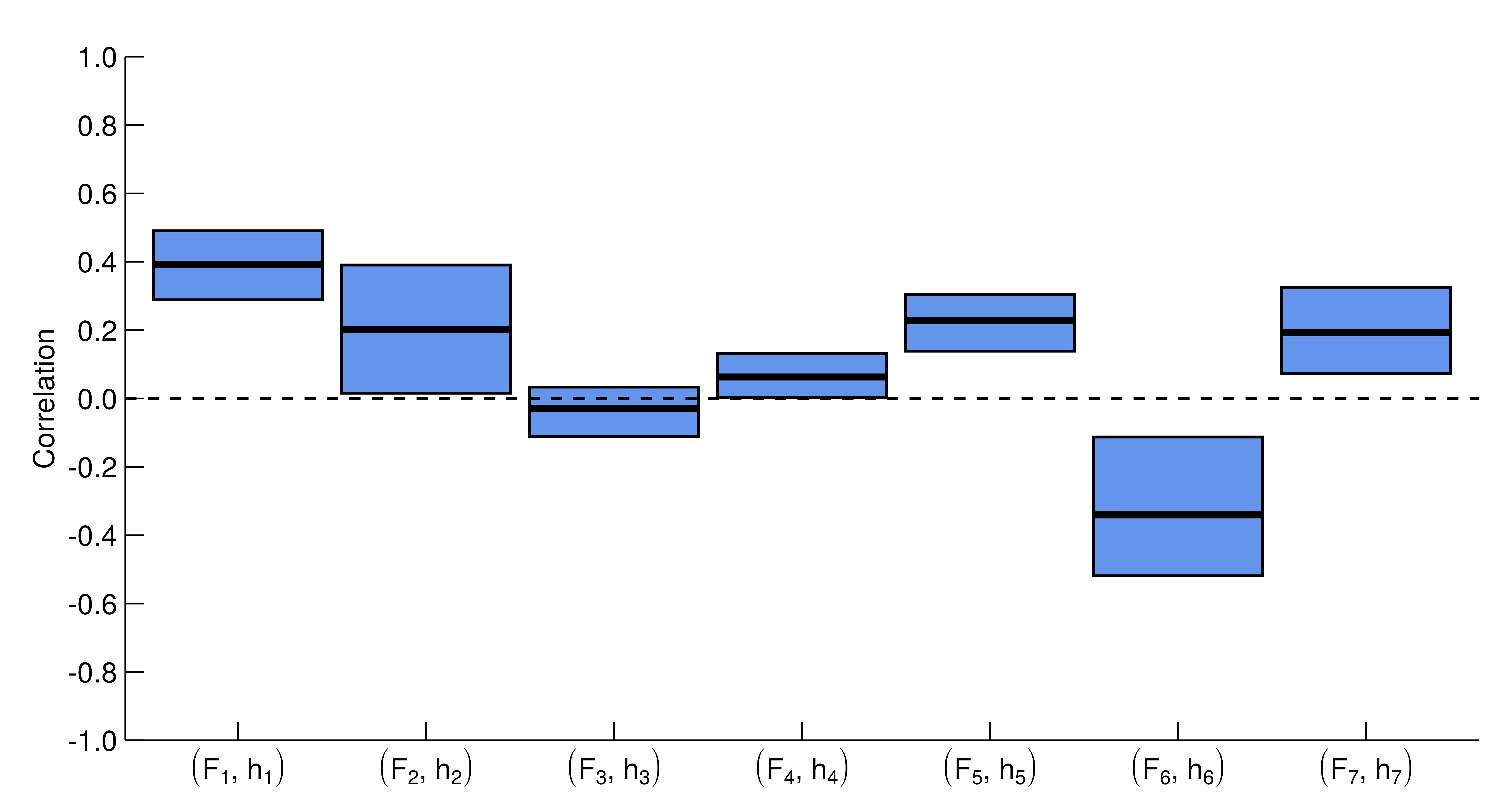}
        \caption{Correlation within Factors: Full Sample.}
        \label{fig:factors-boxplotMeanVol}
    \end{subfigure}
    \hfill
    % Second subplot
    \begin{subfigure}[b]{\textwidth}
        \centering
        \includegraphics[width=\textwidth]{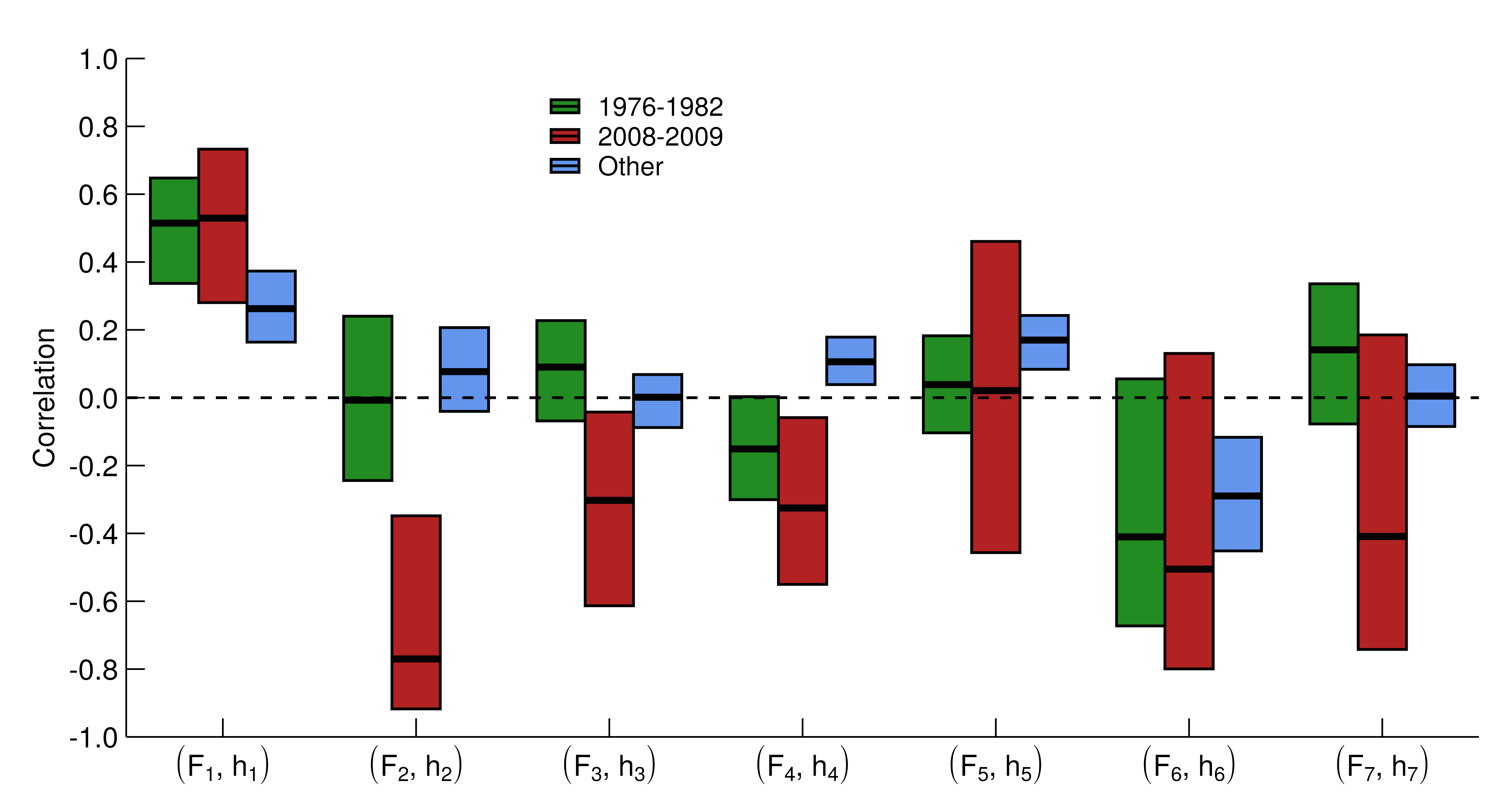}
        \caption{Correlation within Factors: Subsamples.}
        \label{fig:factors-heatmap-meanvol}
    \end{subfigure}
    \end{center}
    \small{Note: Panel~(a) shows the pairwise correlation coefficients of the estimated factors $F_x$ and the associated log volatilities $h_x$. The black lines depict the median correlation, while the boxes depict the min and max elements of the posterior set. Panel~(b) shows pairwise correlations across three subsamples: July 1976--Dec 1982, GFC (Dec 2007- June 2009), and remaining months. The boxes depict 70~percent posterior credible sets.}
\end{figure}

\begin{figure}[b]
\caption{Sectoral Factor Loadings for Industrial Production}
\label{fig:factorload_ip_appendix}
\begin{center}
\includegraphics[width = \textwidth]{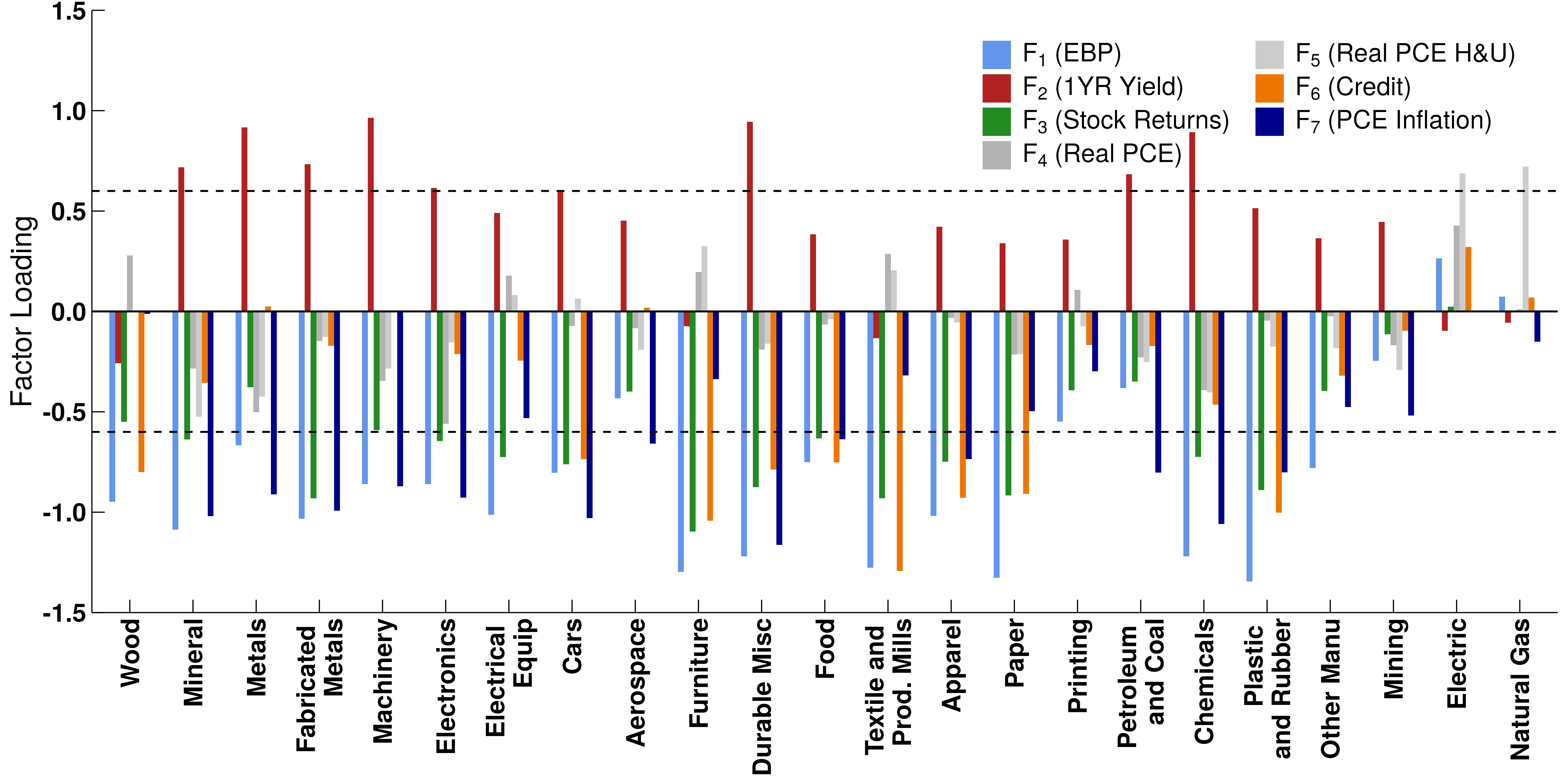}
\end{center}
\vspace{0.5cm}

\small{Note: This figure shows the median factor loading estimates for each IP sector. Factors 1, 2, 3, 6, and 7 are colored, representing the five most important factors The two dashed lines denote $0.6$ and $-0.6$.} 
\end{figure}

This appendix presents additional tables and figures referenced in the main text: Comparison of aggregate risk indices through 2019 (Figure \ref{fig:agg_risk_indices_through2019}); Top 5 Aggregate Indicators by Tail Asymmetry (Table \ref{tab:top_variablesrisk}); smoothed estimates and factor volatilities for factors 2, 3, and 5 (Figure~\ref{fig:factor_estim_appendix}); correlation of factors and log volatilities for full sample and subsamples (Figure~\ref{fig:corr-structure-factors}); and sectoral factor loadings for industrial production (Figure~\ref{fig:factorload_ip_appendix}).

\clearpage

\clearpage

\begin{table}[htbp]
\begin{center} 
\caption{Top 5 Aggregate Indicators by Tail Asymmetry}
\label{tab:top_variablesrisk}
\begin{tabular}{lcc}
    \toprule
    Category & Indicator & Asymmetry (Log($\sigma_{P95}/\sigma_{P5}$))  \\
    \midrule
     Growth & IP Index & -0.98 \\
     & Real Manufacturing and Trade Industries Sales & -0.93   \\
     & Initial Claims & -0.90  \\
     & New Orders for Durable Goods & -0.78 \\
     & All Employees: Total nonfarm & -0.69  \\ 
    \midrule
    Financial & VIX & 0.79 \\
     & Excess Bond Premium & 0.57  \\
     & Fed Res 10-year High Quality Corporate Bond Spread & 0.31  \\
     & 1-Year Treasury Rate & 0.29  \\
     & Baa - 10-Year Treasury & 0.26  \\ 
    \bottomrule
\end{tabular}

\end{center}

\small{Note: The table shows the top 5 aggregate growth and financial indicators with the most pronounced tail asymmetries, measured as the log ratio of the conditional 95th and 5th percentile standard deviations from July 1976 - June 2019. For growth variables, more negative asymmetry values indicate more downside risk. For financial variables, more positive values indicate more upside (adverse) risk.}
\end{table}

\clearpage

\clearpage

\clearpage

\clearpage

\clearpage

\section{\label{appsec: fit robustness}Additional Results for Model Fit and Robustness}

This appendix reports the detailed results underlying the model fit and calibration analysis in Section~\ref{sec: robustness}. Table~\ref{tab:sddecomps} reports the variance decomposition discussed in Section~\ref{subsec: sddecomp}. Table~\ref{tab:factor_qr_qwcrps_ratios} reports, variable by variable, the quantile-weighted CRPS ratios and tail coverage rates plotted in Figure~\ref{fig:factor_qr_qwcrps_ratios}. Table~\ref{tab:qwcrps} reports a detailed comparison with quantile regressions for four representative series, in sample and out of sample, across horizons and both tails.

\clearpage

\begin{table}[htbp]
\begin{center}
\caption{Volatility Decomposition: Common and Idiosyncratic Components\label{tab:sddecomps}}
\footnotesize
\resizebox{0.9\textwidth}{!}{
\begin{tabular}{lccccc}
    \toprule
    &  &  \multicolumn{4}{c}{Standard Deviation (Percent)} \\
    \midrule
    Variable & Category & Data & Common & Idio & R$^2$ \\
    \midrule
    \multicolumn{6}{l}{\textit{Panel A: Aggregate Series}} \\
    \midrule
     IP &  Total &  0.97 &  0.92 &  0.28 &  0.90 \\    
    \midrule 
    Real PCE & Total & 0.86 & 0.67 & 0.32 & 0.62 \\
    \midrule
    PCE Inflation & Total & 0.25 & 0.17 & 0.18 & 0.48 \\
     & Services & 0.21 & 0.18 & 0.09 & 0.74 \\
    \midrule
    Excess bond premium & -  & 0.55 & 0.25 & 0.43 & 0.21 \\
    Equity Prices & S\&P Common & 3.61 & 2.13 & 2.69 & 0.35 \\    
    Credit & Total Non-Rev. &  0.68 & 0.21 & 0.65 & 0.10  \\
    \midrule
    \multicolumn{6}{l}{\textit{Panel B: Sectoral Series}} \\
    \midrule
     IP & Average & 2.17 & 1.20 & 1.70 & 0.34 \\
     & Median & 1.77 & 0.96 & 1.30 & 0.38 \\
     & Minimum & 0.95 & 0.47 & 0.65 & 0.08 \\
     & Maximum & 8.49 & 5.57 & 5.49 & 0.72 \\
     & Standard Dev & 1.68 & 1.03 & 1.16 & 0.17 \\ 
    \midrule
    Real PCE & Average & 2.07 & 0.89 & 1.65 & 0.20 \\
     & Median & 2.03 & 0.50 & 1.65 & 0.15 \\
     & Minimum & 0.72 & 0.13 & 0.63 & 0.01 \\
     & Maximum & 5.74 & 3.59 & 4.27 & 0.44 \\
     & Standard Dev & 1.23 & 0.85 & 0.89 & 0.14 \\ 
    \midrule
    PCE Inflation & Average & 0.72 & 0.25 & 0.67 & 0.19 \\
     & Median & 0.47 & 0.18 & 0.44 & 0.17 \\
     & Minimum & 0.24 & 0.06 & 0.19 & 0.05 \\
     & Maximum & 4.29 & 0.95 & 4.06 & 0.33 \\
     & Standard Dev & 0.98 & 0.21 & 0.93 & 0.09 \\ 
    \midrule
    Equity Prices & Average & 5.28 & 4.37 & 2.76 & 0.70 \\
     & Median & 5.09 & 4.71 & 2.59 & 0.76 \\
     & Minimum & 3.94 & 2.52 & 1.22 & 0.38 \\
     & Maximum & 6.56 & 5.47 & 5.26 & 0.92 \\
     & Standard Dev & 0.79 & 0.91 & 1.15 & 0.20 \\
    \bottomrule
\end{tabular}
}
\end{center}
{\small Note: Standard deviations are in percent. For each series, this table shows the standard deviation of (i) the data, (ii) implied by conditioning only on the common component spanned by the factors, (iii) implied by only the idiosyncratic component, and the $R^2$ of the common component, calculated as the variance of the common component divided by variance of the data. Panel A reports selected aggregate series; Panel B reports summary statistics across sectoral series.}
\end{table}

\clearpage

% Requires \usepackage{booktabs,longtable,array,siunitx}
% Citation keys can be changed in the MATLAB user settings.
\begingroup\let\footnotesize\scriptsize
\setlength{\tabcolsep}{3pt}
\begin{longtable}{@{}>{\footnotesize\raggedright\arraybackslash}p{0.325\textwidth} >{\footnotesize}S[table-format=1.2,table-number-alignment=center]@{\hspace{0.12em}}>{\footnotesize}l >{\footnotesize}S[table-format=1.2,table-number-alignment=center,detect-weight=true]@{\hspace{1.4em}}>{\footnotesize\raggedright\arraybackslash}p{0.325\textwidth} >{\footnotesize}S[table-format=1.2,table-number-alignment=center]@{\hspace{0.12em}}>{\footnotesize}l >{\footnotesize}S[table-format=1.2,table-number-alignment=center,detect-weight=true]@{}}
\caption{Factor model to quantile-regression qwCRPS ratios and factor model empirical coverage by variable}\label{tab:factor_qr_qwcrps_ratios}\\
\toprule
Full name & \multicolumn{2}{c}{\footnotesize\shortstack{qwCRPS\\ratio}\hspace{-0.6em}} & {\footnotesize\shortstack{Cover-\\age}} & Full name & \multicolumn{2}{c}{\footnotesize\shortstack{qwCRPS\\ratio}\hspace{-0.6em}} & {\footnotesize\shortstack{Cover-\\age}} \\
\midrule
\endfirsthead
\multicolumn{8}{c}{\footnotesize \tablename\ \thetable{} -- continued}\\
\toprule
Full name & \multicolumn{2}{c}{\footnotesize\shortstack{qwCRPS\\ratio}\hspace{-0.6em}} & {\footnotesize\shortstack{Cover-\\age}} & Full name & \multicolumn{2}{c}{\footnotesize\shortstack{qwCRPS\\ratio}\hspace{-0.6em}} & {\footnotesize\shortstack{Cover-\\age}} \\
\midrule
\endhead
\midrule
\multicolumn{8}{r}{\footnotesize Continued on next page}\\
\endfoot
\bottomrule
\endlastfoot
\addlinespace[0.5em]
\multicolumn{8}{l}{\footnotesize\textit{Growth variables: 12-month lower tail}}\\*
Real personal income ex transfer receipts & 1.00 & {} & 0.04 & IP: Chemicals (NAICS = 325) & 1.06 & {} & 0.04 \\
Real Manu. and Trade Industries Sales & 1.00 & {} & 0.05 & IP: Plastics and Rubber Products (NAICS = 326) & 0.97 & {} & 0.05 \\
Initial Claims & 0.94 & {} & \bfseries 0.00 & IP: Other Manufacturing (Non-NAICS) (NAICS = 1133, 5111) & 0.97 & {} & 0.06 \\
All Employees: Total nonfarm & 0.95 & {} & 0.03 & IP: Mining (NAICS = 21) & 0.97 & {} & 0.04 \\
Housing Starts: Total New Privately Owned & 1.06 & {} & 0.08 & IP: Utilities (NAICS = 2211, 2212) & 1.11 & {} & \bfseries 0.00 \\
New Orders for Durable Goods & 0.95 & {} & 0.03 & IP: Electric (NAICS = 2211) & 1.03 & {} & \bfseries 0.01 \\
Total Business: Inventories to Sales Ratio & 1.06 & {} & \bfseries 0.00 & IP: Natural Gas (NAICS = 2212) & 1.14 & $^{*}$ & \bfseries 0.01 \\
IP Index & 0.94 & {} & 0.05 & Goods & 0.98 & {} & 0.05 \\
Real personal consumption expenditures & 1.00 & {} & 0.05 & Durable goods & 1.00 & {} & 0.04 \\
IP: Manufacturing (SIC) & 0.94 & {} & 0.04 & Motor vehicles and parts & 1.06 & {} & 0.03 \\
IP: Manufacturing (NAICS) & 0.94 & {} & 0.04 & Furnishings and durable household equipment & 0.95 & {} & 0.05 \\
IP: Durable Manufacturing & 0.94 & {} & 0.04 & Recreational goods and vehicles & 1.05 & {} & 0.05 \\
IP: Wood Products (NAICS = 321) & 1.01 & {} & 0.09 & Other durable goods & 0.98 & {} & 0.02 \\
IP: Nonmetallic Mineral Products (NAICS = 327) & 1.01 & {} & 0.04 & Nondurable goods & 0.95 & $^{*}$ & 0.05 \\
IP: Primary Metals (NAICS = 331) & 1.17 & {} & 0.06 & Food and beverages purchased for off-premises consumption & 0.99 & {} & 0.04 \\
IP: Fabricated Metal Products (NAICS = 332) & 0.97 & {} & 0.08 & Clothing and footwear & 1.00 & {} & 0.02 \\
IP: Machinery (NAICS = 333) & 1.00 & {} & 0.06 & Gasoline and other energy goods & 1.01 & {} & 0.03 \\
IP: Computer and Electronic Products (NAICS = 334) & 0.99 & {} & 0.05 & Other nondurable goods & 1.03 & {} & 0.06 \\
IP: Electrical Equip, Appliances, and Components (NAICS = 335) & 1.02 & {} & 0.05 & Services & 0.91 & {} & 0.04 \\
IP: Motor Vehicles and Parts (NAICS = 3361, 3362, 3363) & 1.12 & {} & \bfseries 0.02 & Household consumption expenditures (for services) & 0.92 & {} & 0.03 \\
IP: Aerospace and Misc Transportation Equipment (NAICS = 3364-3369) & 1.06 & {} & 0.03 & Housing and utilities & 1.09 & {} & \bfseries 0.01 \\
IP: Furniture and Related Products (NAICS = 337) & 0.94 & $^{*}$ & 0.04 & Health care & 1.18 & {} & 0.02 \\
IP: Durable Misc (NAICS = 339) & 0.95 & {} & 0.02 & Transportation services & 0.96 & {} & 0.06 \\
IP: Nondurable Manufacturing & 0.98 & {} & 0.04 & Recreation services & 0.93 & {} & 0.03 \\
IP: Food, Beverage, and Tobacco Products (NAICS = 311, 312) & 1.06 & {} & \bfseries 0.02 & Food services and accommodations & 1.01 & {} & 0.03 \\
IP: Textile and Product Mills (NAICS = 313, 314) & 0.97 & {} & 0.06 & Financial services and insurance & 1.00 & {} & 0.05 \\
IP: Apparel and Leather (NAICS = 315, 316) & 0.96 & {} & 0.07 & Other services & 1.03 & {} & 0.08 \\
IP: Paper (NAICS = 322) & 0.97 & {} & 0.03 & Final consumption expenditures of NPISHs & 0.99 & {} & 0.03 \\
IP: Printing and Support (NAICS = 323) & 0.85 & $^{*}$ & 0.04 & Gross output of nonprofit institutions & 1.20 & $^{*}$ & 0.07 \\
IP: Petroleum and Coal Products (NAICS = 324) & 1.06 & {} & 0.04 & Less: Receipts from sales of goods and services by nonprofit inst & 1.17 & $^{*}$ & \bfseries 0.02 \\
\addlinespace[0.5em]
\multicolumn{8}{l}{\footnotesize\textit{Inflation variables: 12-month upper tail}}\\*
PCE: Chain Index Prices & 0.83 & $^{*}$ & 0.94 & Services & 0.63 & $^{*}$ & \bfseries 0.99 \\
Goods & 0.80 & $^{*}$ & 0.94 & Household consumption expenditures (for services) & 0.62 & $^{*}$ & \bfseries 1.00 \\
Durable goods & 0.83 & $^{*}$ & 0.94 & Housing and utilities & 0.89 & $^{*}$ & \bfseries 0.99 \\
Motor vehicles and parts & 0.86 & {} & 0.96 & Health care & 0.92 & {} & \bfseries 0.99 \\
Furnishings and durable household equipment & 0.68 & $^{*}$ & 0.96 & Transportation services & 0.76 & $^{*}$ & 0.97 \\
Recreational goods and vehicles & 0.85 & $^{*}$ & 0.95 & Recreation services & 0.61 & $^{*}$ & 0.97 \\
Other durable goods & 0.82 & $^{*}$ & 0.93 & Food services and accommodations & 0.75 & $^{*}$ & 0.94 \\
Nondurable goods & 0.86 & {} & 0.95 & Financial services and insurance & 1.27 & {} & \bfseries 1.00 \\
Food and beverages purchased for off-premises consumption & 0.88 & $^{*}$ & 0.92 & Other services & 0.79 & $^{*}$ & \bfseries 1.00 \\
Clothing and footwear & 1.01 & {} & 0.96 & Final consumption expenditures of NPISHs & 0.95 & {} & 0.97 \\
Gasoline and other energy goods & 0.98 & {} & 0.97 & Gross output of nonprofit institutions & 0.74 & $^{*}$ & 0.96 \\
Other nondurable goods & 0.78 & $^{*}$ & 0.97 & Less: Receipts from sales of goods and services by nonprofit inst & 0.93 & {} & \bfseries 0.99 \\
\addlinespace[0.5em]
\multicolumn{8}{l}{\footnotesize\textit{Financial variables: 3-month upper tail}}\\*
Excess Bond Premium & 1.06 & {} & 0.96 & Barron's Best Grade Bond Yield - 10-Year Treasury & 1.04 & {} & 0.96 \\
1-Year Treasury Rate & 1.72 & $^{*}$ & \bfseries 1.00 & Equities: Consumer Nondurables & 1.01 & {} & \bfseries 0.98 \\
Commercial and Industrial Loans & 0.93 & {} & 0.96 & Equities: Consumer Durables & 0.91 & {} & \bfseries 0.99 \\
Real Estate Loans at All Commercial Banks & 0.90 & $^{*}$ & \bfseries 0.98 & Equities: Manufacturing & 0.98 & {} & \bfseries 0.99 \\
Total Nonrevolving Credit & 0.83 & $^{*}$ & \bfseries 0.98 & Equities: Energy & 1.02 & {} & \bfseries 0.99 \\
Nonrevolving consumer credit to Personal Income & 0.93 & {} & \bfseries 0.99 & Equities: Business Equipment & 0.93 & $^{*}$ & \bfseries 0.98 \\
Consumer Motor Vehicle Loans Outstanding & 0.88 & $^{*}$ & \bfseries 0.99 & Equities: Telecommunications & 0.99 & {} & \bfseries 0.98 \\
Total Consumer Loans and Leases Outstanding & 0.86 & $^{*}$ & \bfseries 0.98 & Equities: Shops & 0.96 & {} & \bfseries 0.98 \\
Securities in Bank Credit at All Commercial Banks & 0.90 & $^{*}$ & \bfseries 0.97 & Equities: Healthcare & 0.99 & {} & \bfseries 0.98 \\
S\&P's Common Stock Price Index: Composite & 0.95 & $^{*}$ & \bfseries 0.98 & Equities: Utilities & 1.04 & {} & \bfseries 0.98 \\
VIX & 0.96 & {} & 0.94 & Equities: Other & 0.97 & {} & \bfseries 0.99 \\
3-Month Treasury Bill & 0.89 & {} & \bfseries 0.98 & Equities: Size, Bottom 20\% & 0.94 & $^{*}$ & \bfseries 0.98 \\
10-Year Treasury C Minus FEDFUNDS & 0.99 & {} & 0.97 & Equities: Size, 21\%-40\% & 0.96 & {} & \bfseries 0.99 \\
Baa - 10-Year Treasury & 0.98 & {} & 0.95 & Equities: Size, 41\%-60\% & 0.96 & {} & \bfseries 0.99 \\
Dow Jones Corporate Bond - 10-Year Treasury & 1.00 & {} & 0.96 & Equities: Size, 61\%-80\% & 0.96 & {} & \bfseries 0.99 \\
Federal Reserve 10-year High Quality Corporate Bond - 10-Year Treasury & 0.99 & {} & 0.95 & Equities: Size, Top 20\% & 0.96 & {} & \bfseries 0.99 \\
\end{longtable}
\setlength{\tabcolsep}{6pt}
\endgroup
{\footnotesize\noindent Notes: Growth uses the 12-month lower tail, inflation uses the 12-month upper tail, and financial variables use the 3-month upper tail. Results are based on in-sample conditional distributions computed monthly from July 1976 through September 2023 (3 months ahead) and December 2022 (12 months ahead). The qwCRPS ratio is mean qwCRPS for the factor model divided by mean qwCRPS for quantile regression; values below one favor the factor model. $^{*}$ denotes rejection of equal predictive accuracy at the 5\% level using the two-sided Diebold--Mariano test \citep{DieboldMariano1995}. The long-run variance for that test uses the VAR(1)-prewhitened quadratic-spectral HAC estimator with automatic bandwidth selection \citep{Andrews1991,AndrewsMonahan1992}. Factor coverage is the empirical frequency with which the realization falls below the factor model quantile at the same tail and horizon. The nominal coverage rates are 0.05 for the lower tail and 0.95 for the upper tail. Bold factor-coverage entries have 95\% confidence intervals that exclude the corresponding nominal coverage rate. These are Wilson score intervals adjusted for serial correlation induced by overlapping forecasts. The adjustment estimates the long-run variance of the quantile-hit sequence using a Bartlett HAC estimator with lag $h-1$ and uses it to reduce the effective sample size. When the hit sequence is uniformly zero or one, the adjustment borrows the median adjustment estimated from nondegenerate hit sequences, first from other quantiles on the same tail for the same variable, then from variables in the same category and tail, and finally from all variables on the same tail; a horizon-based adjustment is used only if no valid donor is available.}

\clearpage

\begin{landscape}
\begin{table}[htbp]
    \begin{center}
    \caption{Quantile-Weighted CRPS Ratios: Factor Model vs.\ Quantile Regressions}
    \label{tab:qwcrps}

    \begin{tabular}{llccc@{\hskip 1.5em}ccc}
        \hline\hline
        & & \multicolumn{3}{c}{In-Sample} & \multicolumn{3}{c}{Out-of-Sample} \\
        \cmidrule(lr){3-5} \cmidrule(lr){6-8}
        & & Full Sample & Recession & Ex-Recession & Full Sample & Crisis & Post-Crisis \\
        & & (1976--2023) & & & (2007--2019) & (2007--2009) & (2009--2019) \\
        \hline
        \multicolumn{8}{l}{\textit{Panel A: 3 Months Ahead}} \\
        \hline
        IP Growth       & Left  & $0.84$ & $0.80$ & $0.85$ & $0.89$ & $0.91$ & $0.87$ \\
                        & Right & $0.77$ & $0.69$ & $0.80$ & $0.87$ & $0.83$ & $0.89$ \\
        \addlinespace
        Real PCE Growth & Left  & $0.92$ & $0.94$ & $0.91$ & $0.98$ & $0.93$ & $1.02$ \\
                        & Right & $0.82$ & $0.68$ & $0.86$ & $0.90$ & $0.93$ & $0.88$ \\
        \addlinespace
        Inflation       & Left  & $0.83$ & $0.83$ & $0.83$ & $0.88$ & $0.83$ & $0.94$ \\
                        & Right & $0.82$ & $0.86$ & $0.81$ & $0.94$ & $0.93$ & $0.96$ \\
        \addlinespace
        EBP             & Left  & $1.07$ & $1.08$ & $1.07$ & $1.01$ & $0.99$ & $1.03$ \\
                        & Right & $1.06$ & $1.06$ & $1.06$ & $1.14$ & $1.21$ & $1.07$ \\
        \hline
        \multicolumn{8}{l}{\textit{Panel B: 12 Months Ahead}} \\
        \hline
        IP Growth       & Left  & $0.94$ & $0.98$ & $0.93$ & $0.93$ & $1.03$ & $0.80$ \\
                        & Right & $0.93$ & $0.82$ & $0.96$ & $0.97$ & $1.07$ & $0.89$ \\
        \addlinespace
        Real PCE Growth & Left  & $1.00$ & $1.02$ & $1.00$ & $1.01$ & $1.09$ & $0.91$ \\
                        & Right & $0.89$ & $0.72$ & $0.95$ & $0.97$ & $1.03$ & $0.92$ \\
        \addlinespace
        Inflation       & Left  & $0.84$ & $0.85$ & $0.84$ & $0.76$ & $0.64$ & $0.92$ \\
                        & Right & $0.83$ & $0.74$ & $0.85$ & $0.77$ & $0.61$ & $0.89$ \\
        \addlinespace
        EBP             & Left  & $1.07$ & $1.03$ & $1.08$ & $0.98$ & $0.89$ & $1.15$ \\
                        & Right & $1.05$ & $0.98$ & $1.07$ & $1.03$ & $1.02$ & $1.06$ \\
        \hline\hline
    \end{tabular}
    \end{center}

    {\scriptsize Note: The table reports the ratio of the dynamic factor model's quantile-weighted continuous ranked probability scores (CRPS) to that of the quantile regression baseline (Factor/QR). Values below 1 indicate the factor model forecasts better (lower loss). Left and Right refer to left-tail and right-tail weighted CRPS, emphasizing downside and upside forecast accuracy, respectively. For the in-sample results, full sample spans July 1976 through September 2023 (3 months ahead) and December 2022 (12 months ahead). Recession and Ex-Recession split the sample by NBER recession months. Out-of-sample forecasts use a recursive expanding window with the first forecasts made using data estimated from July 1976 to December 2006, moving forward in three-month steps. The final forecasts use data through September 2019 (3 months ahead) and December 2018 (12 months ahead). Crisis is defined as forecasts made using models estimated through June 2009; Post-Crisis is thereafter.}
\end{table}
\end{landscape}

\clearpage

\end{document}